\documentclass[12pt,a4paper]{article}
\usepackage{geometry}
\usepackage[english]{babel}
\usepackage[labelfont=bf]{caption}
\RequirePackage[sort&compress,square,comma,numbers]{natbib}
\usepackage{bbm}
\usepackage{empheq}
\usepackage{comment}
\usepackage[utf8]{inputenc}
\usepackage{mathrsfs}
\usepackage[mathscr]{eucal}
\usepackage{amsmath}
\usepackage{amsfonts}
\usepackage[most]{tcolorbox}
\usepackage{amssymb}
\usepackage{graphicx}
\usepackage{scalerel}
\usepackage{multirow}
\usepackage{cancel}
\usepackage{slashed}
\usepackage{booktabs}
\usepackage{braket}
\usepackage{hyperref}

\allowdisplaybreaks 
\usepackage{fancyhdr}
\newcommand{\nn}{\nonumber}
\newcommand{\be}{\begin{equation}}
\newcommand{\ee}{\end{equation}}
\newcommand{\bea}{\begin{eqnarray}}
\newcommand{\eea}{\end{eqnarray}}

\newcommand{\ci}[2]{C^{#1}_{#2}}
\newcommand{\hatci}[2]{\hat{C}^{#1}_{#2}}
\newcommand{\ct}[2]{\xi^{#1}_{#2}}
\newcommand{\ccoeff}[2]{c^{#1}_{#2}}
\newcommand{\ctcoeff}[2]{\chi^{#1}_{#2}}
\newcommand{\cL}{\mathcal{L}}
\newcommand{\ba}{\begin{eqnarray}}
\newcommand{\ea}{\end{eqnarray}}
\newcommand{\no}{\nonumber}

\usepackage[shortlabels]{enumitem}
\setlist{itemsep=.1em,topsep=.5em}
\SetEnumerateShortLabel{i}{\textit{\roman*}}

\numberwithin{equation}{section}
\numberwithin{figure}{section}
\numberwithin{table}{section}

 \usepackage[noindentafter]{titlesec} 
\DeclareMathAlphabet{\mathsfit}{OT1}{lmss}{m}{sl}
\DeclareMathAlphabet{\mathsfbf}{OT1}{lmss}{bx}{n}
\DeclareMathAlphabet{\mathsfbfit}{OT1}{lmss}{bx}{sl}
\DeclareMathVersion{chaptermath}
\SetSymbolFont{operators}{chaptermath}{OT1}{cmbr}{m}{n}
\SetSymbolFont{letters}{chaptermath}{OML}{cmbrm}{b}{it}
\SetSymbolFont{symbols}{chaptermath}{OMS}{cmbrs}{m}{n}
\DeclareMathVersion{subsectionmath}
\SetSymbolFont{operators}{subsectionmath}{OT1}{cmbr}{m}{n}
\SetSymbolFont{letters}{subsectionmath}{OML}{cmbrm}{m}{it}
\SetSymbolFont{symbols}{subsectionmath}{OMS}{cmbrs}{m}{n}

\titleformat{\chapter}{\bfseries \huge  }{\thechapter. }{-1pt}{}{}
\titlespacing{\chapter}{0pt}{10pt plus 1pt minus 1pt}{10pt plus 1pt minus 1pt}

\titleformat{\section}{\large \bfseries   }{\thesection}{10pt}{}{}
\titlespacing{\section}{0pt}{15pt plus 1pt minus 1pt}{5pt plus 1pt minus 1pt}
\titleformat{\subsection}{\normalsize \bfseries  }{\thesubsection}{10pt}{}{}
\titlespacing{\subsection}{0pt}{15pt plus 1pt minus 1pt}{5pt plus 1pt minus 1pt}
\titleformat{\subsubsection}{\normalsize \bfseries    }{\thesubsubsection}{10pt}{}{}
\titlespacing{\subsubsection}{0pt}{15pt}{5pt}

\begin{document}
\renewcommand{\headrulewidth}{0pt}
\renewcommand{\footrulewidth}{0pt}

\begin{flushright}
MITP-22-036\\
\end{flushright}

\thispagestyle{empty}
\begin{center}
\vspace{3 cm}
{\Large\bf Gauge Invariance and Finite Counterterms\\ [0.2cm] in Chiral Gauge Theories}\\[1.0cm] 
{Claudia Cornella$^a$, Ferruccio Feruglio$^b$, Luca Vecchi$^b$}\\[0.2cm]
{\em ${}^a$ PRISMA$^+$ Cluster of Excellence {\em \&} MITP, 
Johannes Gutenberg University, 55099 Mainz, Germany\\
${}^b$ INFN, Sezione di Padova, Via Marzolo 8, I-35131 Padua, Italy}
\end{center}
\vspace{1 cm}
\centerline{\large\bf Abstract}
\begin{quote}
We derive the finite one-loop counterterm required to restore the Ward Identities broken by the regularization scheme in chiral gauge theories. Our result is an analytic expression applicable to a wide class of regularizations satisfying a few general properties.
We adopt the background field method, which ensures background gauge invariance in the quantized theory, and focus on renormalizable chiral theories with arbitrary gauge group and fermions in general representations. Our approach can be extended to theories involving scalars, such as the Standard Model, or to non-renormalizable theories, such as the SMEFT.
As a concrete application, we work out the finite counterterm at one loop in the Standard Model,
within dimensional regularization and the Breitenlohner-Maison-'t Hooft-Veltman prescription for $\gamma_5$.
\end{quote}
\vspace{5 cm}
\newpage
\thispagestyle{empty}
\setcounter{page}{0}
\tableofcontents
\newpage
\section{Introduction}

Chiral gauge theories play a key role in the description of fundamental interactions. 
For example, the Standard Model (SM) of strong and electroweak interactions exhibits a chiral fermion content with respect to the gauge group SU(3)$\times$SU(2)$\times$U(1).
While there are good reasons to believe that the SM is incomplete in several respects, the absence of confirmed
signals of new physics suggests charting possible SM extensions in terms of effective chiral gauge theories, such as the SM effective field theory (SMEFT). 

Quantization and renormalization of chiral gauge theories, defined by their symmetry and field content, are well-understood today. In particular, the framework of algebraic renormalization \cite{Stora:1973dka,Becchi:1974md,Becchi:1975nq,Becchi:1981jx,Piguet:1995er,Kraus:1997bi,Ferrari:1998jy,Grassi:1999tp,Grassi:2001zz}, relying on general properties of perturbative quantum field theories
such as the Power Counting Theorem \cite{Weinberg:1959nj,Zimmermann:1968mu} and the Quantum Action Principle \cite{Lowenstein:1971jk,Lam:1972mb,Clark:1976ym,Brennecke:2008zr}, allows to show how symmetries
(local or rigid) are preserved\footnote{Or violated by anomalies \cite{Adler:1969gk,Bell:1969ts}.} in perturbation theory. The great advantage of algebraic renormalization is its independence from the particular regularization used.

A regularization scheme should nonetheless be specified for practical computational purposes. The most convenient choice is provided by schemes preserving as many symmetries as possible of the underlying theory. 
However, the very existence of gauge anomalies prevents adopting a scheme where chiral gauge symmetries are maintained. Even when the field content is anomaly-free, any consistent regulator leads to a breaking of gauge invariance, 
which manifests itself in the amplitudes evaluated in perturbation theory.\footnote{In a path-integral formulation, the breaking of gauge invariance can come from the non-invariance of either the classical action or the integration measure or both.}

Such amplitudes are required to satisfy the Ward Identities (WI) arising from the gauge symmetry of the theory. However, these identities are spoiled by contributions introduced by the regularization procedure.
To remove the unwanted terms, different approaches are possible. The most elementary one is to disregard the undesired contributions, thus enforcing the WI by hand.
This procedure has the disadvantage of requiring the identification of the correct set of WI amplitude by amplitude.
Moreover, since the resulting subtraction is defined up to gauge-invariant contributions, independently for each process, ambiguities may arise when comparing different processes.

In a more comprehensive approach we can analyze (and repair) the breaking of gauge invariance induced by the regularization procedure directly at the level of the effective action, the generating functional of the one-particle irreducible (1PI) Green's functions, thus effectively handling all possible amplitudes at once \cite{Bonneau:1990xu,Martin:1999cc,Grassi:1999tp,Grassi:2001zz,Sanchez-Ruiz:2002pcf,Belusca-Maito:2020ala,Belusca-Maito:2021lnk}. Owing to symmetries, the effective action is bound to satisfy WI in the form of functional identities.\footnote{These are the non-linear Slavnov-Taylor (ST) identities associated to the rigid BRST symmetry of the quantized theory, or else WI related to ordinary gauge invariance if the Background Field Method and the Background Field Gauge are adopted.} These identities are violated in perturbation theory by terms that are severely constrained. In particular,
the Quantum Action Principle requires such terms to be finite local polynomials in the fields and their derivatives, of bounded dimensionality, order by order in perturbation theory. Moreover, if the theory is anomaly free, they are trivial solutions to the Wess-Zumino (WZ) consistency conditions \cite{Wess:1971yu}.  As a consequence, they can be expressed as gauge (or BRST) variations of integrated local polynomials that provide viable counterterms to recover the WI.

Each regularization scheme, combined with a subtraction procedure to remove divergences, requires
its own set of WI-restoring finite counterterms. In fact, the above strategy has already been pursued in the context of dimensionally regularized (DR) \cite{Bollini:1972ui,tHooft:1972tcz} chiral gauge theories. The specific cases that have been analyzed feature charged fermions of a single chirality \cite{Martin:1999cc,Sanchez-Ruiz:2002pcf,Belusca-Maito:2020ala,Belusca-Maito:2021lnk}. To the best of our knowledge, however, a general procedure allowing to identify the whole set of counterterms, independently from the adopted regularization scheme and for arbitrary chiral fermion charges and general (non-simple) gauge group, has not yet been formulated. In this work we discuss this general problem and show how it can be solved in the one-loop approximation. Explicit general expressions for WI-restoring counterterms, adaptable to a wide class of chosen regularization schemes,
can be of great utility for automated computations, such as those carried out today within the SMEFT \cite{Bonnefoy:2020tyv,Feruglio:2020kfq,Passarino:2021uxa}.
As described in Section \ref{sec:setup}, in this paper we deal with a renormalizable chiral gauge theory depending on gauge bosons and fermions only, though there is no obstacle in extending our method to theories involving
scalars, such as the SM, or to nonrenormalizable theories, such as the SMEFT. Indeed we consider this work as the first step of an approach meant to cover a wider range of applications. We assume an arbitrary regularization scheme, required to satisfy a few very general
requirements, such as the Quantum Action Principle, Lorentz invariance, and gauge invariance in the limit where the theory is vector-like.
Our treatment of fermions is completely general: we include fermions of both chiralities, which can transform under arbitrary representations of the gauge group, the latter being associated with a general (non-simple) compact Lie algebra.
Only physical fields (apart from ghosts) are present. In this sense our approach is minimal. 

We find it useful to quantize the theory within the background field method and to adopt the background field gauge fixing \cite{Kluberg-Stern:1974iel,Kluberg-Stern:1974nmx,Abbott:1981ke,Ichinose:1981uw,Capper:1982tf}. The latter preserves gauge invariance at the level of background fields, up to anomalies and regularization effects. The effective action is therefore bound to be a gauge-invariant functional of the background fields. As a consequence of the Quantum Action Principle, the gauge variation
of the one-loop effective action (evaluated in perturbation theory within a given regularization) is a four-dimensional, Lorentz-invariant, finite local polynomial in the fields and their derivatives, that vanishes when the theory is vector-like. Moreover, by treating CP and P as spurious symmetries\footnote{Formal invariance under CP and P is achieved if the generators of the group behave as spurions with well-defined transformation properties, as described below.}, the gauge variation of the one-loop effective action turns out to be P-even and CP-odd. 

It is then straightforward to expand such gauge variation in a basis of local operators with the desired symmetry properties. 
This expansion is characterized by a redundant set of coefficients. We can lift this redundancy
by requiring the gauge variation of the one-loop effective action to satisfy the WZ consistency conditions, which hold for any gauge theory, whether anomalous or not. This request translates into a set of linear equations relating the coefficients of the expansion and reduces the initial set of coefficients to an irreducible one.
As shown in Section \ref{sec:WZgeneral}, these first steps allow to parametrize in the most general and non-redundant way the gauge variation of the effective action at the one-loop order, independently from the adopted regularization. Similarly, we can build the most general parametrization of the one-loop finite counterterm necessary to restore the WI as a linear combination of integrated local operators with the correct symmetry properties. We finally require that, up to gauge anomalies, the gauge variation of the finite counterterm reproduces the gauge variation of the effective action. This allows to uniquely determine the parameters describing the counterterm in terms of those entering the variation of the effective action. As expected, we find that restoring the WI by means of a finite counterterm is always possible as long as the fermion field content is non-anomalous.  We stress that, for non-anomalous theories, our result unambiguously
determines the counterterm that reestablishes gauge invariance, for the entire class of regularizations satisfying the properties outlined above.  

Nowadays the most widely used regularization in practical calculations is dimensional regularization. Within DR, only the Breitenlohner-Maison/’t Hooft-Veltman (BMHV) scheme \cite{Breitenlohner:1977hr}  has been shown to provide a consistent treatment of $\gamma_5$ at all orders in perturbation theory. In Section \ref{sec:DR} we derive explicit expressions for the gauge variation of the effective action and the necessary counterterm at one loop, using DR and the BMHV scheme, which has already been implemented in tools for automated computations, such as \texttt{FeynCalc} or \texttt{Tracer}. Our formalism allows to determine the full set of counterterms needed to cast one-loop results in a fully gauge-invariant form. The calculation is performed via path integral techniques and checked diagrammatically. The outcome is of course consistent with the general results of Section \ref{sec:WZgeneral}.  
 
A paradigmatic example of chiral gauge theory is the Standard Model. Indeed, to illustrate our results, in Section \ref{sec:SM} we work out the counterterms needed at one loop using DR and the BHMV scheme, in the limit of vanishing Yukawa couplings. 

This paper is structured as follows. In Section \ref{sec:setup} we recall the classical and effective action for a chiral gauge theory and discuss three important ingredients of algebraic renormalization, namely the Ward Identities, the Wess-Zumino conditions, and the Quantum Action principle.  In Section \ref{sec:WZgeneral} we put these to use to determine the gauge variation of the effective action and the WI-restoring counterterm at the one-loop order for any regularization scheme respecting the Quantum Action Principle, Lorentz invariance, hermiticity of the action, vectorial gauge symmetry, and P and CP. Section~\ref{sec:DR} is dedicated to deriving the gauge variation of the effective action and the WI-restoring counterterm at one loop for the specific case of Dimensional Regularization. Finally, in Section~\ref{sec:SM} we apply our results to the SM. In the Appendices, we provide some auxiliary expressions used in Sections~\ref{sec:WZgeneral} and \ref{sec:DR}. Appendix~\ref{app:WZgeneral} contains results relevant to the general solution of the WZ conditions of Section~\ref{sec:WZgeneral}.  Appendix~\ref{app:heatkernel} provides details about the computation in Section \ref{sec:DR}.  
\section{The theory}
\label{sec:setup}
We consider a theory based on a compact gauge group $\mathcal{G}$, with gauge fields $A_{\mu}^a$ ($a=1\dots \mathrm{dim} (\mathcal{G})$), and fully antisymmetric structure constants $f_{abc}$. In general the gauge group is the direct product of $N_G$ simple groups ${\cal G}=\prod_G{\cal G}_G$ (with $G=1,\dots,N_G$), possibly including U(1) factors. In this case the index $a$ runs over the adjoint representation of each simple group, and similarly $f_{abc}$ is the direct sum of the structure constants $f^G_{abc}$ of each ${\cal G}_G$. {Throughout sections 1, 2 and 3, Lorentz indices run from 0 to 3 and are denoted by Greek letters $\mu$, $\nu$, etc.
In Section 4, when using DR to exemplify our results, this notation will be slightly modified.}

The matter content consists of two sets of massless chiral fermions, $f_L$ and $f_R$, transforming under $\mathcal{G}$ according to representations characterized by hermitian generators $T^a_{L}$ and $T^a_R$:
\bea\label{algebra}
[T_{X}^a,T_X^b] = i f_{abc} T_X^c\,,  && X=L,R \,. 
\eea
We are interested in chiral gauge theories, where $T^a_{L}$ and $T^a_R$ describe inequivalent
representations. An example is provided by theories where $T^a_{L(R)}=0$ and $T^a_{R(L)}$ is nontrivial, as 
in the case of the $SU(2)$ component of the Standard Model gauge group. Yet, our formalism encompasses all possible (chiral as well as vector-like) gauge theories with fermions.

In general, the representations described by $T^a_{L}$ and $T^a_R$ are reducible and their decomposition
in irreducible representations contains trivial components. We exploit this possibility to describe the
generators $T^a_{L}$ and $T^a_R$ using matrices of the same dimension. As a concrete example, consider hypercharge in the Standard Model. Its action on left-handed fermions can be described via a single generator acting on eight left-handed spinors per generation (six in the quark sector and two in the lepton sector). Its right-handed analogous instead acts non-trivially only on seven right-handed spinors per generation (six quarks and one lepton).  
 Nevertheless, we can formally extend the matrix describing the right-handed generator by one trivial row and column per generation, to match
the dimensionality of the left-handed one. Similarly, the multiplet $f_R$ may be extended to include a dummy degree of freedom, a right-handed neutrino,
which however does not play any role in our discussion and can be safely set to zero. 

While our focus is on theories with matter and gauge fields, fundamental scalars can be discussed along similar lines. This extension is left for future work.
\subsection{Classical action before regularization}
\label{sec:classicalaction}
The most general renormalizable bare action describing the dynamics of a set of fermionic fields $f$ charged under the gauge group $\mathcal{G}$ is:
\be
S[A, f_X, \bar f_X] = \int d^4 x \, ( \cL^\mathrm{YM}  + \cL_{\mathrm{Fermions}} )\,, 
\label{eq:action_classic_unregularized}
\ee
where $X=L, R$, $\cL_{\mathrm{YM}} $ is the usual Yang-Mills Lagrangian, and $\cL_{\mathrm{Fermions}}$ includes kinetic terms and gauge interactions of the fermions. Since we allow the gauge group to be the direct product of simple groups ${\cal G}=\prod_G{\cal G}_G$, the kinetic term of the gauge fields is controlled by a diagonal matrix $1/G_{ab}=\sum_G\delta_G^{ab}/g^2_{G}$, where $g_G$ and $\delta^{ab}_G$ are the gauge coupling and the identity in the adjoint representation of ${\cal G}_G$, respectively. Explicitly, we write:
\begin{align}
\cL_{\mathrm{YM}}  &=  - \frac{1}{4\,G_{ab}} F_{\mu \nu}^a F^{b \mu \nu} \,, \label{eq:classicalYM} \\
\cL_{\mathrm{Fermions}} &=  \bar{f}_L i \slashed{D} f_L + \bar{f}_R i \slashed{D} f_R \label{eq:classicalfermions}  \,,
\end{align}
where the left- and right-handed fermions are defined as
\ba\label{fLfR}
f_L=P_Lf,~~~~~~~f_R=P_Rf,
\ea
with $P_L=\frac{1}{2}(1-\gamma_5)$ and $P_R=\frac{1}{2}(1+\gamma_5)$ the hermitian chirality projectors, satisfying $P_{L(R)}^2=P_{L(R)}$ and $P_L+P_R=1$. The field strength of the gauge fields and the fermion covariant derivatives are defined for $X=L,R$ as 
\begin{align}
F_{\mu \nu}^a &= \partial_\mu A_\nu^a  -  \partial_\nu A_\mu^a - f_{abc}A^{b}_\mu A^c_{\nu}\,,\nn\\
D_\mu f_X &= (\partial_\mu + i A_\mu^a T_X^a)f_X\,,
\end{align}
and $\slashed{D}=\gamma^\mu D_\mu$.\footnote{Note the conventional sign of the vector field in the covariant derivative.}
The bare action is left invariant by the continuous local gauge transformations:
\begin{align}
\delta_\alpha A_{a\mu}&=\partial_\mu \alpha_a+f_{abc} \alpha_b A_{c\mu}\,,\nn\\
\delta_\alpha f_X&=-i\alpha_a T^a_X f_X \,,
\label{eq:gauge_transformations}
\end{align}
$\alpha_a$ being an infinitesimal gauge parameter. 
Given an arbitrary functional $F[A, f_X,\bar f_X]$ of the fermions and the gauge fields, we can write its gauge variation as
\be
\label{anyfun}
\delta_\alpha F[A, f_X,\bar f_X]  \equiv \int d^4 x  \,  \alpha_a(x) L_a(x) F[A, f_X,\bar f_X]\,.
\ee
where the differential operator $L_a$ is
\ba\label{Loperator}
L_a(x)&=&-\partial_\mu\frac{\delta}{\delta A_{a\mu}(x)}+f_{abc}A_{b\mu}(x)\frac{\delta}{\delta A_{c\mu}(x)}\\\no
&+&\sum_{X=L,R}
-i\frac{\overleftarrow{\delta}}{\delta f_X(x)}T^a_X f_X(x)+i\bar f_X(x)T^a_X\frac{\delta}{\delta \bar f_X(x)} \,. 
\ea
With this notation, the gauge invariance of the action, and similarly of any gauge-invariant functional, 
reads $\delta_\alpha S[A, f_X,\bar f_X] = 0$. Because this holds for any value of the gauge parameters, it is equivalent to writing the local relation
 \be
 \label{wi1}
 L_a(x)S[A, f_X,\bar f_X]=0\,.
 \ee
In the following, we will refer to the identity $L_a(x) F[A, f_X,\bar f_X] =0$ as to the {Ward Identity} for the functional $F[A, f_X,\bar f_X]$. 

From the algebra \eqref{algebra} of the gauge group, it follows that any functional $F[A, f_X,\bar f_X]$ of the fields and their derivatives satisfies the Wess-Zumino consistency conditions  \cite{Wess:1971yu}:
\begin{align}
\label{eq:WZ}
\left[ L_a(y) , L_b(x)  \right] F[A, f_X,\bar f_X] =  - \delta^{(4)}(x -y) f_{abc} L_{c}(x) F[A, f_X,\bar f_X]\,.
\end{align}
If $F[A, f_X,\bar f_X]$ is gauge invariant, these equations are trivially satisfied, since both sides vanish identically. If instead $F[A, f_X,\bar f_X]$ is not gauge invariant, Eq.~\eqref{eq:WZ} becomes a non-trivial constraint, which will play an important role in our analysis.

A chiral gauge theory featuring only gauge bosons and fermions is always invariant under CP, provided CP transformations are conveniently defined~\cite{Grimus:1995zi}. On the other hand, P is not a symmetry unless the theory is vector-like. Nevertheless, we can always define a generalized, spurious P symmetry that leaves the bare action invariant. Such a generalized P formally acts on the fields as ordinary P and on the generators, viewed as spurions, in an appropriate way. The resulting combined action reproduces ordinary P in any P-invariant theory, but is formally conserved even in theories that do not respect P, like chiral theories. 
Actually, in order to fully exploit the selection rules associated to both discrete symmetries we find it convenient to define both CP and P as spurious transformations, acting on the gauge and fermion fields as
\begin{align}
\begin{aligned}
\label{candcp}
&x^\mu\xrightarrow{\text{CP}} x_\mu\,,&&x^\mu\xrightarrow{\text{P}} x^\mu_P = x_\mu \,, \\
&\partial_\mu\xrightarrow{\text{CP}}\partial^\mu \,, &&\partial_\mu\xrightarrow{\text{P}}\partial^\mu \,, \\
& A_{a\mu}(x)\xrightarrow{\text{CP}}-A_{a}^\mu(x_P)\,, &&A_{a\mu}(x)\xrightarrow{\text{P}}A_{a}^\mu(x_P)\,, \\
&f_{L,R}(x)\xrightarrow{\text{CP}} C {f^*_{L,R}}(x_P)\,, &&f_{L,R}(x)\xrightarrow{\text{P}} \gamma^0 f_{R,L}(x_P)\,,
\end{aligned}
\end{align}
where $C$ denotes the well-known charge conjugation matrix, and on the generators as
\be
\label{spurion}
T^a_{L(R)}\xrightarrow{\text{CP}}T^{aT}_{L(R)}\,, \qquad T^a_{L(R)}\xrightarrow{\,\,\text{P}\,\,}T^{a}_{R(L)} \,.
\ee
We emphasize that the latter relation implies that the structure constants transform as
\be
\label{spurion2}
f_{abc}\xrightarrow{\text{CP}}-f_{abc}\,, \qquad f_{abc}\xrightarrow{\,\,\text{P}\,\,} f_{abc} \, .
\ee
The transformations in Eqs (\ref{candcp})-(\ref{spurion}) are formally symmetries of any theory defined by a classical action of the type \eqref{eq:action_classic_unregularized}. This restricts the structure of the counterterms needed to enforce the WI of the theory,  provided one adopts a regularization respecting these symmetries. As a final remark, we note that the operator $L_a$ is CP-odd and P-even. Indeed, the CP and P transformations of Eq.~ \eqref{eq:gauge_transformations}, together with Eq.~\eqref{spurion}, demand that
$\alpha_a$ be formally treated as a CP-odd and P-even spurion. Thus,
Eq.~\eqref{anyfun} implies that $L_a$ is CP-odd and P-even.
\subsection{Regularization: the need of local counterterms}
\label{sec:regu}
Going beyond the tree level, a regularization is needed. It is well known that in chiral gauge theories there is no consistent regularization procedure capable of 
preserving gauge invariance at the quantum level. This fact is at the origin of physical anomalies \cite{Adler:1969gk,Bell:1969ts}. The absence of gauge anomalies is guaranteed if the fermion content of the theory satisfies the well-known condition \cite{Georgi:1972bb}: 
\be
\label{georgi}
D^{abc}=\mathrm{tr}(T^a_L\{T^b_L,T^c_L\})-\mathrm{tr}(T^a_R\{T^b_R,T^c_R\})=0\,.
\ee
Yet, even if this condition holds, amplitudes computed in perturbation theory do not generally satisfy the WI. This is because the regularization procedure introduces scheme-dependent contributions 
to amplitudes beyond those removed by Eq.~\eqref{georgi}. Such sources of spurious, unphysical breaking of gauge invariance can always be removed by adding appropriate local counterterms to the classical action in Eq.~\eqref{eq:action_classic_unregularized}.\footnote{No counterterm can repair the breaking of gauge invariance induced by a violation of Eq. \eqref{georgi}.}~
Our analysis provides a general characterization of the counterterms required at the one-loop level in a chiral gauge theory, which applies to a large class of regularization schemes. Explicit expressions for such counterterms are then derived using dimensional regularization (DR). 

Let us explain our plan in some detail. The quantization of a gauge theory requires the introduction of a gauge-fixing term and a Faddeev-Popov term.
Independently from the chosen regularization, these terms necessarily break the original gauge invariance, leaving the classical action invariant under BRST transformations. As a result, the effective 1PI action, as well as all Green's functions of the theory, no more obey linear Ward Identities of the type shown in Eq.~\eqref{wi1}, but rather non-linear Slavnov-Taylor identities. Whenever a non-symmetric regulator is adopted, the identification of the counterterm that must be added to the bare action in order to restore the ST identities is unavoidably complicated by the non-linearity of such identities, as well as by the involved structure of the BRST symmetry \cite{Belusca-Maito:2020ala,Martin:1999cc}. 

Here we follow a different path and quantize the theory with the background field method \cite{Kluberg-Stern:1974iel,Kluberg-Stern:1974nmx,Abbott:1981ke,Ichinose:1981uw,Capper:1982tf}. Concretely, within the background field method the 1PI effective action is obtained by re-writing any field, including ghosts, as the sum of a classical background $\phi$ plus a quantum fluctuation $\tilde{\phi}$, and then integrating over the quantum fluctuations including only one-particle irreducible diagrams. In particular, the regularized 1PI effective action can be written as
\be
\label{1PI}
e^{i\Gamma^{\rm reg}[\phi]}=
\int_{\rm 1PI}{\cal D}{\tilde \phi}~e^{iS_{\rm full}^{\rm reg}[\phi+\tilde \phi]}\,,
\ee
where $S^{\rm reg}_{\rm full}\equiv S^{\rm reg}+S^{\rm reg}_{\rm g.f.}+S^{\rm reg}_{\rm ghost}$ is the sum of the regularized action, an appropriate gauge-fixing term, and the associated ghost action. The gauge-fixing Lagrangian is chosen to be
\be
\label{GaugeFixingGG}
{\cal L}_{\rm g.f.}[\phi+{\tilde \phi}]=-\frac{1}{2\xi} f_a f_a\,,
\ee
where  
\be\label{GFGen}
f_a=\partial_\mu {\tilde{A}_{a}^\mu} -f_{abc} A_{b\mu}{\tilde{A}_c^\mu}\,.
\ee
The gauge-fixing action $S^{\rm reg}_{\rm g.f.}$ serves its standard purpose of breaking gauge invariance. In particular, it is not invariant under gauge transformations of the quantum field. Yet $S^{\rm reg}_{\rm g.f.}$ (and, as a consequence, $S^{\rm reg}_{\rm ghost}$) is manifestly invariant under background gauge transformations. The latter act as a standard gauge transformation on the background $A^\mu_a$, and as a linear re-definition of the integration variable $\tilde{A}^\mu_a$. For all fields transforming linearly under the original gauge symmetry, both the quantum fluctuation and the classical background transform exactly as the original field, and the distinction between standard and background transformations is not relevant. 

As mentioned above, the invariance of the gauge-fixed action under background gauge transformations is the main advantage of the background field method. If one introduces sources for the quantum fields only, all generating functionals are also manifestly background-gauge invariant and satisfy linear Ward Identities as in Eq.~\eqref{wi1}, up to the regularization-dependent effects mentioned earlier. In particular, the background gauge symmetry, along with \eqref{georgi}, guarantees that the unique source of violation of the WI 
is the regularization procedure. The linearity of such relations significantly simplifies the search for the WI-restoring counterterms because, as opposed to the non-linear Slavnov-Taylor equations, the linear WI relate only Green's functions of the same order in perturbation theory \cite{Grassi:1999tp}. 

In our treatment we adopt a regularization scheme preserving the vectorial gauge transformations, four-dimensional Lorentz invariance, the generalized P and CP symmetries defined in Eqs.~\eqref{candcp} and \eqref{spurion} and the Quantum Action Principle \cite{Lowenstein:1971jk,Lam:1972mb,Piguet:1980nr,Piguet:1995er}.  
As a starting point, we assume that a consistent subtraction procedure is defined, making it possible to evaluate the
renormalized functional $\Gamma[\phi]$ from $\Gamma^{\rm reg}[\phi]$. 
At this stage we do not need to specify either how this subtraction is performed or which renormalization conditions are imposed; we will do so in Section \ref{sec:DR}, when performing explicit calculations within DR. Here we simply assume that this subtraction renders $\Gamma[\phi]$ finite order by order in perturbation theory. As we now show, the proof that finite counterterms can be added such that $\Gamma[\phi]$ satisfies the WI
proceeds by induction.
Suppose that we have successfully identified an action $\Gamma[\phi]$ that satisfies the WI of the theory up to loop order $n-1$ (included): 
\begin{align}
\left. L_a(x)\Gamma [\phi]\right|_{(k)}=0&&k\le n-1\,,
\end{align}
where $\left.\Gamma [\phi]\right|_{(k)}$ stands for the $k$-order in the loop expansion of $\Gamma [\phi]$.
Although in general the WI will be broken at order $n$, the Quantum Action Principle guarantees that 
\ba\label{QAexplained}
\left.L_a(x)\Gamma [\phi]\right|_{(n)}=(\Delta_a \cdot \Gamma)(x)= {\left.\Delta_a(x) \right|_{(n)}} +\mathcal{O}(\hbar^{n+1})\,.
\ea
Here $\Delta_a \cdot \Gamma$ 
is the generating functional of the amputated 1PI Green's functions with one insertion of a local polynomial in the fields, $\Delta_a |_{(n)}$, formally of order $\hbar^n$.\footnote{In the last step of Eq.~\eqref{QAexplained} we used the fact that at tree-level the only non-vanishing correlator functions involving ${\Delta_a|_{(n)}}$ are those that contain precisely the fields appearing in $\Delta_a|_{(n)}$, and the corresponding contribution to the one-particle irreducible action reads $\Delta_a$, where by a slight abuse of notation the latter is now interpreted as being a functional of the background fields.} In the rest of the paper, the expressions $L_a(x)\Gamma |_{(n)}$ and $\Delta_a |_{(n)}$ will be used interchangeably. By power counting follows that $\Delta_a |_{(n)}$ is a dimension-four polynomial. According to our assumptions, it should be CP-odd and P-invariant as well as invariant under the four-dimensional Lorentz symmetry and should vanish when $T^a_L=T^a_R$.  

Moreover ${\Delta_a |_{(n)}}$ must satisfy the WZ consistency conditions \eqref{eq:WZ}:
\begin{align}
\label{eq:WZDelta}
L_a(y)  \left. \Delta_b(x) \right|_{(n)}- L_b(x)  \left. \Delta_a(y) \right|_{(n)}  =  - \delta^{(4)}(x -y) f_{abc} \left. \Delta_c(x) \right|_{(n)} \,.
\end{align}
Theories complying with the criterion \eqref{georgi} have no anomalies, and the most general solution of Eq.~\eqref{eq:WZDelta} at order $n$ is:
\begin{align}
\label{defCT}
\left. \Delta_a(x)\right|_{(n)}  = - L_a(x) \left.S_{\rm ct}[\phi]\right|_{(n)}\,,
\end{align}
where $\left.S_{\rm ct}[\phi]\right|_{(n)}=\int d^4y~ \cL_{\mathrm{ct}}(y)|_{(n)}$ is an integrated local polynomial of order $\hbar^n$ in the fields and their derivatives
invariant under the four-dimensional Lorentz group, CP and P, and the vectorial gauge symmetry. We can next define:
\begin{align}
\label{WIagain}
\Gamma_{\mathrm{inv}}[\phi]|_{(n)}=\Gamma[\phi]|_{(n)}+S_{\rm ct}[\phi]|_{(n)}\,,
\end{align}
obtaining:
\begin{align}
L_a(x)\Gamma_{\mathrm{inv}}[\phi]|_{(n)}=\mathcal{O}(\hbar^{n+1})\,.
\end{align}
The spurious noninvariant contributions induced by the regularization procedure are now removed, and gauge invariance is restored at order $\mathcal{O}(\hbar^n)$. After adding the $n+1$-loop contributions and implementing the subtraction procedure, we get a new functional $\Gamma [\phi]|_{(n+1)}$
and we can repeat the above steps to enforce the WI at $\mathcal{O}(\hbar^{n+1})$. 

One of our main results is the determination of the counterterm within the DR scheme at the one-loop order. We will see that DR can be made to comply with our symmetry requirements; in particular, it satisfies the Quantum Action Principle \cite{Breitenlohner:1977hr}. 
It is important to stress that the explicit form of the gauge variation of the effective action, as well as the countertem, does depend on the regularization scheme. Yet, as we show in the following section, several important features can be deduced solely from the general considerations presented in the previous paragraph and apply to all regularization schemes that preserve Lorentz invariance, hermiticity of the action, vectorial gauge transformations as well as generalized P and CP.  Explicit results for DR will be presented in Section \ref{sec:DR}. 
\section{One-loop analysis for generic regularization schemes}
\label{sec:WZgeneral}
As discussed above, whenever the theory is anomaly free the WI identities can be restored order by order by adding a counterterm to the classical action. The goal of this section is to determine the structure of the gauge variation of the effective action and the counterterm at the one-loop order, i.e. $\Delta_a |_{(1)}$ and $S_{\rm ct}[\phi]|_{(1)}$, for any regularization scheme respecting: 
\begin{itemize}
\item[i)] the Quantum Action Principle,
\item[i)] four-dimensional Lorentz-invariance,
\item[ii)] hermiticity of the action,
\item[iii)] vectorial gauge symmetry, \label{item:vectorsymmetry}
\item[iv)] the generalized P and CP symmetries of Eqs. \eqref{candcp}, \eqref{spurion}. 
\end{itemize}
As we show in the following, these rather general hypotheses significantly constrain the form of $\Delta_a |_{(1)}$ and $S_{\rm ct}[\phi]|_{(1)}$.  
\subsection{A basis for the gauge variation and the counterterm}
\label{sec:basis}
We start by providing a convenient representation for both $\Delta_a|_{(1)}$ and $S_{\rm ct}[\phi]|_{(1)}$.  
As discussed above, the former is a finite local polynomial of dimension four in the gauge and fermionic fields, and their derivatives.\footnote{We neglect a possible dependence of ${\Delta_a|_{(1)}}$ on ghosts. As will be discussed in Section \ref{sec:DR}, these do not contribute to $\Delta_a$ at the one-loop level.} We can thus expand it in a basis of monomials involving only gauge and fermion fields:
\be
\label{var1}
{\left. \Delta_a (x)\right |_{(1)} }= \sum_{k =0}^{14} \ci{k}{aA} I^k_{A}(x)\,.
\ee
where a sum over $X=L, R$ is understood. The monomials $I^k_{A}$, where the label $A$ collectively denotes the relevant set of indices, are collected in Table \ref{tab:anomaly_building_blocks}, along with their CP and P properties. The resulting basis coincides with the one already identified in Ref.~\cite{Martin:1999cc}. Observable quantities are basis-independent, thus any other choice of basis would be equally good.
\begin{table}[t]
\centering
\renewcommand{\arraystretch}{1.3} 
\begin{tabular}{|c|c|c|c|}
\hline
 monomial & explicit expression &CP&P\\
\hline
\hline
$I^0_{a} $ & $ \Box\partial^\mu A_{a\mu}$&$-$&$+$ \\
\hline
$I^1_{ab}$ &  $\epsilon^{\mu\nu\alpha\beta}(\partial_{\alpha}A_{a \mu} )(\partial_{\beta}A_{b \nu})$&$-$&$-$ \\
\hline
$I^2_{ab}$ & $A_{a \mu} (\partial^\mu\partial^\nu-\Box g^{\mu\nu})A_{b \nu}$&$+$&$+$ \\
\hline
$I^3_{ab}$  & $A_{a \mu} \Box A_{b}^\mu$&$+$&$+$  \\
\hline
$I^4_{ab}$  & $(\partial_{\nu}A_{a \mu})(\partial^\nu A_{b}^\mu)$&$+$&$+$\\
\hline
$I^5_{ab}$  & $(\partial_\nu A_{a \mu})(\partial^\mu A_b^{\nu})$&$+$&$+$ \\
\hline
$I^6_{ab}$  & $(\partial^\mu A_{a \mu})( \partial^\nu A_{b \nu})$&$+$&$+$ \\
\hline
$ I^7_{abd}$ & $ ( \partial_\mu A^\mu_a)   A_{b\nu} A^\nu_{d}$&$-$&$+$ \\
\hline
$ I^8_{abd}$ & $ (\partial_\mu A_a^\nu)   A_{b\mu} A^\nu_d$&$-$&$+$ \\
\hline
$ I^9_{abd}$ & $\epsilon^{\mu \nu \alpha \beta}   ( \partial_\beta A_{a\mu})   A_{b\nu} A_{d\alpha}$&$+$&$-$ \\
\hline
 $I^{10}_{abde}$ & $ A_{a\mu} A_{b}^\mu A_{d\nu} A_e^\nu$&$+$&$+$  \\
 \hline
 $I^{11}_{abde}$ & $\epsilon^{\mu \nu \rho \sigma} A_{a\mu} A_{b\nu} A_{d\rho} A_{e\sigma}$&$-$&$-$ \\
 \hline
 $I^{12}_{Xij}$ & $\bar f_{Xi}  \scaleto{\overrightarrow{\slashed{\partial}}}{14pt} f_{Xj}$&$-\bar f_{Xj}  \scaleto{\overleftarrow{\slashed{\partial}}}{14pt}   f_{Xi}$&$\bar f_{\tilde{X}i}  \scaleto{\overrightarrow{\slashed{\partial}}}{14pt} f_{\tilde{X}j}$ \\
 \hline
 $I^{13}_{Xij}$ &  $\bar f_{Xi}   \scaleto{\overleftarrow{\slashed{\partial}}}{14pt} f_{Xj}$&$-\bar f_{Xj}  \scaleto{\overrightarrow{\slashed{\partial}}}{14 pt} f_{Xi}$&$\bar f_{\tilde{X}i}  \scaleto{\overleftarrow{\slashed{\partial}}}{14 pt} f_{\tilde{X}j}$ \\
 \hline
 $I^{14}_{Xaij}$ &  $\bar f_{Xi} \slashed{A}_a  f_{Xj}$&$+\bar f_{Xj} \slashed{A}_a  f_{Xi}$&$\bar f_{\tilde{X}i} \slashed{A}_a  f_{\tilde{X}j}$\\
\hline
\end{tabular}
\caption{Basis of local, dimension-four operators depending on gauge bosons, fermions and their derivatives entering the decomposition of $\Delta_a|_{(1)}$. Lorentz indices $\mu$, $\nu$,$\dots$ run from 0 to 3. Also shown are the transformation properties under CP and P. For the fermion bilinears $I^{12}_{Xij}$, $I^{13}_{Xij}$, and $I^{14}_{Xaij}$, $i,j$ being flavour indices, we explicitly display their CP- and P-transformed versions, with $\tilde{L}(\tilde{R})=R(L)$.}
\label{tab:anomaly_building_blocks}
\vspace{30pt}
\end{table}
\begin{table}[t]
\centering
\renewcommand{\arraystretch}{1.3} 
\begin{tabular}{|c|c||c|c|c|}
\hline
monomial & explicit expression &coefficient&CP&P \\
\hline
\hline
$\mathcal{I}^1_{ghl} $ &  $ (\partial^\nu A_g^\mu)  A_{h\nu} A_{l\mu}$&$\ct{1}{ghl}$&$-$&$+$\\
\hline
$\mathcal{I}^2_{gh} $ & $A_{g\mu}\Box A_h^\mu$&$\ct{2}{gh}$&$+$&$+$\\
\hline
$\mathcal{I}^3_{gh} $ & $A_{g\mu}\partial^\mu\partial^\nu A_{h\nu}$&$\ct{3}{gh}$&$+$&$+$\\
\hline
$\mathcal{I}^4_{ghl} $ & $\epsilon^{\mu\nu\rho\sigma} A_{g\mu} A_{h\nu} (\partial_\rho A_{l\sigma}) $&$\ct{4}{[gh]l}$&$+$&$-$\\
\hline
$\mathcal{I}^5_{ghlm} $ & $\epsilon^{\mu\nu\rho\sigma} A_{g\mu} A_{h\nu}~  A_{l\rho} A_{m\sigma}$&$\ct{5}{[ghlm]}$&$-$&$-$\\
\hline
$\mathcal{I}^6_{ghlm} $ & $A_{g\mu} A_{h}^\mu~  A_{l\nu} A_{m}^\nu$&$\ct{6}{(gh)(lm)}$&$+$&$+$\\
\hline
$\mathcal{I}^7_{Xij}$ & $\bar f_{Xi} \scaleto{ \overrightarrow{\slashed{\partial}} }{14 pt} f_{Xj}$&$\ct{7}{Xij}$&$\ct{7}{Xji}$&$\ct{7}{\tilde{X}ij}$\\
\hline
$\mathcal{I}^8_{Xaij}$ &  $\bar f_{Xi} \slashed{A}_a  f_{Xj}$&$\ct{8}{Xaij}$&$\ct{8}{Xaji}$&$\ct{8}{\tilde{X}aij}$\\
\hline
\end{tabular}
\caption{Basis of local, dimension-four operators, depending on gauge bosons, fermions and their derivatives
relevant to build the counterterm $S_{\mathrm{ct}}$. Lorentz indices $\mu$, $\nu$,$\dots$ run from 0 to 3.  Also shown are the corresponding coefficients and their transformation properties under CP and P. For the fermion bilinears we explicitly display their CP- and P-transformed, with $\tilde{L}(\tilde{R})=R(L)$. }
\label{tab:counterterms_building_blocks}
\end{table}

\noindent
The symmetry properties of the $I^k_A$ imply:
\begin{align}
\begin{aligned}
\label{eq:Csymmetries}
C^1_{abc}   = C^1_{a(bc)} \,, \quad  C^4_{abc}   = C^4_{a(bc)} \,,  \quad  C^5_{abc}   = C^5_{a(bc)} \,,    \quad C^6_{abc}   = C^6_{a(bc)} \,,  \quad C^7_{abcd}   = C^7_{ab(cd)} \,,  \\
C^9_{abcd}  =    C^9_{ab[cd]} \,, \quad   C^{10}_{abcde}  =  C^{10}_{a(bc)(de)} = C^{10}_{a(de)(bc)}\,, \quad    C^{11}_{abcde}  =  C^{11}_{a[bcde]}\,,
\end{aligned}
\end{align}
where $(a_1 \dots a_n)$ and $[a_1 \dots a_n]$ denote symmetrization and antisymmetrization over the indices inside the parenthesis. 
For example, $C^1_{a(bc)}=(C^1_{abc}+C^1_{acb})/2$, $C^9_{ab[cd]}=(C^9_{abcd}-C^9_{abdc})/2$, whereas $C^{11}_{a[bcde]}$ involves the anti-symmetrization of the four indices $bcde$. 
The decomposition of Eq.~\eqref{var1} is general, and applies to any regularization scheme satisfying the properties i)-iv). 
We can further constrain this parametrization by observing that
the effective action must fulfill the WZ conditions, hence its variation $\Delta_a|_{(1)}$ must satisfy Eq.~\eqref{eq:WZDelta}. 
Plugging the decomposition \eqref{var1} in \eqref{eq:WZDelta}, a set of relations among the coefficients $\ci{k}{aA}$ is obtained. We collectively denote them as
\be
\label{eq:WZc}
\mathrm{WZ}[\ci{k}{aA}]=0\,,
\ee
and provide their explicit expressions in Appendix \ref{app:equations_ci}. 
It is worth stressing that the mutual dependence among
the coefficients implied by Eq.~\eqref{eq:WZc} is not related to the linear dependence among the elements of the chosen basis, but is rather a consequence of the Lie algebra satisfied by the group generators.

Also the polynomial $\cL_{\mathrm{ct}}$ defining the counterterm  $S_{\rm ct}[\phi]|_{(1)}$
can be expanded in a basis. At variance with the elements $I^k_{A}$, which always occur unintegrated, $\cL_{\mathrm{ct}}$ is integrated over spacetime. Since 
monomials related by integration by parts do not produce independent terms in $S_{\rm ct}[\phi]|_{(1)}$, we can expand $\cL_{\mathrm{ct}}$ in a basis consisting in a subset of the one introduced above: 
\be
\label{gencount}
S_{\rm ct}[\phi]|_{(1)} = \int d^4y~\cL_{\mathrm{ct}}(y) = \int d^4 y \sum_{j=1}^8 \, \ct{j}{B} \mathcal{I}^j_{B} (y)\,,
\ee
where the label $B$ collectively denotes the relevant set of indices and a sum over $X=L, R$ is understood. The monomials $\mathcal{I}^j_{B}$ and their CP and P properties are displayed in Table  \ref{tab:counterterms_building_blocks}. Exchanging the gauge indices we deduce the following constraints on the coefficients:
\be
\ct{4}{ghl} = \ct{4}{[gh]l} \,, \qquad \ct{5}{ghlm} = \ct{5}{[ghlm]}\,, \qquad  \ct{6}{(gh)(lm)} = \ct{6}{(lm)(gh)}\,.
\ee
By computing the gauge variation of $S_{\rm ct}[\phi]|_{(1)}$, we find: 
\begin{align}\label{ctvar}
L_a(x) S_{\rm ct}[\phi]|_{(1)}  =& -(\ct{2}{ba}+\ct{2}{ab}+\ct{3}{ba}+\ct{3}{ab})~I^0_{a}(x) +2\ct{4}{[ab]c}I^1_{bc}(x) \nn \\
&+\left( \ct{1}{abc}+\ct{1}{acb}-\ct{1}{cba} +(\ct{3}{cd}+\ct{3}{dc}) f_{dab}\right) ~ I^2_{bc}(x)\nn \\
&+[(\ct{1}{abc}+\ct{1}{acb}-\ct{1}{cba}-\ct{1}{cab})+(\ct{2}{cd}+\ct{2}{dc}+\ct{3}{cd}+\ct{3}{dc}) f_{dab}]~ I^3_{bc}(x)\nn \\
&-\ct{1}{cab}I^4_{bc}(x)
+(\ct{1}{abc}-\ct{1}{cba})I^5_{bc}(x)\nn \\
&+\ct{1}{abc}I^6_{bc}(x) \nn \\
&-(f_{ace}\ct{1}{ebd}+4~\ct{6}{(ab)(cd)})I^7_{bcd}(x)\nn \\
&+(f_{ade}\ct{1}{bce}-f_{ade}\ct{1}{ecb}+f_{ace}\ct{1}{bed}-8~\ct{6}{(ac)(bd)})I^8_{bcd}(x)\nn\\
&+4~f_{abg}\ct{6}{(gc)(de)}I^{10}_{bcde}\nn\\
&+\left( 12~ \ct{5}{[abcd]} + 2 f_{ace}(\xi^4_{[de]b} - \xi^4_{[bd]e}) \right)~I^{9}_{bcd}(x) \nn\\
&+4~ f_{abg}\ct{5}{[gcde]}~I^{11}_{bcde}(x) \nn\\ 
& + i  (T^a_X \ct{7}{X}  +i\ct{8}{Xa} )_{ij} I^{12}_{Xij}(x) \nn\\
& + i (\ct{7}{X}  T^a_X  +i\ct{8}{Xa} )_{ij}  I^{13}_{Xij}(x) \nn \\
&+ i (T^A_X \ct{8}{Xb}  -\ct{8}{Xb}T^a_X -if_{abc}\ct{8}{Xc} )_{ij}  I^{14}_{Xbij}(x)  \nn \,.\\
=&\sum_{k =0}^{14} \hatci{k}{aA}(\ct{}{}) I^k_{A}(x)\,.
\end{align}
Again, a sum over $X=L,R$ is understood.
Explicit expressions for the coefficients $\hatci{k}{aA}$ as a function of the coefficients $\xi^i_B$ appearing in the counterterm are provided in Table \ref{tab:hatcoeff}.  Note that, since Eq. \eqref{ctvar} describes a gauge variation, the $\hatci{k}{aA}$ automatically satisfy the WZ conditions. 

Using \eqref{ctvar} and \eqref{var1}, the gauge variation of the sum of the 1-loop effective action and the counterterm can be written as
\be
\label{var3}
\left. \Delta_a(x) \right|_{(1)}+  L_a(x) \left. S_{\rm ct}[\phi]\right|_{(1)}  =\sum_{k =0}^{14} \left[\ci{k}{aA}+\hatci{k}{aA}(\ct{}{})\right] I^k_{A}(x)\,. 
\ee
In an anomaly-free theory, the WI can be enforced by requiring the right-hand side of this equation to vanish. If instead the fermion content of the theory is anomalous, we can generalize this requirement by splitting the gauge variation of the effective action into two contributions, only one of which can be removed by a counterterm. The remaining piece represents the anomaly. Since the anomaly of a gauge theory is an equivalence class, where two elements related by
adding an integrated local polynomial of the fields and derivatives are equivalent, such a separation is ambiguous unless we pick up a specific representative element $\mathcal{A}_a$ in the class. When this choice is made, we can write:  
\be
\label{inhom}
\sum_{k =0}^{14} \left[\ci{k}{aA}+\hatci{k}{aA}(\ct{}{})\right] I^k_{A}(x)= \mathcal{A}_a(x)\,.
\ee
This defines our master equation. In practice, it is a set of linear equations that determine the counterterm coefficients $\ct{j}{B}$ as a function of the coefficients $\ci{k}{aA}$ describing the breaking of gauge invariance induced by the regularization. If the theory is anomaly free, Eq. \eqref{inhom} simplifies to:
\be
\label{hom}
\ci{k}{aA}+\hatci{k}{aA}(\ct{}{})=0\,.
\ee
Even in an anomalous theory, Eq. \eqref{hom} can be enforced for a convenient subset of coefficients by appropriately choosing the representative element $\mathcal{A}_a$.  For instance, one can always choose $\mathcal{A}_a$ to be a combination of P-violating operators. Here we show how this well-known fact can be deduced in full generality from the WZ conditions. 

\begin{table}[t]
\centering
\renewcommand{\arraystretch}{1.35} 
\begin{tabular}{|c|c|c|c|}
\hline
coefficient & explicit expression &CP&P\\
\hline
\hline
$\hatci{0}{ab}$&$-(\ct{2}{ba}+\ct{2}{ab}+\ct{3}{ba}+\ct{3}{ab})$&$+$&$+$  \\
\hline
$\hatci{1}{a(bc)}$&$\ct{4}{[ab]c}+\ct{4}{[ac]b}$&$+$&$-$\\
\hline
$\hatci{2}{abc}$&$(\ct{1}{abc}+\ct{1}{acb}-\ct{1}{cba})+(\ct{3}{cd}+\ct{3}{dc}) f_{dab}$&$-$&$+$\\
\hline
$\hatci{3}{abc}$&$(\ct{1}{abc}+\ct{1}{acb}-\ct{1}{cba}-\ct{1}{cab})+(\ct{2}{cd}+\ct{2}{dc}+\ct{3}{cd}+\ct{3}{dc}) f_{dab}$&$-$&$+$\\
\hline
$\hatci{4}{a(bc)}$&$-\frac{1}{2}(\ct{1}{cab}+\ct{1}{bac})$&$-$&$+$\\
\hline
$\hatci{5}{a(bc)}$&$\frac{1}{2}(\ct{1}{abc}-\ct{1}{cba}+\ct{1}{acb}-\ct{1}{bca})$&$-$&$+$\\
\hline
$\hatci{6}{a(bc)}$&$\frac{1}{2}(\ct{1}{abc}+\ct{1}{acb})$&$-$&$+$\\
\hline
$\hatci{7}{a b(cd)}$&$-\frac{1}{2}(f_{ace}\ct{1}{ebd}+f_{ade}\ct{1}{ebc})-4~\ct{6}{(ab)(cd)}$&$+$&$+$\\
\hline
$\hatci{8}{abcd}$&$f_{ade}\ct{1}{bce}-f_{ade}\ct{1}{ecb}+f_{ace}\ct{1}{bed}-8~\ct{6}{(ac)(bd)}$&$+$&$+$\\
\hline
$\hatci{9}{ab[cd]}$&$12~ \ct{5}{[abcd]} + f_{ace}(\xi^4_{[de]b} - \xi^4_{[bd]e})- f_{ade}(\xi^4_{[ce]b} - \xi^4_{[bc]e})$&$-$&$-$\\
\hline
$\hatci{10}{a(bc)(de)}$&$ f_{abg}\ct{6}{(gc)(de)}+f_{acg}\ct{6}{(gb)(de)}+f_{adg}\ct{6}{(ge)(bc)}+f_{aeg}\ct{6}{(gd)(bc)} $&$-$&$+$\\
\hline
$\hatci{11}{a[bcde]}$&$f_{abg}\ct{5}{[gcde]}-f_{acg}\ct{5}{[gbde]}+f_{adg}\ct{5}{[gbce]}-f_{aeg}\ct{5}{[gbcd]}$&$+$&$-$\\
\hline
$\hatci{12}{aX} $&$i(T^a_X \ct{7}{X} +i\ct{8}{Xa} )$&$\hatci{13}{aX \vphantom{\tilde{X}}}\left. \right.^T$&$\hatci{12}{a\tilde{X}}$\\
\hline
$\hatci{13}{aX}$&$i(\ct{7}{X}  T^a_X +i\ct{8}{Xa} )$&$ \hatci{12}{aX \vphantom{\tilde{X}}} \left.\right.^T$&$\hatci{13}{a\tilde{X}}$\\
\hline
$\hatci{14}{abX}$ &$i (T^a_X \ct{8}{Xb}  -\ct{8}{Xb}T^a_X -if_{abc}\ct{8}{Xc} )$&$-\hatci{14}{ab X \vphantom{\tilde{X}}}\left.\right.^T$&$\hatci{14}{ab\tilde{X}}$\\
\hline
\end{tabular}
\caption{Coefficients appearing in the gauge variation of the general counterterm $L_a \Gamma_{\rm ct}$ once it is decomposed
in the basis of Table \ref{tab:anomaly_building_blocks}.  Also shown are the transformation properties under CP and P. For the fermionic coefficients $\hatci{12}{aX}$, $\hatci{13}{aX}$ and $\hatci{14}{abX}$, we display their CP- and P-transformed, with $\tilde{L}(\tilde{R})=R(L)$.}
\label{tab:hatcoeff}
\end{table}
\subsection{Solution to the master equation}
\label{sec:master}
We now wish to simultaneously solve the WZ conditions \eqref{eq:WZc} and the master equation \eqref{inhom}. To this end, we first determine the most general form of the $\ci{k}{aA}$ satisfying $\eqref{eq:WZc}$, and then find
the counterterm coefficients $\ct{j}{B}$ such that \eqref{inhom} is fulfilled.
We do not need to specify the regularization scheme, which is only required to satisfy the general assumptions spelled out at the beginning of Section \ref{sec:WZgeneral}.  
An explicit determination of the coefficients $\ci{k}{aA}$ and of the corresponding counterterms $\ct{j}{B}$ is performed in Sec. \ref{sec:1-loop} using DR. 

Our task is considerably facilitated by the observation that both
$\ci{k}{aA}$ and $\ct{j}{B}$ have definite transformation properties under CP and P. For the $\ci{k}{aA}$ these properties can be deduced from Eq.~\eqref{var1}, recalling that $\Delta_a$ is CP-odd and P-even and
that the operators $I^k_{A}$ transform as shown in Table \ref{tab:anomaly_building_blocks}. Similarly, the transformations of $\ct{j}{B}$ under CP and P, displayed in Table \ref{tab:counterterms_building_blocks}, can be deduced from Eq. \eqref{gencount}, wWWhere each side is invariant under both CP and P. 
For consistency the coefficients $\hatci{k}{aA}$ and $\ci{k}{aA}$ must transform in the same way (see Table \ref{tab:hatcoeff}).

Since gauge transformations do not mix operators with fermions with those containing only bosons, we can treat them independently. We start by solving the set of equations \eqref{eq:WZc} and \eqref{inhom} involving purely bosonic operators and then discuss the fermionic sector.

\begin{table}
\centering
\renewcommand{\arraystretch}{1.35} 
\begin{tabular}{|c|c|c|}
\hline
trace combination &CP&P\\
\hline
\hline
$(T^{a_1...a_n}_{X_1...X_n}+T^{a_n...a_1}_{X_n...X_1})+(T^{a_1...a_n}_{\tilde X_1...\tilde X_n}+T^{a_n...a_1}_{ \tilde X_n...\tilde X_1})$&$+$&$+$\\
\hline
$(T^{a_1...a_n}_{X_1...X_n}+T^{a_n...a_1}_{X_n...X_1})-(T^{a_1...a_n}_{\tilde X_1...\tilde X_n}+T^{a_n...a_1}_{\tilde X_n...\tilde X_1})$&$+$&$-$\\
\hline
$(T^{a_1...a_n}_{X_1...X_n}-T^{a_n...a_1}_{X_n...X_1})+(T^{a_1...a_n}_{\tilde X_1...\tilde X_n}-T^{a_n...a_1}_{\tilde X_n...\tilde X_1})$&$-$&$+$\\
\hline
$(T^{a_1...a_n}_{X_1...X_n}-T^{a_n...a_1}_{X_n...X_1})-(T^{a_1...a_n}_{\tilde X_1...\tilde X_n}-T^{a_n...a_1}_{\tilde X_n...\tilde X_1})$&$-$&$-$\\
\hline
\end{tabular}
\caption{Combinations of single traces eigenstates of CP and P. }
\label{singletraces}
\end{table}
\subsubsection{Bosonic sector}
 The coefficients associated to the bosonic operators are $\ci{k=0-11}{aA}$ and $\ct{j=1-6}{B}$. In this sector the Wess-Zumino conditions \eqref{eq:WZc} and the master equation \eqref{inhom} split into two decoupled sets of equations,
according to the parity of the operators involved. The P-even and P-odd sets are defined by $k=0,2-8,10$ (in short: $k\in$ P-even) and $k=1,9,11$ (in short: $k\in$ P-odd), respectively.  The WZ conditions in the P-even and P-odd sectors are given in Eq. \eqref{eq:WZcPeven} and \eqref{eq:WZcPodd}. The master equation \eqref{inhom} involves the counterterm coefficients $\ct{j=1,2,3,6}{B}$  in the P-even sector, and $\ct{j=4,5}{B}$ in the P-odd sector. 

At the one-loop order the coefficients $\ci{k}{aA}$ and $\ct{j}{B}$ can be written as linear combinations of single traces of the generators\footnote{At higher loops also products of traces can appear.}: 
\begin{align}
\begin{aligned}
\label{traces}
C^k_{a_1...a_n} &=\sum \ccoeff{k}{X_1...X_n}  T^{a_1...a_n}_{X_1...X_n}   \\
\ct{j}{a_1...a_n}& =\sum \ctcoeff{j}{X_1...X_n}  T^{a_1...a_n}_{X_1...X_n}\,,  \\
\end{aligned} 
\end{align}
where
\be
T^{a_1...a_n}_{X_1...X_n}  = \mathrm{tr}(T^{a_1}_{X_1}...T^{a_n}_{X_n}) \,.
\ee
and $ \ccoeff{k}{X_1...X_n}$ and $ \ctcoeff{k}{X_1...X_n}$ are numerical coefficients. Given the assumptions iii) and iv) stated at the beginning of the section and the decompositions in \eqref{var1} and \eqref{gencount},  the coefficients $\ci{k}{aA}$ and $\ct{j}{B}$ must have the following properties:
\begin{enumerate}
\item They transform under CP and P as indicated in Tables \ref{tab:counterterms_building_blocks} and \ref{tab:hatcoeff}.
\item
 Under exchange $a_1 \dots a_n$ they behave as indicated in Table \ref{tab:hatcoeff}.
 \item
$\ci{k}{Aa}$ and $\hatci{k}{Aa}(\xi)$ vanish for vector-like theories, i.e. if $T^a_L=T^a_R$.
\end{enumerate} 
This strongly restricts the form of $\ci{k}{aA}$ and $\ct{j}{B}$. In particular, the first requirement implies that the traces of Eq. \eqref{traces} can only appear in the combinations with definite transformation properties under CP and P listed in Table \ref{singletraces}.  Once the remaining conditions are imposed, we are left with a general, regularization-independent parametrization of the $\ci{k}{aA}$ and $\ct{j}{B}$ at the one-loop order. For example, for elements $\mathcal{I}^j_{A}$
linear or quadratic in the gauge fields, the coefficients $\ci{k}{aA}$ read:
\begin{align}
\ci{0}{ab}&= \ccoeff{0}{} (T^{ab}_{LL}+T^{ab}_{RR}-T^{ab}_{LR}-T^{ab}_{RL})\,,\nonumber \\
\ci{1}{a(bc)}&=~\ccoeff{1}{LLL} \left(T^{abc}_{LLL}+ T^{acb}_{LLL}-T^{abc}_{RRR}-T^{acb}_{RRR}) \right. \nonumber \\
& +\ccoeff{1}{RLL} \left. (T^{abc}_{RLL}+ T^{acb}_{RLL} -T^{abc}_{LRR}- T^{acb}_{LRR}\right)\nonumber \\
&+ \ccoeff{1}{LLR} \left(T^{abc}_{LLR}+ T^{acb}_{LRL} -T^{abc}_{RRL}  -T^{acb}_{RLR} + T^{abc}_{LRL}+ T^{acb}_{LLR}-T^{abc}_{RLR}-T^{acb}_{RRL}\right) \,, \label{eq:genparC}   \\
\ci{k =2,3}{abc}&=\ccoeff{k}{LLL} (T^{abc}_{LLL}- T^{acb}_{LLL} +T^{abc}_{RRR}- T^{acb}_{RRR}- T^{abc}_{LRL}+ T^{acb}_{LLR} -T^{abc}_{RLR} + T^{acb}_{RRL})  \nonumber \\
&+ \ccoeff{k}{LLR} (T^{abc}_{LLR}- T^{acb}_{LRL} +T^{abc}_{RRL}-T^{acb}_{RLR} - T^{abc}_{LRL}+ T^{acb}_{LLR} -T^{abc}_{RLR} + T^{acb}_{RRL})  \nonumber \\
&+ \ccoeff{k}{RLL} (T^{abc}_{RLL}- T^{acb}_{RLL}+T^{abc}_{LRR}- T^{acb}_{LRR} - T^{abc}_{LRL}+ T^{acb}_{LLR} -T^{abc}_{RLR} + T^{acb}_{RRL}) \,, \nonumber \\
\ci{k =4,5,6}{a(bc)}&=~\ccoeff{k}{}  (T^{abc}_{LLR}- T^{abc}_{LRL} +T^{abc}_{RRL}-T^{abc}_{RLR}- T^{abc}_{LRL} +  T^{abc}_{LLR} -T^{abc}_{RLR} +  T^{abc}_{RRL} ) \,. \nonumber
\end{align}

The parametrization for the remaining $\ci{k}{aA}$ can be found in Appendix \ref{app:parametrizatrions}.  

Analogously, the $\ct{j}{B}$ can be parametrized as:
\begin{align}
\ct{1}{abc}&=\ctcoeff{1}{LLL}(T^{abc}_{LLL} - T^{acb}_{LLL} +T^{abc}_{RRR}- T^{acb}_{RRR}) + \ctcoeff{1}{LLR} (T^{abc}_{LLR}- T^{acb}_{LRL} + T^{abc}_{RRL}- T^{acb}_{RLR})  \no  \\
&+\ctcoeff{1}{LRL} (T^{abc}_{LRL} - T^{acb}_{LLR} +T^{abc}_{RLR} - T^{acb}_{RRL}) + \ctcoeff{1}{RLL}(T^{abc}_{RLL} - T^{acb}_{RLL} + T^{abc}_{LRR} - T^{acb}_{LRR}) \,, \no  \\
\ct{j=2,3}{ab}&= \ctcoeff{k}{LL}(T^{ab}_{LL}+T^{ab}_{RR}) + \ctcoeff{k}{LR} ( T^{ab}_{LR}-T^{ab}_{RL}) \,,\no
\\
\ct{4}{[ab]c}&=\ctcoeff{4}{} \left(T^{abc}_{LRL}+T^{abc}_{LRR}-T^{abc}_{RLL}-T^{abc}_{RLR} -
T^{bac}_{LRL}- T^{bac}_{LRR}+T^{bac}_{RLL}+T^{bac}_{RLR}  \right) \,,\no \\
\ct{5}{abcd}& = \ctcoeff{5}{LRLR}\left(  -T^{abcd}_{RLRL}+T^{abdc}_{RLRL}+T^{acbd}_{RLRL} +T^{bacd}_{RLRL} -T^{bcad}_{RLRL}+T^{bcda}_{RLRL} \right. \no \\
&\qquad \qquad   \left. - T^{cabd}_{RLRL}+T^{cbad}_{RLRL}-T^{cbda}_{RLRL}-T^{dbac}_{RLRL}+T^{dbca}_{RLRL}-T^{dcba}_{RLRL} \right) \no  \\
& +  \ctcoeff{5}{LLLR}\left( T^{abcd}_{LLLR}- T^{abcd}_{RLLL}-T^{abcd}_{RLRR}+T^{abcd}_{RRLR}+T^{abdc}_{LLRL}+T^{abdc}_{RLLL}+T^{abdc}_{RLRR}  \right. \no \\
& \qquad \qquad - T^{abdc}_{RRLR} -T^{acbd}_{LLLR}+T^{acbd}_{RLLL}+T^{acbd}_{RLRR} -T^{acbd}_{RRLR}  -T^{acdb}_{RLRR} -T^{adbc}_{RLRR}  \no \\ 
&\qquad \qquad + T^{adcb}_{RLRR}+T^{bacd}_{RLLL}-T^{bcad}_{RLLL} +T^{bcda}_{RLLL} -T^{cabd}_{RLLL}-T^{cabd}_{RLRR}+T^{cadb}_{RLRR}  \no \\
& \qquad \qquad  + T^{cbad}_{RLLL} -T^{cbad}_{RRLR}  -T^{cbda}_{RLLL} +T^{cbda}_{RRLR}-T^{cdab}_{RLRR}+T^{cdba}_{RLRR}-T^{dabc}_{LLLR}  \no \\
&  \qquad \qquad + T^{dabc}_{RLRR}+T^{dacb}_{LLLR}-T^{dacb}_{RLRR} +T^{dbac}_{LLLR}-T^{dbac}_{LLRL}-T^{dbac}_{RLLL}-T^{dbac}_{RLRR}    \no\\
& \qquad \qquad + T^{dbac}_{RRLR} -T^{dbca}_{LLLR}+T^{dbca}_{LLRL}+T^{dbca}_{RLLL}+T^{dbca}_{RLRR}-T^{dbca}_{RRLR} -T^{dcab}_{LLLR}  \no \\
& \qquad \qquad \left. + T^{dcab}_{RLRR} +T^{dcba}_{LLLR}-T^{dcba}_{LLRL}-T^{dcba}_{RLLL}-T^{dcba}_{RLRR}+T^{dcba}_{RRLR}  \right)\,, \no \\
\ct{6}{abcd} = & \ctcoeff{6}{LLLL} \left(T^{abcd}_{LLLL} + T^{cabd}_{LLLL}+T^{cbad}_{LLLL}+T^{dbac}_{LLLL}+T^{cabd}_{RRRR} +
   T^{cbad}_{RRRR} +T^{dabc}_{RRRR} + T^{dbac}_{RRRR} \right)   \no\\
&   +  \ctcoeff{6}{RLLL} \left( T^{abcd}_{RLLL}+T^{abdc}_{RLLL}+T^{bacd}_{RLLL}+T^{bacd}_{RRLR} +T^{badc}_{RLLL}+T^{badc}_{RRLR}+T^{bcda}_{RLRR} +T^{bdca}_{RLRR}  \right.\no  \\
& \qquad \qquad + T^{cabd}_{LLLR}+T^{cabd}_{RLLL}+T^{cabd}_{RLRR}+T^{cabd}_{RRLR}+T^{cbad}_{LLLR}+T^{cbad}_{RLLL}+T^{cbad}_{RLRR}+T^{cbad}_{RRLR}  \no \\
& \qquad \qquad + T^{cdab}_{RLRR}+T^{cdba}_{RLRR}+T^{dabc}_{LLLR} +T^{dabc}_{LLRL}+T^{dabc}_{RLLL}+T^{dabc}_{RLRR}+T^{dabc}_{RRLR}+T^{dbac}_{LLLR}  \no \\
& \left. \qquad  \qquad  +  T^{dbac}_{LLRL} +T^{dbac}_{RLLL} +T^{dbac}_{RLRR} +T^{dbac}_{RRLR}+T^{dcab}_{LLLR} +T^{dcab}_{RLRR} +T^{dcba}_{LLLR}+T^{dcba}_{RLRR}  \right. ) \no  \\
&  + \ctcoeff{6}{LLRR} \left( T^{abcd}_{LLRR}+ T^{cdba}_{LLRR}+T^{dcba}_{LLRR}+T^{cdab}_{LLRRR}+T^{dcab}_{LLRR}+T^{abdc}_{LLRR}+T^{badc}_{LLRR}+T^{bacd}_{LLRR} \right)   \no\\
& + \ctcoeff{6}{LRLR}  \left( T^{abcd}_{LRLR} + T^{bcda}_{LRLR}+T^{bdca}_{LRLR}+T^{dbac}_{LRLR}+T^{cbad}_{LRLR}+T^{dcab}_{LRLR}+T^{dcba}_{LRLR}+T^{cdba}_{LRLR}  \right) \no \\
& + \ctcoeff{6}{RLLR} \left( T^{abcd}_{RLLR}+T^{abdc}_{RLLR}+T^{bacd}_{RLLR}+T^{badc}_{RLLR}+T^{cdab}_{RLLR}+T^{cdba}_{RLLR}+T^{dcab}_{RLLR}+T^{dcba}_{RLLR} \right)  \no \\
&  + \ctcoeff{6 \prime}{LLLL} \left(  T^{cadb}_{LLLL}+T^{dacb} _{LLLL}+T^{cadb}_{RRRR}+T^{dacb}_{RRRR} \right)   \no \\
&+  \ctcoeff{6 \prime}{RLLL} \left(T^{cadb}_{RLLL}+  T^{cadb}_{LLRL}+ T^{acbd}_{RLLL}+T^{bcad}_{RLLL}+T^{bcad}_{RLRR}+T^{bdac}_{RLRR}+T^{cadb}_{RLRR} +T^{cadb}_{RRLR}   \right. \no \\
&\left.  \qquad  \qquad  + T^{cbda}_{RLRR}+T^{dacb}_{LLLR} +T^{dacb}_{LLRL}+T^{dacb}_{RLLL}+T^{dacb}_{RLRR}+T^{dacb}_{RRLR}+T^{dbca}_{LLLR}+T^{dbca}_{RLRR} \right)   \no  \\
   &+  \ctcoeff{6 \prime}{RLLR} \left( T^{cadb}_{RLLR}+T^{acbd}_{RLLR}+T^{adbc}_{RLLR}+T^{bcad}_{RLLR}+T^{bdac}_{RLLR}+T^{cbda}_{RLLR}+T^{dacb}_{RLLR}+T^{dbca}_{RLLR} \right)   \no   \\
   &+  \ctcoeff{6 \prime}{LRLR} \left( T^{cadb}_{LRLR}+ T^{cbda}_{LRLR}+T^{adbc}_{LRLR}+T^{acbd}_{LRLR} \right)\,.  \no
\end{align}
Given the length of these expressions, we also provide the parametrizations of all $\ci{k}{A}$ and $\ct{j}{B}$ coefficients in a \texttt{Mathematica} notebook attached to the arXiv preprint of this article as an ancillary file. 

With our parametrization, $\ci{k=0-11}{Aa}$ automatically vanish for $T_L^a=T_R^a$ . For the same to hold for their hatted counterparts, the counterterm coefficients must obey four additional conditions: 
\begin{align}
\begin{aligned}
\label{eq:ct_vector_symmetry}
&\ctcoeff{2}{LL}+ \ctcoeff{2}{LR} + \ctcoeff{3}{LL} + \ctcoeff{3}{LR} = 0  \,, \\
&\ctcoeff{1}{LLL} + \ctcoeff{1}{RLL} + \ctcoeff{1}{LLR} + \ctcoeff{1}{LRL} - 2 i (\ctcoeff{2}{LL}  + \ctcoeff{2}{LR} )  = 0 \,,\\
& \ctcoeff{2}{LL}+ \ctcoeff{2}{LR} + 4(\ctcoeff{6}{LLLL}+ 4\ctcoeff{6}{RLLL}+\ctcoeff{6}{LLRR}+\ctcoeff{6}{LRLR} + \ctcoeff{6}{RLLR} ) = 0\,, \\
& \ctcoeff{2}{LL}+\ctcoeff{2}{LR}- 2 (\ctcoeff{6 \prime}{LLLL}+4 \ctcoeff{6 \prime}{RLLL}+2 \ctcoeff{6 \prime}{RLLR}+\ctcoeff{6 \prime}{LRLR})  =0 \,.
\end{aligned}
\end{align}
These allow us to express four coefficients, e.g. $\ctcoeff{1}{LRL}$, $\ctcoeff{3}{LR}$, $\ctcoeff{6}{RLLR}$ and $\ctcoeff{6 \prime}{LRLR}$, as a function of the others. We, therefore, conclude that the $\ci{k}{aA}$ are described by a total of 61 real parameters in the P-even sector and 27 in the P-odd one, while the $\ct{j}{B}$ (hence the  $\hatci{k}{aA}(\xi)$) depend on 13 real parameters in the P-even sector and 4 in the P-odd one. 
Note that at this stage the parameters describing $\ci{k}{aA}$ are still redundant, because -- as mentioned above --- the various $\ci{k}{aA}$ are related by the WZ conditions $\mathrm{WZ}[C^k_{cA}]=0$. In contrast to this, the $\hatci{k}{aA}(\xi)$ automatically satisfy $\mathrm{WZ}[\hatci{k}{cA}]=0$, hence there are no further restrictions on the $\ct{j}{B}$. 
 In order to remove the redundancy in the above parametrization of $\ci{k}{aA}$, we proceed to solve the constraints $\mathrm{WZ}[C^k_{cA}]=0$.
\subsubsection*{P-even sector}
We start from the P-even sector. Plugging the parametrizations for $\ci{k \in \mathrm{P-even}}{aA}$ into Eq.~\eqref{eq:WZcPeven}, we obtain 49 independent conditions on the coefficients entering the parametrizations (see Appendix \ref{app:solWZPeven} for the full expressions). This leaves us with $61-49=12$ free parameters, which we choose to be: 
\be
c^0, c^2_{LLL}, \,c^2_{RLL}, \, c^4,  \, c^6, c^7_{LLLR},  \,  c^7_{LRLR},  \, c^7_{LLRR},  \, c^7_{LRRL},  \, c^{7\prime}_{LLLR},  \, c^{7\prime}_{LRLR},  \, c^{7\prime}_{LLRR}\,. 
\label{eq:freeparamPeven}
\ee
From the conditions in \ref{app:solWZPeven} we can also conclude that, independently from the choice of free parameters, the conditions $\mathrm{WZ}[C^k_{cA}]=0$ fully determine $\ci{10}{abcd}$ as a combination of other coefficients in the P-even sector. Making use of the expressions for $\ci{k \in \mathrm{P-even}}{aA}$ in terms of the parameters in \eqref{eq:freeparamPeven}, we 
can solve the homogeneous master equation \eqref{hom}.
The solution
\begin{align}
\ctcoeff{1}{LLL}&= i \ccoeff{0}{} - \ccoeff{2}{LLL} - 2 i \ctcoeff{2}{LL}  \,, \qquad  
\ctcoeff{1}{LLR} =- \ccoeff{2}{RLL} \,,   \qquad 
\ctcoeff{1}{RLL}= 2 \ccoeff{4}{} -  \ccoeff{2}{RLL} \,,  \no \\
\ctcoeff{2}{LR}&= -  \frac{ 1 }{2} \ccoeff{0}{} +  i \ccoeff{4}{} + i  \ccoeff{6}{}  - \frac{i}{2} \ccoeff{2}{LLL}   - \frac{3 i}{2} \ccoeff{2}{RLL}   \,, \no \\
\ctcoeff{3}{LL}  &= \frac12  \ccoeff{0}{} -  \ctcoeff{2}{LL}   \,, \no  \\
\ctcoeff{6}{RLLL}  & = \frac14 \ccoeff{7}{LLLR}  \,,   \qquad   \ctcoeff{6}{LLRR}   = \frac14 \ccoeff{7}{LLRR}  \,,   \qquad  
\ctcoeff{6}{LRLR}   =\frac14 \ccoeff{7}{LRLR} \, , \label{eq:solpeven}  \\
 \ctcoeff{6}{LLLL}  &  = 
\frac18( \ccoeff{0}{}  - 2 i \ccoeff{6}{}  + i \ccoeff{2}{LLL} + 2 i  \ccoeff{2}{RLL}  -  8 \ccoeff{7}{LLLR} - 2 \ccoeff{7}{LLRR} - 2 \ccoeff{7}{LRLR} - 
    2 \ccoeff{7}{LRRL} - 2 \ctcoeff{2}{LL}) \,,\no \\
\ctcoeff{6 \prime}{LLLL}  & = \frac14 (-\ccoeff{0}{} + 2 i \ccoeff{6}{} - i \ccoeff{2}{LLL} - 4 \ccoeff{7\prime}{LLLR} - 2 \ccoeff{7\prime}{LLRR} - \ccoeff{7 \prime}{LRLR} + 2 \ctcoeff{2}{LL}) \,, \no \\
\ctcoeff{6 \prime}{RLLL}  & = \frac14 \left( \ccoeff{7 \prime}{LLLR} -\frac{i}{2} \ccoeff{2}{RLL} \right)   \,, \qquad 
\ctcoeff{6 \prime}{RLLR}    = \frac14 \left(i  \ccoeff{4}{} - \frac{i}{2} \ccoeff{2}{RLL} + \ccoeff{7\prime}{LLRR} \right)\,, \no
\end{align}
determines the P-even counterterms $\ct{1,2,3,6}{B}$ and explicitly shows the absence of anomalies in this sector.\footnote{In particular, all counterterms are fixed by using the master equation for $k  \, \in \,  \mathrm{P-even} \setminus \{10\}$. $\ci{10}{abcd}+\hatci{10}{abcd}(\ct{}{})=0$ is automatically satisfied.} In other words, in the P-even sector the gauge variation of the effective action can always be compensated by a counterterm. 

The conditions \eqref{eq:solpeven} fix only 12 out of the 13 available counterterm coefficients. The residual freedom amounts to the possibility of adding to $\cL_{\rm ct}$ the gauge-invariant counterterm:
\begin{align}
\begin{aligned}
\label{eq:gaugeinvcount}
\cL_{\rm ct} &  \supset  \ctcoeff{2}{LL}  \left(  \mathcal{I}_{ab}^2 - \mathcal{I}_{ab}^3 -2 f_{ceb} \mathcal{I}^1_{eca} +\frac12 f_{dga} f_{ecb} \mathcal{I}^6_{egcd} \right)  (T_{LL}^{ab}+ T_{RR}^{ab})   \\
& = -\frac{ \ctcoeff{2}{LL}  }{2}   F_{\mu \nu}^a F^{b \mu \nu}  (T_{LL}^{ab}+ T_{RR}^{ab})  \,.    
\end{aligned}
\end{align}
This term is manifestly gauge invariant because $T_{LL}^{ab}$ and $T_{RR}^{ab}$ can be written as the direct sum of identifies in the adjoint representations of the gauge group, each multiplied by a representation-dependent Casimir.

\subsubsection*{P-odd sector}
We now repeat the same procedure in the P-odd sector. Plugging the parametrizations for $\ci{k \in \mathrm{P-odd}}{aA}$ into Eq.~\eqref{eq:WZcPodd} we obtain 23 conditions, which we list in Appendix \ref{app:solWZPodd}. Hence, only $27-23=4$ out of the $27$ coefficients appearing in $\ci{k \in \mathrm{P-odd}}{aA}$ are truly independent. We choose\footnote{Note that, as in the P-even sector, the WZ conditions fully determine $\ci{11}{abcd}$ as a combination of the coefficients entering $\ci{1,9}{abcd}$}: 
\be
\ccoeff{1}{LLL} , \, \ccoeff{1}{LLR} ,   \, \ccoeff{9}{LRLR},  \,  \ccoeff{9}{LLLR}  \,.
\ee
On the other hand, the P-odd counterterms depend only on three parameters: $\ctcoeff{4}{}$, $\ctcoeff{5}{LLLR}$ and $\ctcoeff{5}{LRLR}$.
We can use them to remove $\ccoeff{1}{LLR}$, $\ccoeff{9}{LRLR}$ and $\ccoeff{9}{LLLR}$ by choosing\footnote{As  in the previous section, all counterterms are fixed by using the master equation for $k =1,9$. $\ci{11}{abcd}+\hatci{11}{abcd}(\ct{}{})=0$  is automatically satisfied.}:
\begin{align}
\label{eq:solpodd}
\ctcoeff{4}{}&={ \frac{\ccoeff{1}{LLR}}{2}} \,, & 
\ctcoeff{5}{LLLR}&=-\frac{\ccoeff{9}{LLLR}}{12} \,, &   
\ctcoeff{5}{LRLR}&=-\frac{\ccoeff{9}{LRLR}}{12}\,.
\end{align}
The extra coefficient, $\ccoeff{1}{LLL} $ is related to the anomaly. In fact, by combining  the parametrizations for $\ci{k \in \mathrm{P-odd}}{aA}$, the constraints from the WZ conditions in Eq.~\eqref{eq:WZcPodd} and the counterterm choice in \eqref{eq:solpodd}, we get
\begin{align}
\begin{aligned}
\label{eq:Pddanomaly}
& \sum_{k \in \mathrm{P-odd}} \left[\ci{k}{aA}+\hatci{k}{aA}(\ct{}{})\right]  I^k_{A}(x) =\\
 & \qquad\qquad\qquad c^1_{LLL}  \left[  2  T_{LLL}^{a(bc)} ~I^1_{(bc)} - i \left( T_{LLLL}^{ab[cd]}  + T_{LLLL}^{a[c|b|d]} +  T_{LLLL}^{ba[cd]} \right) I^9_{b[cd]} -(L\to R) \vphantom{  T_{LLL}^{a(bc)}} \right]  \,,
\end{aligned}
\end{align}
where the vertical bars indicate that indices inbetween them do not get antisymmatrized. Since $\ci{k \in \mathrm{P-even}}{aA}$ satisfy the homogeneous equation \eqref{hom}, the right-hand side of \eqref{eq:Pddanomaly} can be identified with $\mathcal{A}_a$. Using the explicit expressions for $I^9_{(bc)}$ and $I^6_{b[cd]}$, we can write it as
\begin{equation}
\mathcal{A}_a= - c^1_{LLL}    \epsilon^{\mu \nu \rho \sigma}     \partial_\mu \left(A_\nu^b \partial_\rho A_\sigma^e     -  \frac{i}{4}     A_\nu^b A_\rho^c A_\sigma^d (i f^{cde})  \right)  \mathrm{tr} \left( \left[ T_L^a \left\{ T_L^b, T_L^e \right\} \right] -  \left[ T_R^a \left\{ T_R^b, T_R^e \right\} \right] \right)   \,. 
\end{equation}
Because there is no freedom left in choosing the counterterms, the condition $\mathcal{A}_a=0$ can only be satisfied by imposing Eq.~\eqref{georgi}. 
\subsubsection{Fermionic sector}
\begin{figure}[t]
\centering
\begin{minipage}{.24\textwidth}
  \includegraphics[width=1.\linewidth]{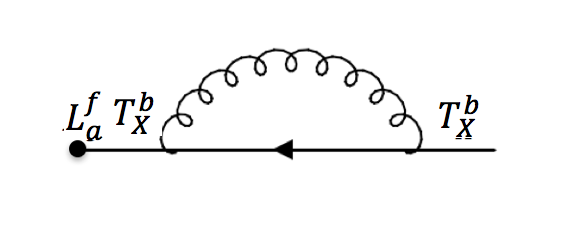}
\end{minipage}%
\begin{minipage}{.24\textwidth}
  \includegraphics[width=1.\linewidth]{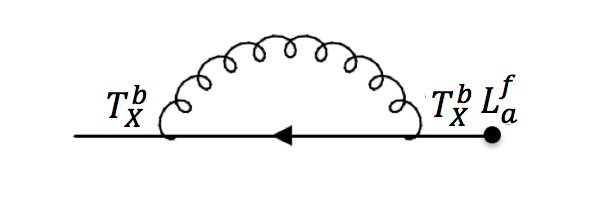}
\end{minipage}
\begin{minipage}{.24\textwidth}
  \includegraphics[width=1.\linewidth]{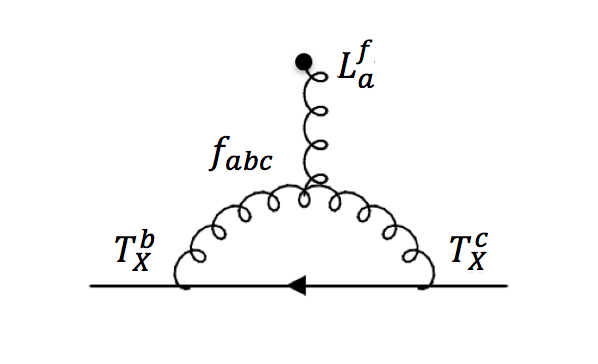}
\end{minipage}
\begin{minipage}{.24\textwidth}
  \includegraphics[width=1.\linewidth]{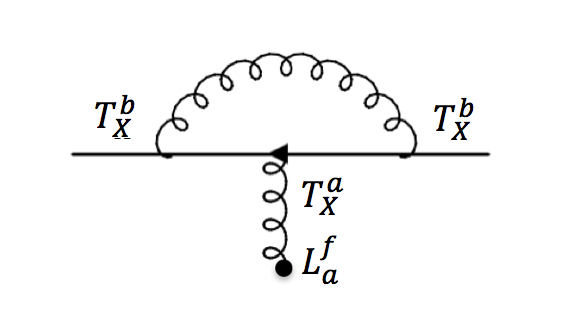}
\end{minipage}
\caption{Diagrams contributing to $\ci{12}{aX} I^{12}_{X}   + \ci{13}{aX} I^{13}_{X}$. The  chirality $X$ is determined by the external fields $\bar f_{X}$ and $f_{X}$.  A dot indicates the action of the operator $L_{a}$ of Eq. \eqref{Loperator}.  }
\label{diagrams1}
\end{figure}

We now turn to the fermionic sector, where it is convenient to first focus on the coefficients $\ci{12}{aX}$ and $\ci{13}{aX}$. Both
are matrices in flavor space that can be parametrized in terms of strings of generators. At the one-loop
order such strings are not completely generic, since the relevant diagrams are the ones depicted in Fig. \ref{diagrams1}, from which we infer the patterns: 
\begin{align}
\begin{aligned}
\ci{12}{aX}&=a_1 T^a_XT^b_XT^b_X + a_2 T^b_XT^b_X T^a_X + a_3 f_{abc} T^b_XT^c_X + a_{4Y} T^b_X T^a_YT^b_X\,, \\
\ci{13}{aX}&=b_1 T^a_XT^b_XT^b_X + b_2 T^b_XT^b_X T^a_X + b_3 f_{abc} T^b_XT^c_X + b_{4Y} T^b_XT^a_YT^b_X\,,
\end{aligned}
\end{align}
where a sum over $Y=L,R$ is understood. The Lie algebra guarantees that the combination $T^a_XT^a_X$ satisfies $[T^a_XT^a_X,T^b_X]=0$
for any $T^b_X$, while $f_{abc} T^b_XT^c_X$ is proportional to $T^a_X$. Without losing generality, we can thus write:
\begin{align}
\ci{12}{aX}&=a'_{1X} T^a_X  + a_{4Y} T^b_XT^a_YT^b_X\nn\\
\ci{13}{aX}&=b'_{1X} T^a_X  + b_{4Y} T^b_XT^a_YT^b_X\,,
\end{align}
where the matrices $a'_{1X}$ and $b'_{1X}$ commute with all generators $T^a_X$. 
We can further refine the parametrization of $\ci{12}{aX}$ and $\ci{13}{aX}$ by imposing invariance under CP. On the one side we have
\be
\ci{12}{aX}\xrightarrow{\text{CP}}\ci{13 \, T}{aX}=b'_{1X} T^{aT}_X  + b_{4Y} T^{bT}_XT^{aT}_YT^{bT}_X\,.
\ee
On the other side we recall that under CP $T^a_X\xrightarrow{\text{CP}} T^{aT}_X$ and we obtain
\be
\ci{12}{aX}\xrightarrow{\text{CP}}a'_{1X} T^{aT}_X  + a_{4Y} T^{bT}_XT^{aT}_YT^{bT}_X\,.
\ee
The two ways lead to the same result provided $a'_{1X}=b'_{1X}$ and $a_{4Y}=b_{4Y}$, resulting in
\be
\ci{12}{aX}=\ci{13}{aX}=a'_{1X} T^a_X  + a_{4Y} T^b_XT^a_YT^b_X\,,\nn\\
\ee
holding at least at one-loop order.
Moreover, by making use of $\ci{12}{cX}=\ci{13}{cX}$, from the WZ consistency conditions (see Appendix A), we 
can express $\ci{14}{abX}$ in terms of $\ci{13}{cX}$:
\be
\ci{14}{abX}=i(\ci{13}{bX}T^a_X-T^a_X \ci{13}{bX}+if_{abc}\ci{13}{cX})\,.
\ee
Therefore the independent coefficients relevant for the  one-loop parametrization of the gauge variation in the fermionic sector are provided by the matrix $\ci{13}{cX}$. We now show that, for any choice of $\ci{13}{cX}$, the homogeneous equation (\ref{hom}) can always be solved, thus proving the absence of anomalies in this sector of the theory.
When $k=12,13,14$, Eq.~\eqref{hom} gives:
\begin{align}
\begin{aligned}
\label{fersys1}
&T^a_X \ct{7}{X} +i\ct{8}{Xa} =i\ci{13}{aX}\,,\\
&\ct{7}{X}  T^a_X +i\ct{8}{Xa} =i\ci{13}{aX}\,, \\
&T^a_X \ct{8}{Xb}  -\ct{8}{Xb}T^a_X -if_{abc}\ct{8}{Xc} =-(\ci{13}{bX}T^a_X-T^a_X \ci{13}{bX}+if_{abc}\ci{13}{cX})\,.
\end{aligned}
\end{align}
By combining the first two equations we see that $\ct{7}{X}$ should commute with all generators $T^a_X$:
\be
\ct{7}{X}  T^a_X-T^a_X \ct{7}{X}=0\,.
\ee
The third equation is automatically satisfied once we eliminate $\ct{8}{Xa}$ in favour of $\ct{7}{X}$ and $\ci{13}{aX}$.
As a consequence, \eqref{fersys1} only determines one combination of $\ct{7}{X}$ and $\ct{8}{Xa}$:
\be
\label{replace}
\ct{8 \prime}{Xa}=-i\ct{7}{X}  T^a_X +\ct{8}{Xa}=\ci{13}{aX}\,.
\ee
By expressing the searched-for counterterm in terms of $\ct{7}{X}$ and $\ct{8 \prime}{Xa}$, we get:
\be
\bar f_{X}  \ct{7}{X}(\slashed{\partial}+iT^a_X\slashed{A}_a) f_{X}+\bar f_{X} \ct{8 \prime}{aX} \slashed{A}_a  f_{X}\,.
\ee
\noindent
Since the matrix $\ct{7}{X}$ commutes with all generators, and thus with all gauge transformations, in the above expression the first term is gauge invariant and can be safely dropped because it does not affect \eqref{defCT}. We end up with
\be
\bar f_{X} \ct{8 \prime }{aX}\slashed{A}_a  f_{X}\,,
\label{eq:fermionicCTgeneral}
\ee
as the unique non-trivial counterterm, where $\ct{8 \prime}{aX}$ is given in Eq. \eqref{replace}. 
\section{One-loop analysis in Dimensional Regularization}
\label{sec:DR}
In this section we present explicit one-loop results for the variation of the effective action and the WI-restoring counterterms in DR, using the BMHV prescription for $\gamma_5$. First, we introduce the conventional dimensionally regularized action, and then we perform the explicit one-loop computation. 
\subsection{Classical action in DR}
\label{sec:reg}
In DR Lorentz indices are analytically extended from $d=4$ to $d= 4-2 \epsilon$ complex dimensions. In this respect it is necessary to slightly modify the notation we used so far. In the present section (only), vector Lorentz indices like $\mu,\nu$ run from $0$ to $d$, and split into a four-dimensional set denoted by $\bar\mu, \bar\nu$ and a $d-4$-dimensional (evanescent) one labeled $\hat\mu,\hat\nu$. As we will discuss more extensively in Section \ref{sec:DRrenscheme}, the gauge transformation is however taken to be purely four-dimensional in nature. Explicitly, the operators $L_a(x)$ in DR is defined as:
\ba\label{LoperatorDR}
L_a(x)&=&-\partial_{\bar\mu} \frac{\delta}{\delta A_{a\bar\mu}(x)}+f_{abc}A_{b\bar\mu}(x)\frac{\delta}{\delta A_{c\bar\mu}(x)}\\\no
&+&\sum_{X=L,R}
-i\frac{\overleftarrow{\delta}}{\delta f_X(x)}T^a_X f_X(x)+i\bar f_X(x)T^a_X\frac{\delta}{\delta \bar f_X(x)}\,. 
\ea
The operators $\{{I}^k_{A} (x)\}$ and $\{\mathcal{I}^j_{B} (y)\}$ of tables \ref{tab:anomaly_building_blocks} and \ref{tab:counterterms_building_blocks} are strictly four-dimensional.

The spurious breaking of gauge invariance in DR arises because chiral fermions cannot be defined for arbitrary $d$. Indeed, as is well known, it is impossible to define a $d$-dimensional Clifford algebra
\be
\{\gamma^\mu,\gamma^\nu\}=2 g^{\mu\nu}\,,
\ee
and a chirality matrix $\gamma_5$ that commutes with all $d-$dimensional Lorentz generators.  More specifically, there is no $d-$dimensional definition of $\gamma_5$ obeying all the familiar four-dimensional properties, namely i) $\{\gamma^\mu,\gamma_5\}=0$, ii) $\mathrm{tr}(\gamma^\mu\gamma^\nu\gamma^\rho\gamma^\sigma\gamma_5)=4 i\epsilon^{\mu\nu\rho\sigma}$, and iii) cyclicity of the trace. Several treatments of $\gamma_5$ retaining i) have been put forward, see for example Refs.~\cite{Chanowitz:1979zu,Jegerlehner:2000dz,Kreimer:1989ke,Korner:1991sx}. Unfortunately, none of them has been proven to be consistent to all orders. Here we adhere to the BMHV prescription, which has been rigorously established to all orders in perturbation theory~\cite{Breitenlohner:1975qe, Breitenlohner:1975hg,Breitenlohner:1977hr, Breitenlohner:1976te}. In this approach the conditions ii) and iii) are preserved while i) is relaxed. In particular, the matrix $\gamma_5$ is taken to be an intrinsically four-dimensional object, and the other $\gamma_\mu$ matrices are split into a four- and a $(d-4)$- dimensional part, denoted by $\gamma_{\bar\mu}$ and $ \gamma_{\hat\mu}$, respectively: 
\be
\gamma_\mu =  \gamma_{\bar\mu} + \gamma_{\hat\mu} \,.
\label{eq:split}
\ee
 An algebraically consistent scheme is then obtained by requiring:
\be\label{CommAnticomm}
 \left \lbrace \gamma_5, \gamma_{\bar\mu} \right \rbrace = 0\,, \qquad  \left[ \gamma_5, \gamma_{\hat\mu} \right] = 0\,.
\ee
Eq. \eqref{CommAnticomm} makes it impossible for $\gamma_5$ to commute with all the $d-$dimensional Lorentz generators. Hence the notion of chirality is lost and, as we will see, a spurious (or genuine) violation of gauge invariance is bound to emerge. 

We now proceed to introduce the dimensionally regularized version of the classical action in Eq.~\eqref{eq:action_classic_unregularized}. While the regularization of Feynman diagrams via DR requires an extension of the kinetic terms to $d$ dimensions, the treatment of the interaction terms is, to a large extent,  arbitrary: the only requirement is that they must reduce to those in \eqref{eq:action_classic_unregularized} for $d\to4$. This leaves open the possibility of defining a large class of regularization schemes. For the bosonic Lagrangian $\cL_{\mathrm{YM}}$, a natural choice is to promote it entirely to $d$ dimensions following the recipe outlined above, i.e. replacing $\cL_{\mathrm{YM}}\to\cL_{\mathrm{YM}}^{ (d)}$. While this choice is obviously not unique, it is by far the most convenient, because it preserves all the symmetries of the unregularized theory. For this reason, it will be adopted in the following. Also the fermionic contribution $\cL^{\mathrm{Fermions}}$ allows for several independent analytic continuations. There is however a fundamental distinction with respect to the bosonic action: because of the absence of $d-$dimensional chirality, there is no way to define a regularized fermionic action that respects chiral gauge invariance. Here we choose the following regularized fermion Lagrangian:
\begin{align}
\label{Lfermions}
\begin{aligned}
\cL_{\mathrm{Fermions}}^{(d)} 
&=i \bar{f} \gamma^\mu \partial_\mu f-A_\mu^a~ \bar{f} \left(P_R\gamma^\mu P_L   T_L^a + P_L\gamma^\mu P_R  T_R^a \right)f\\
&=i \bar{f} \gamma^\mu \partial_\mu f-A_{\bar\mu}^a~ \bar{f} \left(P_R\gamma^{\bar\mu} P_L   T_L^a + P_L\gamma^{\bar\mu} P_R  T_R^a \right)f\,, 
\end{aligned}
\end{align}
with $P_{L,R}$ being the $d-$dimensional versions of the operators introduced around Eq.~\eqref{fLfR} for the (unregularized) four-dimensional theory. Even for arbitrary $d$ $P_{L,R}$ represent hermitian projectors that can be employed to define what we will call $d-$dimensional left- and right-handed fermions, precisely as in \eqref{fLfR}. The crucial difference is that the fermionic kinetic term (which, consistently with DR, is $d-$dimensional) introduces $f_L\leftrightarrow f_R$ transitions, whereas the interaction is purely four-dimensional and does not mediate such regularization-dependent transitions. 
In conclusion, the $d-$dimensional action that replaces \eqref{eq:action_classic_unregularized} is taken to be:  
\be
\label{classreg}
S^{(d)}[A, f_X, \bar f_X] = \int d^d x \, ( \cL_\mathrm{YM}^{(d)}  + \cL_{\mathrm{Fermions}}^{(d)} ) \,.
\ee
Because this definition of $S^{(d)}$ is effectively part of the regularization scheme, all scheme-dependent quantities (including the counterterms derived below) depend on it, and will generally differ if another $S^{(d)}$ is adopted (see also Ref.~\cite{Belusca-Maito:2020ala}). Since the regularized Yang-Mills Lagrangian defined above is widely used in the literature, most of the scheme-dependence (within DR) stems from the fermionic Lagrangian. We will further comment on such scheme-dependence in Section \ref{sec:1loopFerm}. For now let us just stress that any alternative interaction scheme, such as those defined by $-A_\mu^a~ \bar{f} \left(\gamma^\mu P_L   T_L^a + \gamma^\mu P_R  T_R^a \right)f$ or $-A_\mu^a~ \bar{f} \left(P_R\gamma^\mu   T_L^a + P_L\gamma^\mu  T_R^a \right)f$, differs from ours because of the addition of evanescent terms.

The choice in Eq.~\eqref{Lfermions} is motivated by minimality of the resulting gauge variation which, as we will see below, is the central quantity in computing the variation of the 1PI effective action, $\Delta_a|_{(1)}$. In practice, \eqref{Lfermions} minimizes the number of diagrams to be computed in order to identify the WI-restoring counterterms. Perhaps even more importantly, \eqref{Lfermions} preserves P, CP, the vectorial gauge group (see below) and hermiticity of the action, which allow us to perform intermediate checks during the calculations. We also emphasize that, at variance with other approaches \cite{Belusca-Maito:2020ala,Martin:1999cc}, our regularization does not require the introduction of additional fermions. The fermion content of our theory is exactly the same as in the four-dimensional theory. This makes our results directly applicable to theories of interest, like the SM. 
\subsection{Breaking of gauge invariance in DR: general considerations}
\label{sec:DRrenscheme}
Having introduced the regularized action the general results of Section \ref{sec:regu} can be invoked to identify the WI-restoring counterterm $S_{\rm ct}|_{(1)}$. To make contact with the notation of Section \ref{sec:regu} we observe that the quantity \eqref{classreg} represents the tree-level regularized action, $S^{(d)}\equiv\Gamma^{\rm reg}|_{(0)}$, whereas more generally $\Gamma^{\rm reg}|_{(n)}=\Gamma^{(d)}|_{(n)}$. 

At a given perturbative order, the gauge variation of $\Gamma^{(d)}[\phi]|_{(n)}$ contains both purely $4$-dimensional as well as evanescent terms. The evanescent terms are defined as those contributions that are proportional to $d-4$ components of the fields, or contain space-time derivatives in the $(d-4)$-dimensional coordinates. Such contributions to the effective action cannot describe physical processes because the latter are genuinely 4-dimensional. Physical processes are obtained by differentiating the effective action with respect to the 4-dimensional components of the background fields, assumed to carry purely 4-dimensional external momenta. For this reason {{evanescent}} contributions to $\Gamma^{(d)}$ do not have any physical significance.

To avoid any confusion we emphasize that this statement refers to the 1PI effective action, as opposed to the classical action. Evanescent terms actually appear in the classical action, are essential to the regularization procedure and in fact are at the origin of anomalies. Explicitly, performing $4$-dimensional transformations of the fermionic and bosonic fields one finds that $4$-dimensional gauge invariance is indeed explicitly broken by the regularized action \eqref{classreg}:
\ba\label{breaking}
L_a(x)S^{(d)} &=&L_a(x)S^{(d)}_{\rm Fermions}\\\no
&=&-\left[\bar{f}_L\gamma^\mu T^a_L (\partial_\mu f_R)+\bar{f}_R\gamma^\mu T^a_R (\partial_\mu f_L)+(\partial_\mu \bar{f}_L)\gamma^\mu T^a_R  f_R+(\partial_\mu\bar{f}_R)\gamma^\mu T^a_L  f_L\right](x)  \\\no
&=& \mathcal{O}({\rm Eva}),
\ea%
where $\mathcal{O}({\rm Eva})$ indicates that this is an evanescent quantity because it is controlled by terms of the type $\bar f_X\gamma^\mu f_{Y\neq X}$ which do not exist in $d=4$. As already anticipated earlier, the fundamental reason why $\delta_\alpha S^{(d)}[A, f_X,\bar f_X]$ is not exactly zero is that the $d-$dimensional kinetic term characterizing DR necessarily mediates $f_L\leftrightarrow f_R$ transitions.\footnote{At the root of these transitions is that the projectors $P_{L,R}$ do not commute with the $J_{\bar\mu\hat\mu}$ generators of the $d-$dimensional Lorentz group, which is respected by the kinetic term (see Eq.~\eqref{CommAnticomm}). Hence, Lorentz transformations mix $f_L$, $f_R$, as opposed to what happens in $d=4$.} More specifically, the mixed terms $f_L^\dagger f_R, f_R^\dagger f_L$ are not gauge invariant unless the gauge transformation is vector-like, i.e. our regularization \eqref{classreg} explicitly violates gauge invariance unless $T^a_L =T^a_R$. When $T^a_L =T^a_R$ the gauge variation in Eq.~\eqref{breaking} reduces to a total derivative with respect to the $d-4$ coordinates, which can be safely ignored. Only in this case our DR scheme does not break the physical, four-dimensional gauge invariance. Note that our choice $S^{(d)}_{\rm Fermions}$ minimizes the breaking because the four-dimensional nature of the interaction conserves chirality: any other interaction scheme would feature additional terms on the right-hand side of Eq.~\eqref{breaking}.

In DR the gauge invariance is explicitly lost already at tree-level whenever $T^a_L \neq T^a_R$, i.e. whenever the theory is chiral. Any choice of $\cL_{\mathrm{Fermions}}^{(d)}$ would suffer from the same drawback. The dimensionally regularized classical action \eqref{classreg} is nevertheless invariant under the spurious P and CP transformation laws of Eqs.~(\ref{candcp}) and (\ref{spurion}), as its four-dimensional sibling.\footnote{This is a consequence of the properties of the charge conjugation matrix $C$ in $d$-dimensions (see Eqs. (2.15) and (2.16) of Ref. \cite{Belusca-Maito:2020ala}).} The associated selection rules will be heavily exploited in the calculations of the following sections. There is another sacred principle that appears to be violated by \eqref{classreg}: the fermion interaction does not respect $d-$dimensional Lorentz transformations. However, this violation does not have tangible consequences, because the symmetry principle of physical relevance is the four-dimensional Lorentz group, not its $d-$dimensional extension. Indeed, Eq. \eqref{classreg} preserves  four-dimensional Lorentz provided all the $d-4$ indices, e.g. $\gamma^{\hat\mu}$, are viewed as scalars of $SO(1,3)$. As a result, DR does not require the introduction of counterterms to enforce the Ward Identities associated with physical Lorentz invariance. With this in mind, by an abuse of terminology, we will keep referring to \eqref{classreg} as to the regularized ``$d-$dimensional action". The reader should note that the situation is radically different when considering the breaking of gauge invariance, since Eq. \eqref{breaking} reveals that \eqref{classreg} does not respect even the (physically relevant) four-dimensional version of \eqref{eq:gauge_transformations}, where the gauge parameters $\alpha_a$ are assumed to depend only on the coordinates $x^{\bar\mu}$. The very existence of WIs associated to four-dimensional gauge invariance demands the addition of local counterterms to \eqref{classreg}.

As anticipated earlier, evanescent contributions to the 1PI effective action are unphysical. In particular, the breaking \eqref{breaking} has no effect in the tree approximation, since this is an evanescent quantity that does not exist when $\epsilon \to 0$; said differently, the operatorial version of \eqref{breaking} does not have any tree matrix element with (four-dimensional) physical states. For example, tree matrix elements of $\bar{f}_{L}\gamma^\mu \partial_\mu f_{R}=\bar f_L\gamma^{\hat\mu}\partial_{\hat\mu} f_R$ depend on the unphysical momentum along the $d-4$ directions, and similarly for all other terms. However, when going beyond the tree level in the perturbative expansion, the evanescent terms in the classical action may get multiplied by singular integrals, resulting in non-evanescent contributions to the 1PI action that spoil the Ward Identities. This is the origin of the spurious breaking terms that force us to introduce counterterms.

An explicit expression for $\Delta_a$ in DR can be derived order by order in perturbation theory. As anticipated in Eq. \eqref{1PI}, the regularized 1PI effective action in the background field method can be written as:
\be
\label{1PIDR}
e^{i\Gamma^{(d)}[\phi]}=
\int_{\rm 1PI}{\cal D}{\tilde \phi}~e^{iS_{\rm full}^{(d)}[\phi+{\tilde{\phi}} ]}
\ee
where $S^{(d)}_{\rm full}\equiv S^{(d)}+S^{(d)}_{\rm g.f.}+S^{(d)}_{\rm ghost}$ is the sum of the $d-$dimensional action \eqref{classreg}, an invariant gauge-fixing term\footnote{$f_a$ is the $d-$dimensional version of the expression in Eq. \eqref{GFGen}.}:
\be
\label{GaugeFixingGG}
S_{\rm g.f.}[\phi+{\tilde \phi}]=\int d^d x~\left[-\frac{1}{2\xi} f_a f_a\right]\,,
\ee
and the associated ghost action. It is a remarkable property of DR that the non-invariance of the $d-$dimensional action $S^{(d)}$, see Eq. \eqref{breaking}, represents the only source of gauge-symmetry breaking. In particular, under a gauge transformation the measure of the dimensionally-regularized path integral remains invariant because any local transformation of the field is associated to a Jacobian ${\cal J}$ of the form $\ln{\rm det}{\cal J}=\delta^{(d)}(0)\int d^dx~f(x)$, with some function $f(x)$ that depends on the transformation parameters, and in DR $\delta^{(d)}(0)$ identically vanishes, implying that ${\cal J}=1$. Any potential anomaly in local field transformations in DR must therefore come from the non-invariance of the classical action. In particular, the gauge variation of the 1PI effective action reads
\be\label{anomaly1}
L_a\Gamma^{(d)}[\phi]
=\frac{\int_{\rm 1PI}{\cal D}\tilde \phi ~e^{iS^{(d)}_{\rm full}[\phi+{\tilde \phi}]}~L_a S^{(d)}_{\rm Fermions}[\phi+\tilde \phi]}{\int_{\rm 1PI}{\cal D}{\tilde \phi}~e^{iS^{(d)}_{\rm full}[\phi+{\tilde \phi}]}}.
\ee
Thus, the spurious gauge symmetry breaking terms arise from the one-particle irreducible vacuum correlation functions of the gauge variation of the classical fermionic action. This is the regularized version of the Quantum Action Principle of \eqref{QAexplained}. 

According to Eq. \eqref{defCT}, the WI-restoring counterterm $S_{\rm ct}|_{(1)}$ is determined by the variation of renormalized 1PI effective action. We should therefore discuss how this is connected to the variation of the regularized 1PI action in \eqref{anomaly1}. To appreciate this it is necessary to introduce a renormalization scheme. 

In general, there are two types of contributions to the regularized 1PI effective action: (finite as well as divergent) evanescent terms and (finite as well as divergent) non-evanescent terms. In formulas, we may write
\be\label{formGamma1DR}
\Gamma^{(d)}|_{(1)}=\overline\Gamma^{\rm fin}|_{(1)}+\frac{1}{\epsilon}\overline\Gamma^{\rm div}|_{(1)}+\widehat\Gamma^{\rm fin}|_{(1)}+\frac{1}{\epsilon}\widehat\Gamma^{\rm div}|_{(1)},
\ee
where a bar/hat identifies the non-evanescent/evanescent contributions.\footnote{There is some ambiguity in this expression because via space-time integration by parts it is possible to convert a divergent evanescent operator into a finite non-evanescent one. Such ambiguity is however absent at the level of momentum-space Feynman diagrams. The expression in Eq.~\eqref{formGamma1DR} is to be understood as a collection of momentum-space correlators, where no space-time integration by parts is performed.} In this paper we adopt a popular (minimal) subtraction scheme according to which the renormalized effective action is defined by subtracting all divergent terms, both the evanescent and non-evanescent ones, so that it reduces to the sum of finite evanescent and finite non-evanescent terms analogously to the tree-level expression $S^{(d)}=\Gamma|_{(0)}$:
\ba
\Gamma|_{(1)}\equiv\lim_{d\to4}\left\{\overline\Gamma^{\rm fin}|_{(1)}+\widehat\Gamma^{\rm fin}|_{(1)}\right\}.
\ea
The formal $4$-dimensional limit is carried out by discarding $\widehat\Gamma^{\rm fin}|_{(1)}$ and sending all fields and momenta in $\overline\Gamma^{\rm fin}|_{(1)}$ to $d=4$. The gauge variation \eqref{QAexplained} of the renormalized effective action hence coincides with 
\be
\left. \Delta_a \right|_{(1)}=L_a\Gamma|_{(1)}=L_a\overline\Gamma^{\rm fin}|_{(1)}. 
\ee
This is the quantity that determines $S_{\rm ct}|_{(1)}$.

Similarly to $\Gamma^{(d)}|_{(1)}$, the gauge variation $L_a\Gamma^{(d)}|_{(1)}$ of the regularized action is in general the sum of evanescent terms and non-evanescent terms. In evaluating \eqref{anomaly1} we find two contributions:
\be\label{VarGamma}
L_a\Gamma^{(d)}|_{(1)}=\overline{\Delta}^{\rm fin}_a  |_{(1)}+\widehat{\Delta}^{\rm fin}_a |_{(1)}+\frac{1}{\epsilon}\widehat\Delta_a^{\rm div} |_{(1)},
\ee
namely a (finite) $4$-dimensional one and an (finite plus divergent) evanescent one. Crucially, the action of $L_a$ on any finite term remains finite, and similarly the action of $L_a$ on a divergent term remains divergent. Furthermore, $L_a$ cannot turn an evanescent term into a non-evanescent one. These considerations imply that\footnote{Incidentally, \eqref{VarGamma} also implies that the divergent $4$-dimensional terms $\overline\Gamma^{\rm div}|_{(1)}$ are gauge-invariant.}
\be
\label{resultDelta1}
\Delta_a|_{(1)}=\overline{\Delta}^{\rm fin}_a|_{(1)}  \, . 
\ee
This represents an important simplifying result for us: in a 1-loop calculation, and with the subtraction scheme illustrated above, the variation of the renormalized 1PI action is fully determined by the finite $4$-dimensional part of \eqref{anomaly1}. This is the only contribution necessary to identify the corresponding counterterm $S_{\rm ct}|_{(1)}$.

In the next subsection, we will present an explicit one-loop calculation of \eqref{anomaly1}. Because the focus of our paper is $S_{\rm ct}|_{(1)}$, the result summarized in Eq. \eqref{resultDelta1} ensures that in that calculation we can safely neglect the divergent evanescent terms in $L_a\Gamma^{(d)}|_{(1)}$. Yet, were we interested in carrying out a 2-loop computation of $S_{\rm ct}$, an explicit expression of the 1-loop counterterms necessary to subtract the divergences from $\Gamma^{(d)}|_{(1)}$ would also be needed.
\subsection{Breaking of gauge invariance in DR: one-loop calculation}
\label{sec:1-loop}
There are several important simplifications that occur in the computation of \eqref{anomaly1} at the one-loop order. First, we only need the expansion of $S_{\rm full}^{(d)}[\phi+{\tilde \phi}]$ up to quadratic order in the quantum fluctuations ${\tilde \phi}$. Second, since by definition the effective action \eqref{1PI} includes only one-particle irreducible diagrams, terms linear in the quantum fluctuations do not contribute and can be discarded. Furthermore, as we will see shortly, ghosts do not play any role at the order of interest. In particular, we can safely switch off both their classical backgrounds and their quantum fluctuations. As a consequence, the only relevant degrees of freedom in our analysis are the gauge and the fermionic fields, along with their quantum fluctuations.

The central player in our calculation is the fermionic action. Upon performing the shift $A_\mu^a\to A_\mu^a+{\tilde A}_\mu^a$, the covariant derivative becomes $i{\slashed{D}}\to i{\slashed{D}}-\gamma^{\bar\mu} {\tilde A}_{\bar\mu}^a(P_L   T_L^a + P_R  T_R^a)$. Expanding up to quadratic order we obtain
\ba\label{Sfermions1}
S^{(d)}_{\rm Fermions}[\phi+\tilde \phi]&=&\int d^dx\,\bar{f}i{\slashed{D}}f\\\no
&+&\int d^dx~{\bar{\tilde{f}}}i{\slashed{D}} {\tilde f}+{\cal S}^{(d)}_{\rm F}\\\no
&+&{\cal O}({\tilde \phi,\tilde\phi^3}),
\ea
where we defined
\be\label{DeltaSdef}
{\cal S}^{(d)}_{\rm F}\equiv\int d^dx\left[-{\tilde{A}}_{\bar\mu}^a {\bar{\tilde{f}}} \gamma^{\bar\mu} (P_L   T_L^a + P_R  T_R^a)f-{\tilde{A}}_{\bar\mu}^a\bar{f}\gamma^{\bar\mu} (P_L   T_L^a + P_R  T_R^a){\tilde f} \right],
\ee
and, as promised, we neglected terms linear and cubic in the fluctuations. The first term in \eqref{Sfermions1} represents the classical fermionic action, and can be factored out of the path integral \eqref{anomaly1} because it involves no quantum fluctuations. The second line of Eq. \eqref{Sfermions1} consists of the sum of two terms: a non-gauge-invariant one, ${\bar{\tilde{f}}}i \slashed{D}  \tilde f$, which represents the original fermionic Lagrangian with the fermionic field replaced by its quantum fluctuation and the covariant derivative containing only the background gauge field, plus a genuinely four-dimensional gauge-invariant piece we called ${\cal S}^{(d)}_{\rm F}$. At one-loop accuracy it is sufficient to expand $e^{i{\cal S}^{(d)}_{\rm F}}$ up to quadratic order, because \eqref{DeltaSdef} is linear in the background fermionic fields, and $L_a\Gamma^{(d)}[\phi]$ is a dimension-four local operator that contains at most two powers of such fields. Furthermore, the linear term in $e^{i{\cal S}^{(d)}_{\rm F}}=1+i{\cal S}^{(d)}_{\rm F}-\frac{1}{2}[{\cal S}^{(d)}_{\rm F}]^2+\cdots$ does not contribute, because no 1PI diagram can be built out of it. We then conclude that the one-loop approximation of \eqref{anomaly1} reads
\ba\label{anomaly2}
\left.\delta_\alpha\Gamma^{(d)}[A,f_X,\bar f_X]\right|_{(1)}
&=&\delta_\alpha S^{(d)}\\\no
&+&\langle\Omega|\delta_\alpha \left(\int d^dx\,{\bar{\tilde{f}}} i{\slashed{D}}{\tilde{f}} \right)|\Omega\rangle_A\\\no
&-&\frac{1}{2}\langle\Omega|T\left\{\left[{\cal S}^{(d)}_{\rm F}\right]^2~\delta_\alpha \left(\int d^dx\,{\bar{\tilde{f}}} i{\slashed{D}} {\tilde{f}}  \right)\right\}|\Omega\rangle_A\,,
\ea
where the time-ordered Green-functions are vacuum to vacuum correlators in the background gauge $A^a_\mu$:
\be
\langle\Omega|T\left\{O(x)O(y)\right\}|\Omega\rangle_A\equiv\frac{\int_{\rm 1PI}{\cal D}{\tilde \phi}~e^{i\int d^dx\,{\bar{\tilde{f}}}  i{\slashed{D}}{\tilde f}+iS^{(d)}_{\rm gauge}[A+{\tilde A}]}~O(x)O(y)}{\int_{\rm 1PI}{\cal D} {\tilde \phi} ~e^{i\int d^dx\,{\bar{\tilde{f}}} i{\slashed{D}} {\tilde{f}} +iS^{(d)}_{\rm gauge}[A+{\tilde A}]}}\,,
\ee
and we introduced the compact notation $S^{(d)}_{\rm Gauge}\equiv S^{(d)}_{\rm YM}+S^{(d)}_{\rm g.f.}+S^{(d)}_{\rm ghost}$. 

The quantity $\delta_\alpha S^{(d)}$ in \eqref{anomaly2} describes the classical effect \eqref{breaking} and can be ignored because finite evanescent. The second and third terms instead induce contributions that do not vanish for $\epsilon\to0$, because divergent $1/\epsilon$ one-loop effects turn them into finite non-evanescent. In four dimensions the one-loop gauge variation reads
\be\label{4dAnomalyFirst}
\left.\delta_\alpha\Gamma^{(4)}\right|_{(1)}=\left.\delta_\alpha\Gamma^{(4)}\right|_{\rm Gauge}+\left.\delta_\alpha\Gamma^{(4)}\right|_{\rm Fermions},
\ee
where we introduced the notation
\ba\label{1loopGauge}
\left.\delta_\alpha\Gamma^{(d)}_{\rm Gauge} \right|_{(1)}
&=&\langle\Omega|\delta_\alpha \left(\int d^dx\,{\bar{\tilde{f}}} i{\slashed{D}} {\tilde{f}} \right)|\Omega\rangle_A\,,
\\\label{1loopFermion}
\left.\delta_\alpha\Gamma^{(d)}_{\rm Fermions}\right|_{(1)}
&=&-\frac{1}{2}\langle\Omega|T\left\{\left[{\cal S}^{(d)}_{\rm F}\right]^2~\delta_\alpha \left(\int d^dx\,{\bar{\tilde{f}}} i{\slashed{D}}{\tilde{f}} \right)\right\}|\Omega\rangle_A\,.
\ea
The term \eqref{1loopGauge} arises from a single ${\tilde f}$ loop and only depends on the background gauge fields. At one loop the gauge bosons in these diagrams are necessarily non-dynamical, i.e. the gauge field is a purely classical background. The term \eqref{1loopFermion} instead receives contributions from diagrams with both virtual fermions and gauge bosons, and its explicit form depends on the fermionic background.

It is easy to see that at one loop ghosts can be neglected. Indeed, one-loop diagrams contributing to either \eqref{1loopGauge} or \eqref{1loopFermion} cannot simultaneously involve virtual ghosts and the necessary virtual fermions. We can therefore safely neglect ghosts, keeping in mind that they should not be ignored when performing calculations beyond the one-loop approximation. 

\subsubsection{Bosonic sector}
\label{sec:1loopFerm}

The gauge variations in Eqs.~\eqref{1loopGauge} and \eqref{1loopFermion} become significantly more compact when expressed in terms of vector and axial combinations of the gauge fields. These are defined, along with the associated generators, as
\ba\label{VA}
{\cal V}_\mu=T_V^aA^a_\mu,~~~~&&~~~~T_V^a=\frac{1}{2}({T}_R^a+{T}_L^a)\,, \\\no
{\cal A}_\mu=T_A^aA^a_\mu,~~~~&&~~~~T_A^a=\frac{1}{2}({T}_R^a-{T}_L^a).
\ea
We therefore prefer to temporarily switch notation from $T_{L,R}$ to $T_{V,A}$. To avoid confusion we restrict this change of notation to this section. 

Another useful quantity is 
\be\label{defT}
T^a={T}_V^a+T_A^a\gamma_5=P_LT_L^a+P_RT^a_R.
\ee
For clarity, we stress that the matrices $T_{L}, T_R$ do not live in orthogonal spaces and therefore do not commute in general. As a result neither $T_V^a$ nor $T_A^a$ usually form an algebra. Yet, orthogonality of the chirality projectors always implies $[{T}^a,{T}^b]=if_{abc}{T}^c$.\footnote{More explicitly, the reader might want to verify that
\ba\label{AlgebraGeneral}
[{T}^a_V,{T}^b_V]&=&\frac{1}{2}if^{abc}{T}^c_V+\frac{1}{4}[{T}^a_R,{T}^b_L]+\frac{1}{4}[{T}^a_L,{T}^b_R]\,, \\\no
[{T}^a_A,{T}^b_A]&=&\frac{1}{2}if^{abc}{T}^c_V-\frac{1}{4}[{T}^a_R,{T}^b_L]-\frac{1}{4}[{T}^a_L,{T}^b_R]\,, \\\no
[{T}^a_V,{T}^b_A]&=&\frac{1}{2}if^{abc}{T}^c_A-\frac{1}{4}[{T}^a_R,{T}^b_L]+\frac{1}{4}[{T}^a_L,{T}^b_R].
\ea}

To familiarize with the new notation let us begin by re-writing the first term in \eqref{anomaly2}:
\ba\label{SdEVA}
\delta_\alpha S^{(d)}&=&\delta_\alpha\left\{\int dx\,\bar{f}i{\slashed{D}}f\right\}\\\no
&=&\int d^dx\left[\alpha_a\bar{f}{T}^a_A\left\{{\slashed{D}},\gamma_5\right\}f+\partial_{\hat\mu}\alpha_a
\,\bar fT^a\gamma^{\hat\mu}f\right]\\\no
&=&{\rm Eva}.
\ea
It is easy to see that this expression correctly reproduces Eq. \eqref{breaking} after integration by parts. A similar quantity, with the replacement $f\to \tilde f$, is needed to compute the two remaining contributions. We find
\ba\label{anomalyFerm1}
\left.\delta_\alpha \Gamma^{(d)}_{\rm Gauge}\right|_{(1)}
&=&\int d^dx\langle\Omega|\left[\alpha_a {\bar{\tilde f}}{T}^a_A\left\{{\slashed{D}},\gamma_5\right\} {\tilde f}+\partial_{\hat\mu}\alpha_a
\,{\bar{\tilde f}} T^a\gamma^{\hat\mu}{\tilde f} \right]|\Omega\rangle_A\\\no
&=&-{\rm Tr}\left[\alpha_a{T}^a_A\left\{{\slashed{D}},\gamma_5\right\}\frac{1}{{\slashed{D}}}\right] -{\rm Tr}\left[ \partial_{\hat\mu}\alpha_a
T^a\gamma^{\hat\mu}\frac{1}{{\slashed{D}}}\right],
\ea
where the minus sign in the second line arises due to Fermi statistics. The trace ``Tr" differs from the Dirac trace ``tr" because it acts on the Dirac indices as well as space-time, i.e. ${\rm Tr}[O]=\int d^d x\,\langle x|{\rm tr}[O]|x\rangle$. 

As a non-trivial consistency check of \eqref{anomalyFerm1}, we note that this quantity arises from a single fermion loop with gauge bosons evaluated on their classical backgrounds. In this approximation the 1PI effective action reads $-i{\rm det}[{\slashed{D}}]$ and its variation may alternatively be given by $-i{\rm Tr}[{\slashed{D}}^{-1}\delta_\alpha{\slashed{D}}]$. An explicit computation gives $\delta_\alpha(i{\slashed{D}})=-[{\slashed{D}},\alpha_aT_V^a]-[{\slashed{D}},\alpha_aT_A^a]\gamma_5+\partial_{\hat\mu}\alpha_aT^a\gamma^{\hat\mu}$, so that 
\ba
-i{\rm Tr}[{\slashed{D}}^{-1}\delta_\alpha{\slashed{D}}]&=&{\rm Tr}\left[\frac{1}{{\slashed{D}}}[{\slashed{D}},\alpha_aT_V^a]+\frac{1}{{\slashed{D}}}[{\slashed{D}},\alpha_aT_A^a]\gamma_5-\frac{1}{{\slashed{D}}}\partial_{\hat\mu}\alpha_aT^a\gamma^{\hat\mu}\right]\\\no
&=&{\rm Tr}\left[\alpha_aT_V^a-\frac{1}{{\slashed{D}}}\alpha_aT_V^a{\slashed{D}}+\alpha_aT_A^a\gamma_5-\frac{1}{{\slashed{D}}}\alpha_aT_A^a{\slashed{D}}\gamma_5-\partial_{\hat\mu}\alpha_a
T^a\gamma^{\hat\mu}\frac{1}{{\slashed{D}}}\right]\\\no
&=&-{\rm Tr}\left[\alpha_aT_A^a\left\{{\slashed{D}},\gamma_5\right\}\frac{1}{{\slashed{D}}}+\partial_{\hat\mu}\alpha_a
T^a\gamma^{\hat\mu}\frac{1}{{\slashed{D}}}\right],
\ea
where we used ${\rm Tr}[\alpha_aT^a_A\gamma_5]=0$ and the cyclicity of the trace. The above expression exactly agrees with \eqref{anomalyFerm1}, as it should. Incidentally, this consistency check also provides an indirect proof of the gauge invariance of the dimensionally-regularized path integral measure (see discussion above Eq. \eqref{anomaly1}). 

The trace in Eq.~\eqref{anomalyFerm1} may be computed diagrammatically or via other methods. Any of these would lead to the same result because the expression has been already regularized and is unambiguous. In the following, we employ the heat kernel method. The following result was first obtained in Ref.~\cite{Balachandran:1981cs} via this same method. We think that a re-derivation makes our work more complete and self-contained, and at the same time might help clarify a few non-trivial steps of the computation.

Now, the second term in the second line of Eq.~\eqref{anomalyFerm1} is evanescent and can be safely discarded as argued around \eqref{resultDelta1}. The first term in Eq.~\eqref{anomalyFerm1} is however not entirely negligible. In Appendix \ref{app:heatkernel} we show that the first trace in the second line of Eq.\eqref{anomalyFerm1} may be expressed in terms of the heat kernel coefficient $a_2$ plus divergent evanescent terms. Neglecting all evanescent terms, using \eqref{4dfermionAnomaly1} and explicitly evaluating $a_2(x,x)$ via \eqref{a2General}, after a tedious but straightforward computation we arrive at\footnote{Note that the first trace includes both gauge and Lorentz indices, whereas the second only the gauge indices are summed over.}
\ba
\label{anomalyFerm2}
{\lim_{d\to4}\left.L_a\Gamma^{(d)}_{\rm Gauge}\right|_{(1)}}
=\frac{i}{8\pi^2}~{\rm tr}\left[T_A^a\gamma_5a_2(x,x)\right] \equiv -{\rm tr}\left[T_A^a({a}_2^\epsilon(x)+{a}_2^{\slashed\epsilon}(x))\right],
\ea
with
\ba\no
{a}_2^{\epsilon}
&=&\frac{\epsilon^{\mu\nu\alpha\beta}}{16\pi^{2}}\left[{\cal V}_{\mu\nu}{\cal V}_{\alpha\beta}+\frac{1}{3}{\cal A}_{\mu\nu}{\cal A}_{\alpha\beta}-\frac{8}{3}i\left({\cal A}_\alpha {\cal A}_\beta {\cal V}_{\mu\nu}+{\cal A}_\alpha {\cal V}_{\mu\nu}{\cal A}_\beta+{\cal V}_{\mu\nu}{\cal A}_\alpha {\cal A}_\beta\right)-\frac{32}{3}{\cal A}_\mu {\cal A}_\nu {\cal A}_\alpha {\cal A}_\beta\right]\,, \\
\label{A21}
{a}_2^{\slashed\epsilon}&=&\frac{1}{16\pi^{2}}\left[\frac{4}{3}D^{\cal V}_\nu D^{\cal V}_\nu D^{\cal V}_\mu {\cal A}_{\mu}+\frac{8}{3}i[{\cal A}_\mu,D^{\cal V}_\nu {\cal V}_{\mu\nu}]-\frac{2}{3}i[{\cal A}_{\mu\nu},{\cal V}_{\mu\nu}]\right]\\\no
\label{A22}
&+&\frac{1}{16\pi^{2}}\left[-8{\cal A}_\mu (D^{\cal V}_\nu {\cal A}^\nu){\cal A}_\mu-\frac{8}{3}\left\{D^{\cal V}_\mu {\cal A}_\nu+D^{\cal V}_\nu {\cal A}_\mu,{\cal A}_\mu {\cal A}_\nu\right\}+\frac{4}{3}\left\{D^{\cal V}_\mu {\cal A}^\mu,{\cal A}_\nu {\cal A}_\nu\right\}\right]\, .
\ea
In the above expression we introduced the field strengths of the vector and axial components of the four-dimensional gauge fields:
\begin{align}
\begin{aligned}
\label{AVfs}
{\cal V}_{\mu\nu}&=&\partial_{\mu}{\cal V}_{\nu}-\partial_{\nu}{\cal V}_{\mu}+i[{\cal V}_{\mu},{\cal V}_{\nu}]+i[{\cal A}_{\mu},{\cal A}_{\nu}]\\ 
{\cal A}_{\mu\nu}&=&\partial_{\mu}{\cal A}_{\nu}-\partial_{\nu}{\cal A}_{\mu}+i[{\cal V}_{\mu},{\cal A}_{\nu}]+i[{\cal A}_{\mu},{\cal V}_{\nu}].
\end{aligned}
\end{align}
Our result \eqref{anomalyFerm2} agrees with Ref.~\cite{Balachandran:1981cs}, where a different convention for the gauge vectors was adopted. Interestingly, note that the one-loop variation $\delta_\alpha\Gamma^{(4)}|_{\rm Gauge}$ is completely independent from the definition of the interaction in the regularized fermionic action \eqref{Lfermions}. Any alternative regularization of the interaction would differ by evanescent terms involving $\hat\mu$-components of the vector fields, and these would not affect the four-dimensional limit of \eqref{anomalyFerm1}. The mixed fermion-boson loops appearing in \eqref{1loopFermion} are instead sensitive to such definitions and below will be evaluated for our choice \eqref{Lfermions}.

The gauge variation in Equations~\eqref{anomalyFerm2} and \eqref{A21} satisfies all the desired properties. First, since in our convention the generators $T^a_{V,A}$ are hermitian, the factors of $i$ in \eqref{A21} guarantee that $\delta_\alpha \Gamma^{(4)}$ is hermitian. Second, the vector-like component of the gauge symmetry, defined by $T^a_A=0$ (or, equivalently, by $T^a_L=T^a_R$), is manifestly conserved, consistently with what is anticipated below Eq. \eqref{breaking}. Third, expressions \eqref{anomalyFerm2} and \eqref{A21} are consistent with $L_a\Gamma^{(4)}$ being CP-odd and P-even, see below Eq. \eqref{spurion}. In particular, ${a}_2^{\slashed\epsilon}$ is P-odd because it contains an odd number of axial vectors, whereas ${a}_2^{\epsilon}$ is P-odd because it contains an even number of axial vectors contracted with the Levi-Civita tensor. Finally, the expression \eqref{anomalyFerm2} satisfies the WZ conditions, as we will discuss below.

The operators in ${a}_2^{\epsilon},{a}_2^{\slashed\epsilon}$ form a complete set of P-odd, Lorentz-singlet, dimension-four local functions of the vectors and their derivatives compatible with vector gauge invariance. As expected, this operator basis is in one-to-one correspondence with the one presented in Table \ref{tab:anomaly_building_blocks}. We can therefore equally decompose
Eq.~\ref{anomalyFerm2} as we did in Section \ref{sec:WZgeneral} (see Eq. \ref{var1} and text around it). The corresponding coefficients $\ci{k}{A}$ are collected in Section~\ref{sec:DRresults}. 

\subsubsection{Fermionic sector}
\label{sec:DRfermions}

The mixed fermion-gauge contribution to $\delta_\alpha\Gamma^{(4)}$ may be calculated directly from its definition \eqref{1loopFermion} (${\cal S}^{(d)}_{\rm F}$ is given in \eqref{DeltaSdef}):
\ba\label{LaGauge}
&& \left. L_{c} \Gamma^{(d)}_{\rm{Fermions}}  \right|_{(1)} \\\no
&&=-2\langle\Omega|T\left\{\int d^dy_1\left[{\tilde{A}}_{\bar\rho}^a\bar{f}\gamma^{\bar\rho} T^a {\tilde{f}} \right]_{y_1}\left[{\bar{\tilde f}} {T}^c_A\gamma^{\hat\mu}\gamma_5\partial_{\hat\mu}{\tilde f} \right]_x\int d^dy_2\left[{\tilde{A}}_{\bar\sigma}^b {\bar{\tilde f}}\gamma^{\bar\sigma} T^b f\right]_{y_2}\right\}|\Omega\rangle_A+{\rm Eva}\,,
\ea
where we used the variation \eqref{SdEVA}, where $\left\{{\slashed{D}},\gamma_5\right\}=2\gamma^{\hat\mu}\gamma_5\partial_{\hat\mu}$, as well as the definition \eqref{defT}. The numerical factor in front is a multiplicity factor due to the presence of two possible contractions with $\left[{\cal S}^{(d)}_{\rm F}\right]^2$.

The full result of our computation will be presented below. Here, for brevity, we discuss explicitly only the derivation of terms containing two background fermions and a derivative. The remaining ones are of the form $\bar f{\slashed{A}}f$, involving background fermions and a background gauge field, and can be obtained analogously.

In the evaluation of terms containing no gauge fields the average in \eqref{LaGauge} can be interpreted as a vacuum to vacuum transition. We find
\ba
&&\left. L_{c} \Gamma^{(d)}_{\rm{Fermions}} \right|_{(1)} \\\no
&&\supset2\int\frac{d^dk_1}{(2\pi)^d}\int\frac{d^dk_2}{(2\pi)^d}~e^{i(k_1-k_2)x}\int\frac{d^dq}{(2\pi)^d}\\\no
&&\bar f(k_1)\gamma^{\bar\rho}T^a\frac{({\slashed{q}}+{\slashed{k_1}})}{(q+k_1)^2}T^c_A({\hat{\slashed{q}}}+{\hat{\slashed{k_2}}})\gamma_5\frac{({\slashed{q}}+{\slashed{k_2}})}{(q+k_2)^2}\gamma^{\bar\sigma}T^a f(k_2)~G_{aa}\left(\frac{g^{\bar\rho\bar\sigma}-(1-\xi)\frac{q^{\bar\rho}q^{\bar\sigma}}{q^2}}{q^2}\right)\\\no
&&= - \frac{2G_{aa}}{16\pi^2}\left(1+\frac{\xi-1}{6}\right)\int\frac{d^dk_1}{(2\pi)^d}\int\frac{d^dk_2}{(2\pi)^d}~e^{i(k_1-k_2)x}~\bar f(k_1)\gamma_5i\left({\slashed{k_1}}-{\slashed{k_2}}\right){T}^a{T}_A^c{T}^a f(k_2)\\\no
&&=- \frac{2G_{aa}}{16\pi^2}\left(1+\frac{\xi-1}{6}\right)~\bar{f}\gamma_5\left(\overrightarrow{\slashed{\partial}}+\overleftarrow{\slashed{\partial}}\right){T}^a{ T}_A^c{T}^af,
\ea
where we made use of the shorthand notation in Eq.~\eqref{defT}. The couplings $G_{aa}$ arise from the gauge propagator because the kinetic term in \eqref{eq:classicalYM} is non-canonical. This contribution can be expressed as in Eq.~\ref{var1}. The resulting coefficients $C^{12,13}$, along with those associated with the $\bar f{\slashed{A}}f$ terms, are collected in the next section, together with those of the purely bosonic operators. 

\subsubsection{Collecting the results}
\label{sec:DRresults}

The one-loop results derived in Section~\ref{sec:DRfermions} and
\ref{sec:1loopFerm} can all be written in the form \eqref{var1}. The corresponding coefficients $\ci{k}{pA}$ are:
\ba\label{coeffFINAL}
C^0_{pa}&=& \frac{1}{16 \pi^2}{\rm tr}\,{T}_A^p\left\{-\frac{4}{3}\,{T}_A^a\right\}\\\no
C^1_{pab}&=& \frac{1}{16 \pi^2} {\rm tr}\,{T}_A^p\left\{  4\left({T}_V^a{T}_V^b+\frac{1}{3}{T}_A^a{T}_A^b\right)\right\}  \\\no
C^2_{pab}&=&\frac{1}{16 \pi^2} {\rm tr}\,{T}_A^p\left\{ {-}  \frac{8}{3}i\,\left([{T}_A^a,{T}_V^b]+[{T}_V^a,{T}_A^b]\right)\right\}\\\no
C^3_{pab}&=& \frac{1}{16 \pi^2}{\rm tr}\,{T}_A^p\left\{  \frac{4}{3}i\,[{T}_A^a,{T}_V^b]{-}4i\,[{T}_V^a,{T}_A^b]\right\}\\\no
C^4_{iab}&=& \frac{1}{16 \pi^2}{\rm tr}\,{T}_A^p\left\{{-}  4i\,[{T}_V^a,{T}_A^b]\right\}\\\no
C^5_{pab}&=&-\frac{1}{3}C^4_{pab}\\\no
C^6_{pab}&=& \frac{1}{3}C^4_{pab}\\\no
C^7_{pabc}
&=&\frac{1}{16 \pi^2} {\rm tr}\,{T}_A^p\left\{\frac{8}{3}\,[{T}_A^b,[{T}_A^c,{T}_A^a]]+8\,{T}_A^b{T}_A^a{T}_A^c-\frac{4}{3}\left\{{T}_A^a,{T}_A^b{T}_A^c\right\}\right.\\\no
&+&\left.\frac{4}{3}\,[{T}_V^a,[{T}_V^b,{T}_A^c]]+\frac{4}{3}\,[{T}_V^b,[{T}_V^c,{T}_A^a]]+\frac{8}{3}\,[{T}_A^b,[{T}_V^c,{T}_V^a]]\right\}
\\\no
C^8_{pabc}
&=&\frac{1}{16 \pi^2} {\rm tr}\,{T}_A^p\left\{\frac{8}{3}\,[{T}_V^b,[{T}_V^a,{T}_A^c]]+\frac{8}{3}\,[{T}_V^b,[{T}_V^c,{T}_A^a]]\right.\\\no
&+&\frac{8}{3}\,[{T}_A^c,[{T}_A^a,{T}_A^b]]+\frac{8}{3}\left\{{T}_A^a,\left\{{T}_A^b,{T}_A^c\right\}\right\}+\frac{16}{3}\,[{T}_A^c,[{T}_V^a,{T}_V^b]]+\frac{8}{3}\,[{T}_A^b,[{T}_V^c,{T}_V^a]]\\\no
&-&\left.\frac{4}{3}\,[{T}_A^a,[{T}_A^b,{T}_A^c]]+\frac{4}{3}\,[{T}_V^a,[{T}_V^b,{T}_A^c]]+\frac{4}{3}\,[{T}_V^a,[{T}_A^b,{T}_V^c]]-\frac{4}{3}\,[{T}_A^a,[{T}_V^b,{T}_V^c]]\right\}\\\no
C^9_{pabc}&=&\frac{1}{16 \pi^2} {\rm tr}{T}_A^p \left\{+4i \left\{{T}_V^a, {T}_V^b{T}_V^c+{T}_A^b{T}_A^c\right\}+\frac{2}{3}i \left\{{T}_A^a, [{T}_V^b,{T}_A^c]+[{T}_A^b,{T}_V^c]\right\}\right.\\\no
&-&\left.\frac{16}{3}i \left({T}_A^b{T}_A^c{T}_V^a+{T}_A^b{T}_V^a{T}_A^c+{T}_V^a{T}_A^b{T}_A^c\right)\right\}\\\no
C^{10}_{pabcd}
&=& \frac{1}{16 \pi^2}{\rm tr}\,{T}_A^p \left\{  {+}  \frac{8}{3}i\,[{T}_A^a,[{T}_V^c,[{T}_A^b,{T}_A^d]]]{-} \frac{2}{3}i\,[[{T}_V^a,{T}_A^c],[{T}_A^b,{T}_A^d]]{-}  \frac{2}{3}i\,[[{T}_A^a,{T}_V^c],[{T}_A^b,{T}_A^d]]\right.\\\no
&{+} &8i\,{T}_A^a[{T}_V^c,{T}_A^d]{T}_A^b {+}  \frac{8}{3}i\left\{[{T}_V^a,{T}_A^c],\left\{{T}_A^b,{T}_A^d\right\}\right\}{-}  \frac{4}{3}i\left\{[{T}_V^a,{T}_A^b],{T}_A^c{T}_A^d\right\}\\\no
&{+}  &\frac{4}{3}i\,[{T}_V^a,[{T}_V^b,[{T}_V^c,{T}_A^d]]]{+}  \frac{8}{3}i\,[{T}_A^a,[{T}_V^c,[{T}_V^b,{T}_V^d]]]\\\no
&{-} &\left.\frac{2}{3}i\,[[{T}_V^a,{T}_A^c],[{T}_V^b,{T}_V^d]]{-} \frac{2}{3}i\,[[{T}_A^a,{T}_V^c],[{T}_V^b,{T}_V^d]]\right\}\\\no
C^{11}_{pabcd}&=&\frac{1}{16 \pi^2} {\rm tr}\,{T}_A^p \left\{  4 \left({T}_V^a{T}_V^b+{T}_A^a{T}_A^b\right)\left({T}_V^c{T}_V^d+{T}_A^c{T}_A^d\right) \right. \\\no 
&{+} & \frac{1}{3} \left([{T}_V^a,{T}_A^b]+[{T}_A^a,{T}_V^b]\right)\left([{T}_V^c,{T}_A^d]+[{T}_A^c,{T}_V^d]\right)\\\no
&{-}&\frac{16}{3}\left[{T}_A^c{T}_A^d({T}_V^a{T}_V^b+{T}_A^a{T}_A^b)+{T}_A^c({T}_V^a{T}_V^b+{T}_A^a{T}_A^b){T}_A^d+({T}_V^a{T}_V^b+{T}_A^a{T}_A^b){T}_A^c{T}_A^d\right]\\\no
&{+} &\left.\frac{32}{3}{T}_A^a{T}_A^b{T}_A^c{T}_A^d\right\}\,, \\\no
C^{12}_{pLij}&=&-\frac{1}{16 \pi^2} \left(\frac{5+\xi}{6}\right)\left[T^a_L(T_R^p-T_L^p)T^a_L\right]_{ij}\\\no
C^{12}_{pRij}&=&+\frac{1}{16 \pi^2} \left(\frac{5+\xi}{6}\right)\left[T^a_R(T_R^p-T_L^p)T^a_R\right]_{ij}\\\no
C^{13}_{pLij}&=&C^{12}_{pLij}\\\no
C^{13}_{pRij}&=&C^{12}_{pRij}\\\no
C^{14}_{pLaij}&=&+ \frac{i}{16 \pi^2} \left(\frac{5+\xi}{6}\right)\left\{[T^p_L,T_L^mT^a_RT_L^m]-if_{pan}T^m_LT_R^nT_L^m\right\}_{ij}\\\no
C^{14}_{pRaij}&=&-\frac{i}{16 \pi^2} \left(\frac{5+\xi}{6}\right)\left\{[T^p_R,T_R^mT^a_LT_R^m]-if_{pan}T^m_RT_L^nT_R^m\right\}_{ij}\,,
\ea
where ${\rm tr}\, {T}_A^p\left\{\cdots\right\}$ is short for ${\rm tr}[{T}_A^p \left\{\cdots\right\}]$.\footnote{The coefficients $C^{12,13,14}$ of the fermionic operators are written in terms of the $T_{L,R}$ generators because they only carry gauge indices. On the contrary, $T_{V,A}$ also involve Lorentz indices, which are fully contracted in the definitions of $I^{12,13,14}$.} 
Note that these $\ci{k}{A}$ do not automatically satisfy the symmetry properties of Eq.~\eqref{eq:Csymmetries} and need to be (anti)symmetrized accordingly. We nevertheless prefer to report the results without (anti)symmetrization to avoid complicating these already unwieldy expressions. 

The results collected in Eq. \eqref{coeffFINAL} pass a number of highly non-trivial consistency checks. To start, the coefficients $\ci{0-6}{A}$ and $\ci{12,13,14}{A}$ have been independently computed diagrammatically for $\xi =1$. The Feynman diagrams exactly reproduce the coefficients in Eq.~\eqref{coeffFINAL}. Furthermore, we explicitly verified that the $\ci{k}{A}$ in Eq.~\ref{coeffFINAL}, after being properly (anti)symmetrized, satisfy the WZ conditions in \ref{app:equations_ci}. We also computed the corresponding values of the coefficients $\ccoeff{k}{}$ introduced in Eq. \eqref{traces} (see Table \ref{tab:CTcoeffDRresults}) and checked that these satisfy the constraints in \ref{app:solWZ}, as they should.

\begin{table}[t]
\centering
\renewcommand{\arraystretch}{1.2} 
\scalebox{1}{
\begin{tabular}{|c| c|}
\hline
$\ci{k}{A}$ & $\ccoeff{k}{XYZ}$ \\
\hline
\hline
$\ci{0}{ }$ &  $\ccoeff{0}{}= -\frac{1}{3}$  \\
\hline 
$\ci{1}{ }$ &  $\ccoeff{1}{LLL}= -\frac{1}{3}$,  $\ccoeff{1}{LLR}= -\frac{1}{6}$, $\ccoeff{1}{RLL}= \frac{1}{3}$   \\
\hline
$\ci{2}{ }$ & $ \ccoeff{2}{LLL}= -\frac{2i}{3}$,  
 $\ccoeff{2}{LLR}= 0$, $\ccoeff{2}{RLL}= \frac{2 i}{3}$ \\
\hline
$\ci{3}{ }$ & $ \ccoeff{3}{LLL} = -\frac{i}{3}$
$, \ccoeff{3}{LLR} = \frac{2 i}{3}$ $,
\ccoeff{3}{RLL}  = \frac{i}{3}$ \\
\hline
$\ci{4}{ }$  &  $\ccoeff{4}{}= \frac{i}{2}$ \\
\hline
$\ci{5}{ }$  &  $\ccoeff{5}{}= -\frac{i}{6}$ \\
\hline
$\ci{6}{ }$ & $\ccoeff{6}{}= \frac{i}{6}$ \\
\hline
\multirow{3}{50pt}{\centering $\ci{7}{}$ } & $\ccoeff{7}{LLLL}= \ccoeff{7}{LLRL}=\ccoeff{7 \prime} {LRRR}=\ccoeff{7 \prime} {LLRR}=\ccoeff{7}{LRRL}=\ccoeff{7 \prime} {LRLR}=  \frac{1}{6}$  \\
& $ \ccoeff{7 \prime} {LLLL}= \ccoeff{7 \prime} {LLLR}= \ccoeff{7}{LLLR}=\ccoeff{7}{LRRR}=\ccoeff{7}{LRLR}=\ccoeff{7 \prime} {LLRL}= -\frac{1}{6} $ \\
& $\ccoeff{7}{LRLL}=\ccoeff{7}{LLRR} = 0$  \\
\hline
\multirow{3}{50pt}{\centering $\ci{8}{}$ } & $\ccoeff{8 \prime}{LLLL}= \ccoeff{8\prime\prime}{LLLL}= \ccoeff{8 \prime}{LLRL}= \ccoeff{8}{LLLR}=\ccoeff{8 \prime \prime}{LLLR}=\ccoeff{8 \prime \prime}{LLRL} =   \ccoeff{8}{LRLR}= \ccoeff{8 \prime}{LLRR}= \frac{1}{3}$ \\
&  $  \ccoeff{8}{LLLL}=\ccoeff{8}{LLRL} = \ccoeff{8 \prime}{LRLL}=  \ccoeff{8}{LRRL}=  \ccoeff{8 \prime \prime}{LRLR}=  \ccoeff{8 \prime}{LRLR}= \ccoeff{8 \prime \prime}{LRRL}=   - \frac{1}{3} $ \\
& $\ccoeff{8 \prime \prime}{LLRR} = - \ccoeff{8 \prime \prime }{LRLL}=1$, 
$\ccoeff{8}{LRLL}= \ccoeff{8}{LLRR}= 0 $,  
$\ccoeff{8 \prime}{LLLR} = -\ccoeff{8 \prime}{LRRL} = -\frac{2}{3}$ \\
\hline
\multirow{3}{50pt}{\centering $\ci{9}{}$} & $\ccoeff{9}{LLLL}= \ccoeff{9}{LLRL}= \ccoeff{9}{LLLR}= \ccoeff{9}{LRRR} = \ccoeff{9}{LRLR} = \ccoeff{9}{LRRL}= -\frac{i}{6} $ \\
& $ \ccoeff{9}{LRLL}= \ccoeff{9}{LLRR}= 0$ \\
&  $\ccoeff{9\prime}{LLLL}= \ccoeff{9\prime}{LRRR}=  \ccoeff{9\prime}{LRLR}= \ccoeff{9\prime}{LLRL}=- \ccoeff{9\prime}{LLLR}= - \ccoeff{9\prime}{LLRR} = -\frac{i}{6} $ \\
\hline
$\ci{10}{}$ & $ \ccoeff{10}{1,2,3,4,5,6,15,17}  = 0$, 
$ \ccoeff{10}{8,9,11} = -\ccoeff{10}{7,10,12,13,14,16} = -\frac{i}{24}$, $ \ccoeff{10}{18} =\frac{i}{12} $ \\
\hline
$\ci{11}{}$ & $ \ccoeff{11}{1,2,3,7,8,9,10} = 0$, $\ccoeff{11}{4,6} = - \ccoeff{11}{5}   =  -\frac{1}{72}$ \\
\hline
\end{tabular}}
\caption{Explicit results at one loop in DR for the coefficients $\ccoeff{k}{}$ entering the $\ci{k}{A}$ parametrization introduced in Section \ref{sec:basis}, in units of $1/(16 \pi^2)$.}
\label{tab:CTcoeffDRresults}
\end{table}

\subsubsection{Counterterms}

The explicit form of the gauge variation of the effective action induced by DR at one loop, for the specific renormalization scheme of Section \ref{sec:DRrenscheme}, is given by the sum of \eqref{anomalyFerm2} and the fermionic operators discussed in Section \ref{sec:DRfermions}, see \eqref{4dAnomalyFirst}. Its $4$-dimensional limit is unambiguous, and so does the counterterm $S_{\rm ct}|_{(1)}=\int d^4x\,{\cal L}_{\rm ct}|_{(1)}$ in \eqref{defCT}.

We can now write explicitly the counterterm necessary to restore gauge invariance in our renormalization scheme, under the hypothesis that \eqref{georgi} is satisfied. Using the definitions in Eqs.~\eqref{defT} and \eqref{VA}, we find, up to gauge-invariant contributions:
\ba\label{LctA2}
{\cal L}_{\rm ct}|_{(1)}
&=&\frac{\epsilon^{\mu\nu\alpha\beta}}{16\pi^{2}}{\rm Tr}\left\{\frac{8}{3} \partial_\mu {\cal V}_\nu\left\{{\cal V}_\alpha, {\cal A}_\beta\right\}+4i{\cal V}_{\mu}{\cal V}_\nu {\cal V}_\alpha {\cal A}_\beta+\frac{4}{3}i{\cal V}_\mu {\cal A}_\nu {\cal A}_\alpha {\cal A}_\beta\right\}\\\no
&+&\frac{1}{16\pi^{2}}{\rm Tr}\left\{-\frac{4}{3}(D^{\cal V}_\mu {\cal A}_\nu)^2+2(D^{\cal V}_\mu {\cal A}^\mu)^2-\frac{4}{3}[{\cal A}_\mu,{\cal A}_\nu]^2+\frac{4}{3}({\cal A}_\mu {\cal A}_\nu)^2 + {\cal A}_{\mu\nu}^2   \right\}   \\\no
& -  &\frac{2}{16\pi^2}\left(1+\frac{\xi-1}{6}\right)G_{aa}\bar f\gamma_5\gamma^\mu T^a {\cal A}_\mu T^af.
\ea
We emphasize that in our notation (see Eq. \eqref{eq:classicalYM}) a further rescaling $A^a_\mu\to g_G \delta^G_{ab} A_\mu^b$ is needed to canonically normalize the kinetic term for the gauge bosons.

The counterterm is non-gauge-invariant by definition, see \eqref{defCT}, but respects P, CP, as well as Lorentz invariance.\footnote{Possible gauge-invariant operators may be added to $S_{\rm ct}$. However, these would have no role in restoring the WIs. Rather, they would correspond to renormalizations of the couplings of the theory.} In addition, being proportional to the axial vector component, it manifestly vanishes for $T_A^a=0$, namely for $T_L^a=T_R^a$, consistently with the fact that our regularization does not break vector-like gauge symmetries.

\begin{table}[t]
\centering
\renewcommand{\arraystretch}{1.2} 
\begin{tabular}{|c| c|}
\hline 
$\ct{k}{B}$ & $\ctcoeff{k}{XYZ}$ \\
\hline
\hline
$\ct{1}{}$ & $\ctcoeff{1}{LLL} = \frac{i}{3} -2 i \ctcoeff{2}{LL}$,   
$\ctcoeff{1}{LLR} =- \frac{2 i}{3}$, $ \ctcoeff{1}{RLL} = \frac{i}{3}$  \\
 \hline
 $\ct{2}{}$ & $\ctcoeff{2}{LR}  =  \frac{1}{6}$  \\
\hline
$\ct{3}{}$ & $\ctcoeff{3}{LL} = -  \frac{1}{6} - \ctcoeff{2}{LL}$\\
\hline
$\ct{4}{}$ & $\ctcoeff{4}{} = \frac{1}{6}$ \\
\hline
$\ct{5}{}$ & $\ctcoeff{5}{LLLR}=\ctcoeff{5}{LRLR} = \frac{i}{72}$ \\
\hline
\multirow{2}{50pt}{\centering $\ct{6}{}$} & $\ctcoeff{6}{LLLL} = \frac{1}{12} - \frac{\ctcoeff{2}{LL} }{4}$, $  \ctcoeff{6}{RLLL} = \ctcoeff{6}{LRLR} = - \ctcoeff{6 \prime}{RLLL} = - \frac{1}{24} $ \\
& $  \ctcoeff{6}{LLRR}=\ctcoeff{6 \prime}{RLLR} = 0$,  $\ctcoeff{6 \prime}{LLLL} =  - \frac{1}{8}  + \frac{\ctcoeff{2}{LL}}{2}$ \\
\hline
\end{tabular} 
\caption{Explicit results at one loop in DR for the coefficients $\ctcoeff{k}{}$ entering the $\ct{j}{B}$ parametrization introduced in Section \ref{sec:basis}, in units of $1/(16 \pi^2)$.}
\label{tab:CTDRresults}
\end{table} 

The first, second, and third lines of \eqref{LctA2} can be found independently from each other because they do not mix under gauge transformations. The counterterm in the second line, which does not contain the Levi-Civita tensor, can be identified starting from the most general Lagrangian constructed with dimension-four vector operators invariant under P and covariant under the vector transformations. This requirement identifies all operators in the second line of \eqref{LctA2} plus of course, ${\cal V}_{\mu\nu}^2+{\cal A}_{\mu\nu}^2$, which is irrelevant to our analysis because invariant under the full gauge symmetry group and is in one to one correspondence to the term in Eq.~\eqref{eq:gaugeinvcount}. The coefficients of the operators selected via this procedure are finally derived by requiring the gauge variation cancels the part of {$\Delta_a|_{(1)}$} controlled by ${a}_2^{\slashed\epsilon}$.\footnote{In deriving the variation it is useful to note that $D^{\cal V}_\mu$ satisfies the Leibniz rule.} This fixes all coefficients but the one of ${\cal V}_{\mu\nu}^2+{\cal A}_{\mu\nu}^2$, coherently to what was found in Section \ref{sec:WZgeneral}. 

There are only two independent dimension-four operators with Levi-Civita that are invariant under P and built out of combinations that are manifestly singlet of the vector transformations; these are $\epsilon^{\mu\nu\alpha\beta} {\cal A}_{\mu\nu} {\cal V}_{\alpha\beta}$ and $\epsilon^{\mu\nu\alpha\beta} {\cal A}_{\mu\nu}{\cal A}_\alpha {\cal A}_\beta$. However, using the Bianchi identity one finds that both of them are total derivatives. To arrive at \eqref{LctA2} we have to relax the assumption that the building blocks be manifestly invariant, and instead simply demand that the gauge variation vanishes for $T_A^a=0$ (plus as usual invariance under the truly conserved symmetries P and CP as well as hermiticity). This less stringent request leaves us with the three independent operators shown in the first line of \eqref{LctA2} (the complex $i$ follows from hermiticity and invariance under CP). The numerical coefficients may then be obtained demanding that their variation exactly cancel the part of the anomaly controlled by ${a}_2^{\epsilon}$ {{whenever}} \eqref{georgi} holds.

Finally, the last line of \eqref{LctA2} is determined requiring its variation exactly compensates the fermion-dependent part of $\Delta_a$. The most general set of 2-fermion operators would also include a gauge-invariant combination, but that cannot play any role in restoring the WIs and has not been included in \eqref{LctA2}.

The result in Eq.~\eqref{LctA2} is a particular case of the general counterterm derived in Section \ref{sec:WZgeneral}, obtained for the choice $\ctcoeff{2}{LL} =  {-1}/(96 \pi^2)$. To verify this one may use the explicit values of the $\ccoeff{k}{}$ in Table \ref{tab:CTcoeffDRresults} and plug them in \eqref{eq:solpeven}, \eqref{eq:solpodd}, obtaining the $\ctcoeff{k}{}$ in Table \ref{tab:CTDRresults}. Substituting these in \eqref{gencount} one reproduces exactly the bosonic terms in \eqref{LctA2}. Analogously, plugging the expressions of $\ci{12}{cX}=\ci{13}{cX}$ shown in \eqref{coeffFINAL} into \eqref{replace} and \eqref{eq:fermionicCTgeneral}, we arrive at the last line of  \eqref{LctA2}. This is a strong cross check of the validity or our results.

\section{An explicit example: counterterms in the SM}
\label{sec:SM}

As an application of the formalism developed in this paper, we derive the WI-restoring counterterms for the SM gauge group $\mathrm{SU(3)}_\mathrm{c}\times\mathrm{SU(2)}_\mathrm{L} \times \mathrm{U(1)}_\mathrm{Y}$, using DR and the BMHV scheme for $\gamma_5$. Since our calculations do not include scalar loops, the results of this section apply to the SM in the limit of vanishing Yukawa couplings. We postpone to future work the derivation of the additional counterterms such couplings would require. 

Before regularization, the SM gauge bosons and their interactions with the SM fermions are described by the classical action in Eq. \ref{eq:action_classic_unregularized}. The gluon and electroweak gauge fields may be collected in a 12-dimensional tensor
\begin{align}
 A_\mu^a= 
\begin{cases}
      G_\mu^a & \text{for}\,\,\,  a = 1,\dots,8 \\ 
   W_\mu^a & \text{for}\,\,\,  a = 9,10, 11 \\ 
   B_\mu               & \text{for}\,\,\,   a=12
\end{cases}  
\end{align} 
and their gauge couplings in a 12-dimensional tensor given by $G_{aa}=g_{\rm c}^2$ (for $a=1,\cdots8$), $G_{aa}=g^2$ (for $a=9,10,11$), and $G_{aa}=g'^2$ (for $a=12$). For each fermion family, $f_L$ and $f_R$ can be written as vectors with eight components, $f_L=(u_L,d_L,\nu_L,e_L)$ and $f_R=(u_R,d_R,0,e_R)$, with the quarks carrying color index. The generators $T_{L,R}^a$ are eight-dimensional matrices. For example, the hypercharge generators explicitly read 
\ba
T_L^{12} =\left(
\begin{matrix}
\frac{1}{6}{\bf 1}_3 & & & \\
& \frac{1}{6}{\bf 1}_3 & &\\
& &-  \frac{1}{2} &\\
& & &  -  \frac{1}{2} 
\end{matrix}\right) \,, &&  T_R^{12} =\left(
\begin{matrix}
\frac{2}{3}{\bf 1}_3 & & & \\
& -\frac{1}{3}{\bf 1}_3 & &\\
& & 0 &\\
& & & -1
\end{matrix}\right) \,,
\ea
where ${\bf 1}_3$ is the $3 \times 3$ identity matrix in color space. Analogous expressions may be derived for all other generators.

Having specified these conventions, we can compute the counterterm Lagrangian in Eq.~\eqref{LctA2}. Before presenting the result it is useful to anticipate a few features. The vector component of the SM group contains color, which forms an algebra on its own. Invariance under $\mathrm{SU(3)}_{\rm c}$ implies that the counterterm can only depend on gluons via their field strength and covariant derivatives. Yet, it is straightforward to verify that the fully bosonic part of \eqref{LctA2} cannot involve gluons, the reason being that all color-invariant dimension-4 operators are automatically fully gauge-invariant. Similarly, the gluons cannot appear in the fermionic part of the counterterm, since they live in the vectorial components ${\cal V}_\mu$. 

Having established that Eq.~\eqref{LctA2} can only depend on the electroweak gauge bosons we can proceed by presenting its explicit form. To make the invariance under the vector $\mathrm{U(1)}_{\rm em}$ manifest it is convenient to express \eqref{LctA2} in terms of $W^\pm_\mu, Z_\mu$ and the photon $A_\mu$, defined as usual (in the canonically normalized basis) by:
  \be
  W_{\mu}^{\pm} = \frac{W_\mu^1 \mp i W_\mu^2}{\sqrt{2}}\,, \qquad Z_\mu = -s_w B_\mu + c_w W_\mu^2 \,, \qquad A_\mu = c_w B _\mu + s_w W_\mu^2 \,, 
  \ee
 with $c_w$ and $s_w$ cosine and sine of the weak angle, i.e. $c_w = {g}/{\sqrt{g^{\prime\,2}+ g^2}}$, $s_w = {g'}/{\sqrt{g^{\prime\,2}+ g^2}}$.
The complete result, after an integration by parts and having canonically normalized the gauge fields, reads
 \ba
\label{eq:ctSMPeven}
{\cal L}_{\rm ct}&=&
\frac{g^2}{16\pi^{2}} \left[ \frac{2}{3}D_\mu W^{-}_\nu D^\mu W^{+ \nu} +\frac{1}{3c_w^2}\partial_\mu Z_\nu \partial^\mu Z^\nu\right]- \frac{ig^2e}{8\pi^{2}}F^{\mu\nu}W^+_\mu W^-_\nu  \\\no
&-& \frac{ig^3}{48\pi^2c_w}\left[ \vphantom{\frac23} (-4+6s_w^2) D^\mu W^-_\mu W^+_\nu Z^\nu +(8-6s_w^2)D_\mu W^-_\nu W^+_\mu Z_\nu \right.\\\no
&&\left.\quad\quad\quad+(-4+2s_w^2)D_\mu W^-_\nu Z_\mu W^+_\nu-{\rm h.c.}\right] \\\no
&+&\frac{g^4}{16\pi^{2}} \left[ (W^+_\mu W^{- \mu})^2 -\frac{5}{6}W^+_\mu  W^{+ \mu} W^-_\nu  W^{- \nu} +\frac{1}{24c_w^4} (Z_\mu Z^\mu)^2 \right. \\
&& \left.  \qquad  +  \frac{(-5+8s_w^2)}{3c_w^2}W^+_\mu W^-_{\nu} Z^\mu Z^\nu +\frac{(11-16s_w^2+4s_w^4)}{6c_w^2} W^+_\mu W^{- \mu} Z_\nu Z^\nu \right]\\\no
&  -& \frac{g^3}{16 \pi^2} \left\{ \frac{9-t_w^2}{36\sqrt{2}}\left[\bar{u}_L \gamma^\mu W^+_\mu d_L+\bar{d}_L \gamma^\mu W^-_\mu u_L\right] \right. \\\no
&& \qquad + \frac{9-t_w^2}{72 c_w} \left[\bar{u}_L \gamma^\mu Z_\mu u_L-\bar{d}_L \gamma^\mu Z_\mu d_L\right]\\\no
&& \qquad + \frac{1-t_w^2}{4\sqrt{2}}\left[\bar{\nu}_L \gamma^\mu W^+_\mu e_L+\bar{e}_L \gamma^\mu W^-_\mu \nu_L\right]\\\no
&& \qquad +  \frac{1-t_w^2}{8c_w}\left[\bar{\nu}_L \gamma^\mu Z_\mu \nu_L-\bar{e}_L \gamma^\mu Z_\mu e_L\right] \\\no
&& \qquad +  \frac{2 t_w^2}{9\sqrt{2}}\left[\bar{u}_R \gamma^\mu W^+_\mu d_R+\bar{d}_R \gamma^\mu W^-_\mu u_R\right]\\\no
&& \qquad - \frac{t_w^2}{18c_w}  \left[4\bar{u}_R \gamma^\mu  Z_\mu u_R-\bar{d}_R \gamma^\mu  Z_\mu d_R \right] \\\no
&& \qquad + \left. \frac{t_w^2}{2 c_w}   \bar{e}_R \gamma^\mu  Z_\mu e_R   \right\} .
\ea
In this expression $D_\mu W^\pm_\nu=(\partial_\mu\pm ieA_\mu)W^\pm_\nu$ denotes the QED-covariant derivative and $t_w=s_w/c_w$. As required by invariance under $\mathrm{U(1)}_{\mathrm{em}}$,  the dependence of the counterterm on the photon field occurs only via the field strength and the covariant derivative. Interestingly, the bosonic counterterm involving the Levi-Civita tensor, shown in the first line of Eq.~\eqref{LctA2}, exactly vanishes. This turns out to be a special property of the electroweak gauge group and can be traced back to the peculiarity of the $\mathrm{SU(2)}$ algebra. 

\section{Outlook}
Any consistent regularization scheme induces an apparent violation of gauge invariance in non-anomalous chiral gauge theories. This violation shows up in amplitudes evaluated in perturbation theory and can be removed by the
inclusion of finite counterterms. In this context, renormalization is more sophisticated than in a vector-like gauge theory. 
Two steps can be distinguished in the subtraction procedure. A first one is required to remove infinities. 
At a given order in perturbation theory, this can be done by adding a set of local 
divergent counterterms. At this stage, the theory delivers finite results, but the corresponding amplitudes do not preserve gauge invariance in general. Indeed, the latter is broken by finite terms that can be systematically deleted
by adding local finite counterterms. 
 
The two steps can be reiterated at each order of perturbation theory and can be
implemented directly at the level of the generating functional of the 1PI Green's functions of the theory.
Starting from the regularized functional $\Gamma^{\rm reg}[\phi]$, divergencies are canceled by the local
counterterm $\Gamma_{\rm ct}[\phi]$, such that $\Gamma[\phi]=\Gamma^{\rm reg}[\phi]+\Gamma_{\rm ct}[\phi]$ produces finite results. By further adding the local finite counterterm $S_{\rm ct}[\phi]$, we finally get the
functional $\Gamma_{\rm inv}[\phi]=\Gamma[\phi]+S_{\rm ct}[\phi]$ that satisfies the WI of the theory.
Of course, such a separation of the subtraction procedure into two moves is purely conventional. What matters is the overall
combination $\Gamma_{\rm ct}[\phi]+S_{\rm ct}[\phi]$, which can be split into the sum of a divergent term and a finite one in infinitely many ways. In practical computation, however, the two above steps appear to be very convenient and have been adopted in our approach.
 
The main result of this work is a general analytic expression of the finite one-loop counterterm $S_{\rm ct}[\phi]$ for a renormalizable chiral gauge theory including gauge bosons and fermions transforming in arbitrary representations of the gauge group.
 A very appealing feature of this result is that the counterterm $S_{\rm ct}[\phi]$ is determined for any
possible consistent regulator belonging to a wide class. We only require that the chosen regularization scheme obeys the Quantum Action Principle, preserves Lorentz invariance in four dimensions, and gauge invariance when the theory is vector-like.
The physical information is entirely encoded in the gauge variation $L_a \Gamma[\phi]$.\footnote{The dependence of $S_{\rm ct}[\phi]$ on the subtraction procedure is specified by $L_a \Gamma[\phi]=L_a (\Gamma^{\rm reg}[\phi]+\Gamma_{\rm ct}[\phi])$.} This can be expressed
as a linear combination of local operators of dimension four, whose coefficients can be determined by a one-loop computation for each given regularization scheme.
The counterterm $S_{\rm ct}[\phi]$ automatically follows from the knowledge of these coefficients.
 
We started by quantizing the theory with the Background Field Method and by choosing the Background Field Gauge, which guarantees the gauge invariance of the functional $\Gamma_{\rm inv}[\phi]$ at the level of background fields. In this respect, we differ from previous approaches, where the theory is quantized with the help of a traditional gauge fixing
that breaks the gauge symmetry down to the rigid BRST invariance. The WI of the functional $\Gamma_{\rm inv}[\phi]$ resulting from the Background Field Gauge
are easier to deal with compared to the non-linear Slavnov-Taylor identities consequences of the BRST invariance: 
they simply read $L_a \Gamma_{\rm inv}[\phi]=0$.
 
A key ingredient of our derivation is the non-redundant parametrization of the gauge variation $L_a \Gamma[\phi]$ at the one-loop order, which has been established independently from the adopted regularization by exploiting several
properties of the theory. The Quantum Action Principle guarantees that, order by order in perturbation theory,
$L_a  \Gamma[\phi]$ is a finite local polynomial in the fields and their derivatives preserving the symmetries of the regulator. 
Last but not least, the WZ consistency conditions greatly reduce the number of independent coefficients needed to
describe $L_a \Gamma[\phi]$.
Similar considerations restrict the form of the sought-after counterterm $S_{\rm ct}[\phi]$.
Its analytic expression can be fully determined in complete generality -- up to gauge-invariant contributions -- from the equality $L_a (\Gamma[\phi]+S_{\rm ct}[\phi])=0$.
 
One of the most widely used regularization in practical computation is DR and an important part of our work has been devoted to specifying our general results to such a scheme. Within a path-integral formalism, we have computed the gauge variation of the whole one-loop renormalized functional $\Gamma[\phi]$ in the BMHV scheme. The result was also reproduced in several parts via
a diagrammatic computation. The full set of one-loop finite counterterms in DR for the class of theories under investigation has been obtained and is compactly summarized in Eq.~\eqref{LctA2}.   

To exemplify our result, we have computed the one-loop finite counterterm for the SM in the limit of vanishing Yukawa couplings,
when DR and the BMHV scheme for $\gamma_5$ are chosen. This can be seen as a first step toward the automation of one-loop computations in an even more general class of theories such as chiral gauge theories including a scalar sector, like the SM, or non-renormalizable ones, such as the SMEFT.
The need for local counterterms restoring gauge invariance in SMEFT one-loop computations have already been
emphasized \cite{Bonnefoy:2020tyv,Feruglio:2020kfq,Passarino:2021uxa} and we are confident that our approach, suitably generalized, can represent a useful tool in this context.

\section*{Acknowledgements}

The research of C.C. was supported by the Cluster of Excellence \textit{Precision Physics, Fundamental Interactions, and Structure of Matter} (PRISMA$^+$ - EXC 2118/1) within the German Excellence Strategy (project ID 39083149).

\newpage
\appendix

\section{General solution of the Wess-Zumino conditions} 
\label{app:WZgeneral}
\subsection{Wess-Zumino consistency conditions in terms of the $\ci{k}{A}$} \label{app:equations_ci}
%
\subsubsection{Bosonic sector}
\subsubsection*{P-even sector}
\begin{align}
\label{eq:WZcPeven}
& ~~ C^0_{[pc]}  =   0 \no   \\[2pt]
& \left. \left( C_{cbp}^3  + f_{cbe} C^0_{pe}\right)  \right|_{\textrm{symm. in pc}}      = 0 \no  \\[2pt]
& ~(C^3_{pbc}  +  f_{pbe} C^0_{ce} ) +   (C^4_{cpb} +  C^{5}_{cpb} ) -  (C^4_{pcb}  +   C^{5}_{pcb})   = 0 \no \\
& ~~2 (C^3_{pbc}  +   f_{pbe}  C^0_{ce})  + ( C^2_{cpb} + C^2_{pcb})    -2 ( C^4_{pcb}  + C^{5}_{pcb} ) -  4 C^{6 }_{pcb} = 0 \no \\[2pt]
&~( C^3_{pbc} + f_{pbe} C^0_{ce})    - ( C^2_{cpb} + C^2_{pcb}) + C^3_{cpb}+C^3_{pcb}    -2 (C^4_{pcb}  +  C^{ 5}_{pcb})  = 0 \no \\ 
& ~( C^3_{pbc} +  f_{pbe} C^0_{ce})  +  ( C^3_{pcb}   +  f_{pce}  C^0_{eb} )   -2 ( C^4_{pcb}  +  C^{5}_{pcb})   -  2 C^{6}_{pcb}   = 0  \no  \\[2pt]
& \left. \left(  C^3_{cbe}   f_{pde}   -  C^3_{pde}f_{cbe}  -   C^2_{cbe}  f_{pde} +  C^2_{pde} f_{cbe}       \right)  \right|_{\textrm{symm. in bd}}     =   2 C^7_{[pc](bd)}  \no  \\[2pt]
& \left. \left(  C^2_{cbe} f_{pde} - C^2_{pde} f_{cbe} \right)  \right|_{\textrm{symm. in bd}}  =  2  C^8_{[pc](bd)} \no    \\[2pt]
& ~~C^2_{cbe} f_{pde} - C^2_{pde} f_{cbe} + 2 C^{5 }_{c(ed)} f_{pbe} + 2 C^{5}_{p(ed)} f_{cbe}  + 2C^8_{(c|db|p)}   - 2  C^8_{pc(db)}    =   0   \no   \\[5pt] 
& ~~ 2 ( C^3_{cbe} -  C^2_{cbe}) f_{pde}  -  2  (C^3_{pde}  - C^2_{pde}) f_{cbe}   +   2  C^4_{c(ed)} f_{pbe}  + 2 C^4_{p(ed)} f_{cbe} \no \\
& \qquad\qquad\qquad\qquad\qquad\qquad\qquad \qquad\qquad  \qquad\qquad  -  4 C^7_{pc(db) }  + 2 C^8_{(c|d|p)b}  = 0 \no  \\[2pt]
&~~C^2_{cbe} f_{pde} - C^2_{pde} f_{cbe} + 2 C^{6}_{c(ed)} f_{pbe}  + 2  C^{6}_{p(ed)} f_{cbe}  +  2 C^7_{cd(pb)}     + 2 C^7_{pd(cb)} - 2 C^8_{pc(db)} = 0   \\[2pt]
&~~C^2_{ced} f_{pbe} + C^2_{cbe} f_{pde} - 2 C^2_{p(ed)}  f_{cbe} + 2 C^5_{p(ed)} f_{cbe} + 2 C^{ 6 }_{p(ed)} f_{cbe}\no  \\[2pt]
&\qquad\qquad\qquad\qquad\qquad\qquad\qquad\qquad\qquad + 2 C^7_{pd(cb)}-2 C^8_{pc(db)}    + C^8_{pdbc} = - f_{pce} C^2_{ebd} \no \\
&  ~ (C^3_{ced} - C^2_{ced})  f_{pbe}  + (C^3_{cbe} -  C^2_{cbe}) f_{pde} - 2 ( C^3_{p(ed)}   - C^2_{p(ed)} ) f_{cbe}         \no \\[2pt] 
&\qquad\qquad\qquad\qquad \qquad\qquad \qquad   +  2 C^4_{p(ed)} f_{cbe}  - 2 C^7_{pc(db)}     +  C^8_{pdcb}   = -   ( C^3_{ebd}  -    C^2_{ebd} )  f_{pce}   \no \\[2pt]
&  \left.  \left( 2 C^4_{c(eb)} f_{pde}  - 2 C^7_{pc(bd)}  +  C^8_{pbcd}  \right)   \right|_{\textrm{symm. in bd}}  =  -  C^4_{e(bd)} f_{pce}  \no  \\[2pt]
& \left. \left(  2 C^5_{c(eb)} f_{pde} +  C^8_{pdbc} - C^8_{pcbd}  \right)   \right|_{\textrm{symm. in bd}} = -  C^5_{e(bd)} f_{pce}   \no \\[2pt] 
&   \left. \left(   2 C^6_{c(eb)} f_{pde}    +  2 C^7_{pd(bc)}   - C^8_{pcbd}  \right) \right|_{\textrm{symm. in bd}}  =   -  C^6_{e(bd)} f_{pce}  \no  \\[2pt]
&  \left. \left( f_{phe} C^7_{ce(bd)}  +  f_{che} C^7_{pe(bd)} +f_{pbe}  C^8_{cehd}  + f_{cbe} C^8_{pehd}  + 4 C^{10}_{cphbd}   +  4  C^{10}_{pchbd}  \right)  \right|_{\textrm{symm. in bd}}  = 0 \no   \\[2pt]
 &  \left.  \left(  f_{pde} C^7_{ce (bh)} + 2 f_{phe} C^7_{cd(be)}  - 2 f_{che} C^7_{pd(be)} +  4 C^{10}_{pcdbh} + f_{che} C^8_{pedb}  \right)  \right|_{\textrm{symm. in bh}}  = -  f_{pce}   C^7_{ed(bh)}\no \\[2pt]
& ~~2 f_{che} C^7_{pe(bd)}  + f_{pde}  C^8_{cehb}+f_{phe} C^8_{cdeb}  +   C^8_{cdhe}  f_{pbe} +  C^8_{pehd} f_{cbe} \no  \\
& \qquad\qquad\qquad\qquad\qquad\qquad\qquad  -  f_{che} C^8_{pdeb} - f_{cbe}  C^8_{pdhe}   +  8 C^{10}_{pchbd}   =    -  f_{pce}    C^8_{edhb} \no  \\[2pt]
& \left. \left(  4  f_{pha}  C^{10}_{cabdf}   - 4 f_{cha} C^{10}_{pabdf}  \right)  \right|_{\textrm{symm. in hb, df, hb $\leftrightarrow$ df}}    = - f_{pce} C^{10}_{ehbdf}  \no
\end{align}
\subsubsection*{P-odd sector}
\begin{align}
\begin{aligned}
\label{eq:WZcPodd}
& f_{pde} \ci{1}{c(eb)}  +   f_{cde}  \ci{1}{p(eb)}   +  \ci{9}{pb[cd]}  + \ci{9}{cb[pd]}    = 0     \\ 
& \left. (2  f_{pde} \ci{1}{ceb} + 2 \ci{9}{pb[cd]})  \right|_{\textrm{symm. in bd}} =  -    f_{pce}  \ci{1}{e(db)} \\
& \left. (4 \ci{11}{c[pbdf]} + 4 \ci{11}{p[cbdf]}  +  f_{pde} \ci{9}{ce[bf]} + f_{cde} \ci{9}{pe[bf]}  )\right|_{\textrm{antisymm. in bdf}}  = 0 \\
& \left.  \left(  2 f_{pfe} \ci{9}{cb[ed]}   - f_{cfe} \ci{9}{pe[bd]} - f_{cde} \ci{9}{pe[fb]} - 2  f_{cfe} \ci{9}{pb[ed]}  \right) \right|_{\textrm{antisymm. in df}} \\
& \qquad \qquad \qquad \qquad \qquad \qquad   +    12 \ci{11}{p[cbdf]}   +  f_{pbe} \ci{9}{ce[fd] } =   -  f_{pce}   \ci{9}{eb[fd]} \\
& \left.  \left( 4 f_{phb} \ci{11}{c[badf]} - 4 f_{chb}  \ci{11}{p[badf]}  \right) \right|_{\textrm{antisymm. in adfh}}   =  - f_{pce}   \ci{11}{e[afdh]} 
\end{aligned}
\end{align}
\subsubsection{Fermionic sector}
\begin{align}
\begin{aligned}
\label{WZ}
&-i\ci{14}{pcX}-\ci{12}{cX} T^p_X+T^p_X \ci{12}{cX} -T^c_X  \ci{12}{pX} +T^c_X\ci{13}{pX} -i f_{pcb}  \ci{12}{bX} =0\\
&-i\ci{14}{pcX}-\ci{13}{cX} T^p_X+T^p_X \ci{13}{cX} + \ci{13}{pX} T^c_X-\ci{12}{pX}T^c_X-i f_{pcb} \ci{13}{bX} =0\\
&-if_{pqb} \ci{14}{cbX} +if_{cqb} \ci{14}{pbX}+\ci{14}{cqX} T^p_X-T^p_X \ci{14}{cqX} + \ci{14}{pqX} T^c_X-T^c_X \ci{14}{pqX}  -i f_{pcb} \ci{14}{bqX}=0\\
&-i \ci{14}{cpX} -i \ci{14}{pcX}-\ci{12}{cX} T^p_X+T^p_X \ci{13}{cX} - \ci{12}{pX} T^c_X +T^c_X \ci{13}{pX}  =0 
\end{aligned}
\end{align} 
%
\subsection{Parametrizations for the remaining $\ci{k}{A}$}
\label{app:parametrizatrions}
{\footnotesize
\begin{align}
\ci{7}{abcd} =&~\ccoeff {7}{LRRR} (-T^{abcd}_{LLLL}+T^{abcd}_{LRRR}+  T^{abcd}_{RLLL}-T^{abcd}_{RRRR}-T^{abdc}_{LLLL}+T^{abdc}_{LRRR}+T^{abdc}_{RLLL}-T^{abdc}_{RRRR}  \\
&\qquad \qquad -T^{acdb}_{LLLL}+T^{acdb}_{LRRR}+T^{acdb}_{RLLL}-T^{acdb}_{RRRR}-T^{adcb}_{LLLL}+T^{adcb}_{LRRR}+T^{adcb}_{RLLL}-T^{adcb}_{RRRR})+\no \\
&~\ccoeff{7}{L LRR}    (-T^{abcd}_{LLLL}+T^{abcd}_{LLRR}+   T^{abcd}_{RRLL}-T^{abcd}_{RRRR}-T^{abdc}_{LLLL}+T^{abdc}_{LLRR}+T^{abdc}_{RRLL}-T^{abdc}_{RRRR}\no \\
&\qquad \qquad  -T^{acdb}_{LLLL}  +T^{acdb}_{LRRL}+T^{acdb}_{RLLR}-T^{acdb}_{RRRR}-T^{adcb}_{LLLL}+T^{adcb}_{LRRL}+T^{adcb}_{RLLR}-T^{adcb}_{RRRR})+ \no \\
&~\ccoeff{7}{LRLR}  (-T^{abcd}_{LLLL}+T^{abcd}_{LRLR}+   T^{abcd}_{RLRL}-T^{abcd}_{RRRR}-T^{abdc}_{LLLL}+T^{abdc}_{LRLR}+T^{abdc}_{RLRL}-T^{abdc}_{RRRR}\no \\
&\qquad \qquad -T^{acdb}_{LLLL}+T^{acdb}_{LRLR}+T^{acdb}_{RLRL}-T^{acdb}_{RRRR}-T^{adcb}  _{LLLL}+T^{adcb}_{LRLR}+T^{adcb}_{RLRL}-T^{adcb}_{RRRR})+\no \\
&~\ccoeff{7}{LLLR}  (-T^{abcd}_{LLLL}+T^{abcd}_{LLLR}+T^{abcd}_{RRRL}-T^{abcd}_{RRRR}-T^{abdc}_{LLLL}+T^{abdc}_{LLLR}+T^{abdc}_{RRRL}-T^{abdc}_{RRRR}\no \\
&\qquad \qquad-T^{acdb}_{LLLL}+T^{acdb}_{LRLL}+T^{acdb}_{RLRR}-T^{acdb}_{RRRR}-T^{adcb}_{LLLL}+T^{adcb}_{LRLL}+T^{adcb}_{RLRR}-T^{adcb}_{RRRR})+\no \\
&~\ccoeff{7}{LRRL}  (-T^{abcd}_{LLLL}+T^{abcd}_{LRRL}+  T^{abcd}_{RLLR}-T^{abcd}_{RRRR}-T^{abdc}_{LLLL}+T^{abdc}_{LRRL}+T^{abdc}_{RLLR}-T^{abdc}_{RRRR}\no \\
&\qquad \qquad-T^{acdb}_{LLLL}+T^{acdb}_{LLRR}+T^{acdb}_{RRLL}-T^{acdb}_{RRRR}-T^{adcb}_{LLLL}+T^{adcb}_{LLRR}+T^{adcb}_{RRLL}-T^{adcb}_{RRRR})+\no\\
&~\ccoeff{7}{LLRL}(-T^{abcd}_{LLLL}+T^{abcd}_{LLRL}+  T^{abcd}_{RRLR}-T^{abcd}_{RRRR}-T^{abdc}_{LLLL}+T^{abdc}_{LLRL}+T^{abdc}_{RRLR}-T^{abdc}_{RRRR}\no \\
&\qquad \qquad-T^{acdb}_{LLLL}+T^{acdb}_{LLRL}+T^{acdb}_{RRLR}-T^{acdb}_{RRRR}-T^{adcb} _{LLLL}+T^{adcb}_{LLRL}+T^{adcb}_{RRLR}-T^{adcb}_{RRRR})+\no \\
&~\ccoeff{7}{LRLL}(-T^{abcd}_{LLLL}+T^{abcd}_{LRLL}+T^{abcd}_{RLRR}-T^{abcd}_{RRRR}-T^{abdc}_{LLLL}+T^{abdc}_{LRLL}+T^{abdc}_{RLRR}-T^{abdc}_{RRRR}\no \\
&\qquad \qquad -T^{acdb}_{LLLL}+T^{acdb}_{LLLR}+T^{acdb}_{RRRL}-T^{acdb}_{RRRR}-T^{adcb}  _{LLLL}+T^{adcb}_{LLLR}+T^{adcb}_{RRRL}-T^{adcb}_{RRRR}) \no  \\
&~ \ccoeff{7 \prime}{LLRR} (-2  T^{acbd}_{LLLL}+T^{acbd}_{LLRR}+T^{acbd}_{LRRL}+T^{acbd}_{RLLR}+T^{acbd}_{RRLL}-2 T^{acbd}_{RRRR}-2 T^{adbc}_{LLLL} \no \\
&\qquad\qquad +T^{adbc}_{LLRR} +T^{adbc}_{LRRL}+T^{adbc}_{RLLR}+T^{adbc}_{RRLL}-2  T^{adbc}_{RRRR})+\no  \\
& ~\ccoeff{7 \prime}{LLLR}  (-2
   T^{acbd}_{LLLL}+T^{acbd}_{LLLR}+T^{acbd}_{LRLL}+T^{acbd}_{RLRR}+T^{acbd}_{RRRL}-2 T^{acbd}_{RRRR}-2 T^{adbc}_{LLLL}\no \\
  & \qquad\qquad  +T^{adbc}_{LLLR}+T^{adbc}_{LRLL}+T^{adbc}_{RLRR}+T^{adbc}_{RRRL}-2
   T^{adbc}_{RRRR})+\no \\
&~\ccoeff{7 \prime}{LRRR}(-T^{acbd}_{LLLL}+T^{acbd}_{LRRR}+ T^{acbd}_{RLLL}-T^{acbd}_{RRRR}-T^{adbc}_{LLLL}+T^{adbc}_{LRRR}+T^{adbc}_{RLLL}-T^{adbc}_{RRRR}) + \no \\
&~\ccoeff{7 \prime}{LRLR} (-T^{acbd}_{LLLL}+T^{acbd}_{LRLR}+  T^{acbd}_{RLRL}-T^{acbd}_{RRRR}-T^{adbc}_{LLLL}+T^{adbc}_{LRLR}+T^{adbc}_{RLRL}-T^{adbc}_{RRRR}) + \no \\
& \ccoeff{7\prime}{LLRL}(-T^{acbd}_{LLLL}+T^{acbd}_{LLRL}+T^{acbd}_{RRLR}-T^{acbd}_{RRRR}-T^{adbc}_{LLLL}+T^{adbc}_{LLRL}+T^{adbc}_{RRLR}-T^{adbc}_{RRRR}) \,, \no\\[6pt]
\ci{8}{abcd} = &~\ccoeff{8}{LLRR}   (T^{abcd}_{LLRR}-T^{abcd}_{LRRR}-T^{abcd}_{RLLL}+T^{abcd}_{RRLL}+T^{adcb}_{LRRL}-T^{adcb}_{LRRR}-T^{adcb}_{RLLL}+T^{adcb}_{RLLR})+ \\
  &~\ccoeff{8}{LRLR}  (T^{abcd}_{LRLR}-T^{abcd}_{LRRR}-T^{abcd}_{RLLL}+T^{abcd}_{RLRL}+T^{adcb}_{LRLR}-T^{adcb}_{LRRR}-T^{adcb}_{RLLL}+T^{adcb}_{RLRL})+\no\\
  &~\ccoeff{8}{LLLR} (T^{abcd}_{LLLR}-T^{abcd}_{LRRR}-T^{abcd}_{RLLL}+T^{abcd}_{RRRL}+T^{adcb}_{LRLL}-T^{adcb}_{LRRR}-T^{adcb}_{RLLL}+T^{adcb}_{RLRR})+\no\\
  &~\ccoeff{8}{LRRL} (T^{abcd}_{LRRL}-T^{abcd}_{LRRR}-T^{abcd}_{RLLL}+T^{abcd}_{RLLR}+T^{adcb}_{LLRR}-T^{adcb}_{LRRR}-T^{adcb}_{RLLL}+T^{adcb}_{RRLL})+\no\\
  &~\ccoeff{8}{LLRL}  (T^{abcd}_{LLRL}-T^{abcd}_{LRRR}-T^{abcd}_{RLLL}+T^{abcd}_{RRLR}+T^{adcb}_{LLRL}-T^{adcb}_{LRRR}-T^{adcb}_{RLLL}+T^{adcb}_{RRLR})+\no\\
  &~\ccoeff{8}{LRLL}(T^{abcd}_{LRLL}-T^{abcd}_{LRRR}-T^{abcd}_{RLLL}+T^{abcd}_{RLRR}+T^{adcb}_{LLLR}-T^{adcb}_{LRRR}-T^{adcb}_{RLLL}+T^{adcb}_{RRRL})+\no\\
  &~\ccoeff{8}{LLLL}   (T^{abcd}_{LLLL}-T^{abcd}_{LRRR}-T^{abcd}_{RLLL}+T^{abcd}_{RRRR}+T^{adcb}_{LLLL}-T^{adcb}_{LRRR}-T^{adcb}_{RLLL}+T^{adcb}_{RRRR})+ \no \\
&~ \ccoeff{8 \prime}{LRRL}(T^{abdc}_{LLRR}-T^{abdc}_{LRRR}-T^{abdc}_{RLLL}+T^{abdc}_{RRLL}+T^{acdb}_{LRRL}-T^{acdb}_{LRRR}-T^{acdb}_{RLLL}+T^{acdb}_{RLLR}) + \no \\ 
&~\ccoeff{8\prime}{LRLR}  (T^{abdc}_{LRLR}-T^{abdc}_{LRRR}-T^{abdc}_{RLLL}+T^{abdc}_{RLRL}+T^{acdb}_{LRLR}-T^{acdb}_{LRRR}-T^{acdb}_{RLLL}+T^{acdb}_{RLRL})+ \no  \\
&~\ccoeff{8\prime}{LRLL} (T^{abdc}_{LLLR}-T^{abdc}_{LRRR}-T^{abdc}_{RLLL}+T^{abdc}_{RRRL}+T^{acdb}_{LRLL}-T^{acdb}_{LRRR}-T^{acdb}_{RLLL}+T^{acdb}_{RLRR}) + \no \\
&~ \ccoeff{8 \prime}{LLRR} (T^{abdc}_{LRRL}-T^{abdc}_{LRRR}-T^{abdc}_{RLLL}+T^{abdc}_{RLLR}+T^{acdb}_{LLRR}-T^{acdb}_{LRRR}-T^{acdb}_{RLLL}+T^{acdb}_{RRLL}) +\no \\
&~ \ccoeff{8 \prime}{LLRL} (T^{abdc}_{LLRL}-T^{abdc}_{LRRR}-T^{abdc}_{RLLL}+T^{abdc}_{RRLR}+T^{acdb}_{LLRL}-T^{acdb}_{LRRR}-T^{acdb}_{RLLL}+T^{acdb}_{RRLR}) +\no \\
& ~\ccoeff{8 \prime}{LLLR} (T^{abdc}_{LRLL}-T^{abdc}_{LRRR}-T^{abdc}_{RLLL}+T^{abdc}_{RLRR}+T^{acdb}_{LLLR}-T^{acdb}_{LRRR}-T^{acdb}_{RLLL}+T^{acdb}_{RRRL}) +\no \\
 &~\ccoeff{8\prime}{LLLL}(T^{abdc}_{LLLL}-T^{abdc}_{LRRR}-T^{abdc}_{RLLL}+T^{abdc}_{RRRR}+T^{acdb}_{LLLL}-T^{acdb}_{LRRR}-T^{acdb}_{RLLL}+T^{acdb}_{RRRR})+\no \\
  &~\ccoeff{8\prime\prime}{LRRL}(T^{acbd}_{LLRR}-T^{acbd}_{LRRR}-T^{acbd}_{RLLL}+T^{acbd}_{RRLL}+T^{adbc}_{LRRL}-T^{adbc}_{LRRR}-T^{adbc}_{RLLL}+T^{adbc}_{RLLR})+\no \\
  &~\ccoeff{8\prime\prime}{LRLR} (T^{acbd}_{LRLR}-T^{acbd}_{LRRR}-T^{acbd}_{RLLL}+T^{acbd}_{RLRL}+T^{adbc}_{LRLR}-T^{adbc}_{LRRR}-T^{adbc}_{RLLL}+T^{adbc}_{RLRL})+\no \\
  &~\ccoeff{8\prime\prime}{LRLL}(T^{acbd}_{LLLR}-T^{acbd}_{LRRR}-T^{acbd}_{RLLL}+T^{acbd}_{RRRL}+T^{adbc}_{LRLL}-T^{adbc}_{LRRR}-T^{adbc}_{RLLL}+T^{adbc}_{RLRR})+\no \\
  &~\ccoeff{8\prime\prime}{LLRR}(T^{acbd}_{LRRL}-T^{acbd}_{LRRR}-T^{acbd}_{RLLL}+T^{acbd}_{RLLR}+T^{adbc}_{LLRR}-T^{adbc}_{LRRR}-T^{adbc}_{RLLL}+T^{adbc}_{RRLL})+\no \\
  &~\ccoeff{8\prime\prime}{LLRL}(T^{acbd}_{LLRL}-T^{acbd}_{LRRR}-T^{acbd}_{RLLL}+T^{acbd}_{RRLR}+T^{adbc}_{LLRL}-T^{adbc}_{LRRR}-T^{adbc}_{RLLL}+T^{adbc}_{RRLR})+\no \\
  &~\ccoeff{8\prime\prime}{LLLR} (T^{acbd}_{LRLL}-T^{acbd}_{LRRR}-T^{acbd}_{RLLL}+T^{acbd}_{RLRR}+T^{adbc}_{LLLR}-T^{adbc}_{LRRR}-T^{adbc}_{RLLL}+T^{adbc}_{RRRL})+\no \\
  &~\ccoeff{ 8\prime\prime}{LLLL} (T^{acbd}_{LLLL}-T^{acbd}_{LRRR}-T^{acbd}_{RLLL}+T^{acbd}_{RRRR}+T^{adbc}_{LLLL}-T^{adbc}_{LRRR}-T^{adbc}_{RLLL}+T^{adbc}_{RRRR}) \,, \no \\[6pt]
  \ci{9}{abcd} =&~\ccoeff{9 \prime}{LRRR}
   (-T^{acbd}_{LRRR}+T^{acbd}_{RLLL}+T^{adbc}_{LRRR}-T^{adbc}_{RLLL})+  \\
  &~\ccoeff{9\prime}{LRLR}
   (-T^{acbd}_{LRLR}+T^{acbd}_{RLRL}+T^{adbc}_{LRLR}-T^{adbc}_{RLRL})+\no \\
  &~\ccoeff{9\prime}{LLRR}
   (-T^{acbd}_{LLRR}-T^{acbd}_{LRRL}+T^{acbd}_{RLLR}+T^{acbd}_{RRLL}+T^{adbc}_{LLRR}+T^{adbc}_{LRRL}-T^{adbc}_{RLLR}-T^{adbc}_{RRLL})+\no \\
  &~\ccoeff{9\prime}{LLRL}
   (-T^{acbd}_{LLRL}+T^{acbd}_{RRLR}+T^{adbc}_{LLRL}-T^{adbc}_{RRLR})+\no \\
  &~\ccoeff{9\prime}{LLLR}
   (-T^{acbd}_{LLLR}-T^{acbd}_{LRLL}+T^{acbd}_{RLRR}+T^{acbd}_{RRRL}+T^{adbc}_{LLLR}+T^{adbc}_{LRLL}-T^{adbc}_{RLRR}-T^{adbc}_{RRRL})+\no \\
  &~\ccoeff{9\prime}{LLLL}
   (-T^{acbd}_{LLLL}+T^{acbd}_{RRRR}+T^{adbc}_{LLLL}-T^{adbc}_{RRRR})+\no \\
  &~\ccoeff{9\prime}{LRRR}
   (T^{abcd}_{LRRR}-T^{abcd}_{RLLL}-T^{abdc}_{LRRR}+T^{abdc}_{RLLL}+T^{acdb}_{LRRR}-T^{acdb}_{RLLL}-T^{adcb}_{LRRR}+T^{adcb}_{RLLL})+\no \\
  &~\ccoeff{9}{LLRR}
   (T^{abcd}_{LLRR}-T^{abcd}_{RRLL}-T^{abdc}_{LLRR}+T^{abdc}_{RRLL}+T^{acdb}_{LRRL}-T^{acdb}_{RLLR}-T^{adcb}_{LRRL}+T^{adcb}_{RLLR})+\no \\
  &~\ccoeff{9}{LRLR}
   (T^{abcd}_{LRLR}-T^{abcd}_{RLRL}-T^{abdc}_{LRLR}+T^{abdc}_{RLRL}+T^{acdb}_{LRLR}-T^{acdb}_{RLRL}-T^{adcb}_{LRLR}+T^{adcb}_{RLRL})+\no \\
  &~\ccoeff{9}{LLLR}
   (T^{abcd}_{LLLR}-T^{abcd}_{RRRL}-T^{abdc}_{LLLR}+T^{abdc}_{R
   RRL}+T^{acdb}_{LRLL}-T^{acdb}_{RLRR}-T^{adcb}_{LRLL}+T^{adcb}_{RLRR})+\no \\
  &~\ccoeff{9}{LRRL}
   (T^{abcd}_{LRRL}-T^{abcd}_{RLLR}-T^{abdc}_{LRRL}+T^{abdc}_{RLLR}+T^{acdb}_{LLRR}-T^{acdb}_{RRLL}-T^{adcb}_{LLRR}+T^{adcb}_{RRLL})+\no \\
  &~\ccoeff{9}{LLRL}
   (T^{abcd}_{LLRL}-T^{abcd}_{RRLR}-T^{abdc}_{LLRL}+T^{abdc}_{RRLR}+T^{acdb}_{LLRL}-T^{acdb}_{RRLR}-T^{adcb}_{LLRL}+T^{adcb}_{RRLR})+\no \\
  &~\ccoeff{9}{LRLL}
   (T^{abcd}_{LRLL}-T^{abcd}_{RLRR}-T^{abdc}_{LRLL}+T^{abdc}_{RLRR}+T^{acdb}_{LLLR}-T^{acdb}_{RRRL}-T^{adcb}_{LLLR}+T^{adcb}_{RRRL})+\no \\
  &~\ccoeff{9}{LLLL}
   (T^{abcd}_{LLLL}-T^{abcd}_{RRRR}-T^{abdc}_{LLLL}+T^{abdc}_{RRRR}+T^{acdb}_{LLLL}-T^{acdb}_{RRRR}-T^{adcb}_{LLLL}+T^{adcb}_{RRRR}) \,, \no \\[6pt]
  \ci{10}{abcde} = &   ~\ccoeff{10}{1}(T^{abdec}_{RRLRR}-T^{abdec}_{RRRLR}+T^{abedc}_{RRLRR}- T^{abedc}_{RRRLR}-T^{adbce}_{LLLRL}+T^{adbce}_{LLRLL}+T^{adbce}_{RRLRR}- T^{adbce}_{RRRLR}- \\
   & \qquad T^{adcbe}_{LLLRL}+T^{adcbe}_{LLRLL}+T^{adcbe}_{RRLRR}-T^{adcbe}_{RRRLR}+T^{bcdae}_{RLLLL}+T^{cbdae}_{RLLLL}+T^{cdeba}_{RLRRR}+T^{cedba}_{RLRRR}+\no \\
   & \qquad T^{dbace}_{LLLLR}-T^{dbace}_{RLLLL}+T^{dcabe}_{LLLLR}-T^{dcabe}_{RLLLL}-T^{debac}_{RLRRR}+T^{ebacd}_{LLLLR}-T^{ebacd}_{RLLLL}-T^{ebcda}_{LLRLL}+\no \\
   & \qquad T^{ebcda}_{RLRRR}-T^{ebcda}_{RRLRR}+T^{ecabd}_{LLLLR}-T^{ecabd}_{RLLLL}-T^{ecbda}_{LLRLL}+T^{ecbda}_{RLRRR}-T^{ecbda}_{RRLRR}-T^{edbac}_{RLRRR})+\no \\
   &~\ccoeff{10}{2}
   (T^{abdce}_{LLLRL}-T^{abdce}_{LLRLL}-T^{abdce}_{RRLRR}+
   T^{abdce}_{RRRLR}-T^{abecd}_{LLRLL}-T^{abecd}_{RRLRR}+T^{abecd}_{RRRLR}+T^{acdbe}_{LLLRL}-\no\\
   & \qquad T^{acdbe}_{LLRLL}-T^{acdbe}_{RRLRR}+T^{acdbe}_{RRRLR}-T^{acebd}_{LLRLL}-T^{acebd}_{RRLRR}+T^{acebd}_{RRRLR}+T^{bdace}_{RLLLL}-T^{bdcae}_{RLLLL}+ \no\\
   & \qquad T^{cdabe}_{RLLLL}-T^{cdbae}_{RLLLL}+T^{dbeca}_{LLRLL} -T^{dbeca}_{RLRRR}+T^{dbeca}_{RRLRR}+T^{dceba}_{LLRLL}-T^{dceba}_{RLRRR}+T^{dceba}_{RRLRR}-\no\\
   & \qquad T^{ebadc}_{LLLLR}+T^{ebdca}_{LLRLL}-T^{ebdca}_{RLRRR}+T^{ebdca}_{RRLRR}- T^{ecadb}_{LLLLR}+T^{ecdba}_{LLRLL}-T^{ecdba}_{RLRRR}+T^{ecdba}_{RRLRR})+  \no \\
   &~\ccoeff{10}{3}  (T^{abdce}_{RLLRL}-T^{abdce}_{RLRLL}-T^{abecd}_{RLRLL}+T^{acdbe}_{RLLRL}-T^{acdbe}_{RLRLL}-T^{acebd}_{RLRLL}-T^{badce}_{RLRLR}-T^{baecd}_{RLRLR}+ \no \\
   &\qquad T^{bdace}_{RLRLL}-T^{bdcae}_{RLLRL}+T^{bdcae}_{RLRLR}+T^{becad}_{RLRLR}-T^{cadbe}_{RLRLR}-T^{caebd}_{RLRLR}+T^{cdabe}_{RLRLL}-T^{cdbae}_{RLLRL}+\no \\
   &\qquad T^{cdbae}_{RLRLR}+T^{cebad}_{RLRLR}-T^{dabec}_{RLRLR}-T^{daceb}_{RLRLR}+T^{dbeac}_{RLRLR}+T^{dbeca}_{LLRLR}+T^{dceab}_{RLRLR}+T^{dceba}_{LLRLR}- \no \\
   &\qquad T^{eabdc}_{RLRLR}-T^{eacdb}_{RLRLR}-T^{ebadc}_{LLRLR}+T^{ebdac}_{RLRLR}+T^{ebdca}_{LLRLR}-T^{ecadb}_{LLRLR}+T^{ecdab}_{RLRLR}+T^{ecdba}_{LLRLR})+\no \\
   &~\ccoeff{10}{4} (-T^{adbce}_{RLLRL}+T^{adbce}_{RLRLL}-T^{adcbe}_{RLLRL}+T^{adcbe}_{RLRLL}+T^{bacde}_{RLRLR}+T^{baced}_{RLRLR}+ T^{bcdae}_{RLLRL}-T^{bcdae}_{RLRLR}- \no \\
   & \qquad T^{bcead}_{RLRLR}+T^{cabde}_{RLRLR}+T^{cabed}_{RLRLR}+T^{cbdae}_{RLLRL}-T^{cbdae}_{RLRLR}-T^{cbead}_{RLRLR}+T^{daebc}_{RLRLR}+T^{daecb}_{RLRLR}+ \no \\
   & \qquad  T^{dbace}_{LLRLR}-T^{dbace}_{RLRLL}+T^{dcabe}_{LLRLR}-T^{dcabe}_{RLRLL}-T^{debac}_{RLRLR}-T^{decab}_{RLRLR}+T^{eadbc}_{RLRLR}+T^{eadcb}_{RLRLR}+ \no \\
   & \qquad  T^{ebacd}_{LLRLR}-T^{ebacd}_{RLRLL}-T^{ebcda}_{LLRLR}+T^{ecabd}_{LLRLR}-T^{ecabd}_{RLRLL}-T^{ecbda}_{LLRLR}-T^{edbac}_{RLRLR}-T^{edcab}_{RLRLR})+\no \\
   &~\ccoeff{10}{5}
   (T^{abcde}_{RLLRL}-T^{abcde}_{RLRLL}+T^{abced}_{RLLRL}+T^{acbde}_{RLLRL}-T^{acbde}_{RLRLL}+T^{acbed}_{RLLRL}-T
   ^{badec}_{RLRLR}-T^{baedc}_{RLRLR}+\no \\
   & \qquad T^{bcade}_{RLRLL}+T^{bdeac}_{RLRLR}+T^{bedac}_{RLRLR}-T^{cadeb}_{RLRLR}-T^{caedb}_{RLRLR}+T^{cbade}_{RLRLL}+T^{cdeab}_{RLRLR}+T^{cedab}_{RLRLR}-\no \\
   & \qquad T^{dabce}_{RLRLR}-T^{dacbe}_{RLRLR}-T^{dbcae}_{RLLRL}+T^{dbcae}_{RLRLR}-T^{dcbae}_{RLLRL}+T^{dcbae}_{RLRLR}-T^{eabcd}_{RLRLR}-T^{eacbd}_{RLRLR}-\no \\
   & \qquad T^{ebcad}_{RLLRL}+T^{ebcad}_{RLRLR}-T^{ecbad}_{RLLRL}+T^{ecbad}_{RLRLR}-T^{edabc}_{LLRLR}-T^{edacb}_{LLRLR}+T^{edbca}_{LLRLR}+T^{edcba}_{LLRLR})+ \no  \\
   &~\ccoeff{10}{6}
   (-T^{abcde}_{LLLRL}+T^{abcde}_{LLRLL}+T^{abcde}_{RRLRR}
   -T^{abcde}_{RRRLR}-T^{abced}_{LLLRL}+T^{abced}_{RRLRR}- T^{abced}_{RRRLR}-T^{acbde}_{LLLRL}+\no \\
   & \qquad T^{acbde}_{LLRLL}+T^{acbde}_{RRLRR}-T^{acbde}_{RRRLR}-T^{acbed}_{LLLRL}+T^{acbed}_{RRLRR}-T^{acbed}_{RRRLR}-T^{bcade}_{RLLLL}-T^{cbade}_{RLLLL}+\no \\
   & \qquad T^{dbcae}_{RLLLL}+T^{dcbae}_{RLLLL}+T^{debca}_{RLRRR}+T^{decba}_{RLRRR}+T^{ebcad}_{RLLLL}-T^{ebcad}_{RLRRR}+T^{ecbad}_{RLLLL}-T^{ecbad}_{RLRRR}+\no \\
   & \qquad T^{edabc}_{LLLLR}+T^{edacb}_{LLLLR}-T^{edbca}_{LLRLL}+T^{edbca}_{RLRRR}-T^{edbca}_{RRLRR}-T^{edcba}_{LLRLL}+T^{edcba}_{RLRRR}-T^{edcba}_{RRLRR})+\no \\
   &~\ccoeff{10}{7}
   (T^{abcde}_{LLLLR}-T^{abcde}_{RLRRR}+T^{abced}_{LLLLR}-
   T^{abced}_{RLRRR}+T^{acbde}_{LLLLR}-T^{acbde}_{RLRRR}+T^{acbed}_{LLLLR}-T^{acbed}_{RLRRR}+\no \\
   & \qquad T^{adebc}_{LLLLR}-T^{adebc}_{RLRRR}+T^{adecb}_{LLLLR}+T^{aedbc}_{LLLLR}-T^{aedbc}_{RLRRR}+T^{aedcb}_{LLLLR}-T^{bcdea}_{RLLLL}-T^{bceda}_{RLLLL}-\no \\
   & \qquad T^{cbade}_{RRRLR}-T^{cbaed}_{RRRLR}-T^{cbdea}_{RLLLL}+T^{cbdea}_{RRRLR}-T^{cbeda}_{RLLLL}+T^{cbeda}_{RRRLR}+T^{deabc}_{RLRRR}-T^{debca}_{RLLLL}-\no \\
   & \qquad T^{decba}_{RLLLL}+T^{ebcad}_{RRLRR}+T^{ecbad}_{RRLRR}+T^{edabc}_{RLRRR}-T^{edbca}_{RLLLL}+T^{edbca}_{RRRLR}-T^{edcba}_{RLLLL}+T^{edcba}_{RRRLR}) \no \\
   &~\ccoeff{10}{8}
   (-T^{abdec}_{LLLLR}-T^{abedc}_{LLLLR}-T^{acdeb}_{LLLLR}
   -T^{acedb}_{LLLLR}-T^{adbce}_{LLLLR}+T^{adbce}_{RLRRR}-
   T^{adcbe}_{LLLLR}+T^{adcbe}_{RLRRR}-\no \\
   &\qquad  T^{aebcd}_{LLLLR}+T^{aebcd}_{RLRRR}-T^{aecbd}_{LLLLR}+T^{aecbd}_{RLRRR}+T^{bdeca}_{RLLLL}+T^{bedca}_{RLLLL}+T^{cdeba}_{RLLLL}+T^{cedba}_{RLLLL}- \no \\
   &\qquad  T^{dbace}_{RLRRR}+T^{dbace}_{RRRLR}+T^{dbcea}_{RLLLL}-T^{dbcea}_{RRRLR}-T^{dcabe}_{RLRRR}+T^{dcabe}_{RRRLR}+T^{dcbea}_{RLLLL}-T^{dcbea}_{RRRLR}- \no \\
   &\qquad T^{ebacd}_{RLRRR}+T^{ebacd}_{RRRLR}+T^{ebcda}_{RLLLL}-T^{ebcda}_{RRRLR}-T^{ecabd}_{RLRRR}+T^{ecabd}_{RRRLR}+T^{ecbda}_{RLLLL}-T^{ecbda}_{RRRLR})+\no \\
   &~\ccoeff{10}{9} (-T^{abdce}_{LLLLR}+T^{abdce}_{RLRRR}-T^{abecd}_{LLLLR}+T^{abecd}_{RLRRR}-T^{acdbe}_{LLLLR}+T^{acdbe}_{RLRRR}-T ^{acebd}_{LLLLR}+T^{acebd}_{RLRRR}-\no\\
   & \qquad  T^{adbec}_{LLLLR}+T^{adbec}_{RLRRR}-T^{adceb}_{LLLLR}-T^{aebdc}_{LLLLR}+T^{aebdc}_{RLRRR}-T^{aecdb}_{LLLLR}+T^{bdcea}_{RLLLL}+ T^{becda}_{RLLLL}-\no\\
   & \qquad T^{cdabe}_{RLRRR}+T^{cdbea}_{RLLLL}-T^{ceabd}_{RLRRR}+T^{cebda}_{RLLLL}+T^{dbaec}_{RRRLR}-T^{dbeac}_{RRLRR}+T^{dbeca}_{RLLLL}-T^{dbeca}_{RRRLR}+\no\\
   & \qquad T^{dceba}_{RLLLL}-T^{dceba}_{RRRLR}+T^{ebadc}_{RRRLR}-T^{ebdac}_{RRLRR} + T^{ebdca}_{RLLLL}-T^{ebdca}_{RRRLR}+T^{ecdba}_{RLLLL}-T^{ecdba}_{RRRLR})+\no \\
    &~\ccoeff{10}{10}
   (T^{abcde}_{LLRLR}-T^{abcde}_{RLRLR}+T^{abced}_{LLRLR}- T^{abced}_{RLRLR}+T^{acbde}_{LLRLR}-T^{acbde}_{RLRLR}+T^{acbed}_{LLRLR}-T^{acbed}_{RLRLR}+\no \\
   & \qquad T^{adebc}_{LLRLR}-T^{adebc}_{RLRLR}+T^{adecb}_{LLRLR}-T^{adecb}_{RLRLR}+T^{aedbc}_{LLRLR}-T^{aedbc}_{RLRLR}+T^{aedcb}_{LLRLR}-T^{aedcb}_{RLRLR}-\no \\
   & \qquad T^{bcdea}_{RLRLL}+T^{bcdea}_{RLRLR}-T^{bceda}_{RLRLL}+T^{bceda}_{RLRLR}-T^{cbdea}_{RLRLL}+T^{cbdea}_{RLRLR}-T^{cbeda}_{RLRLL}+T^{cbeda}_{RLRLR}-\no \\
   & \qquad T^{debca}_{RLRLL}+T^{debca}_{RLRLR}-T^{decba}_{RLRLL}+T^{decba}_{RLRLR}-T^{edbca}_{RLRLL}+T^{edbca}_{RLRLR}-T^{edcba}_{RLRLL}+T^{edcba}_{RLRLR})+\no \\
   &~\ccoeff{10}{11} ( 
    T_{LLRLR}^{abdce}  + T_{LLRLR}^{abecd} + 
 T_{LLRLR}^{acdbe} + T_{LLRLR}^{acebd} + 
 T_{LLRLR}^{adbec}+ T_{LLRLR}^{adceb}+ 
 T_{LLRLR}^{aebdc} + T_{LLRLR}^{aecdb} - \no \\
 & \qquad T_{RLRLL}^{bdcea} - T_{RLRLL}^{becda} - 
 T_{RLRLL}^{cdbea} - T_{RLRLL}^{cebda}- 
 T_{RLRLL}^{dbeca} - T_{RLRLL}^{dceba}- 
 T_{RLRLL}^{ebdca} - T_{RLRLL}^{ecdba} - \no \\
 & \qquad T_{RLRLR}^{abdce} - T_{RLRLR}^{abecd}- 
 T_{RLRLR}^{acdbe}- T_{RLRLR}^{acebd} - 
 T_{RLRLR}^{adbec}- T_{RLRLR}^{adceb}- 
  T_{RLRLR}^{aebdc} - T_{RLRLR}^{aecdb} + \no \\
& \qquad T_{RLRLR}^{bdcea} + T_{RLRLR}^{becda} + 
 T_{RLRLR}^{cdbea} + T_{RLRLR}^{cebda}+ 
 T_{RLRLR}^{dbeca} + T_{RLRLR}^{dceba}+ 
 T_{RLRLR}^{ebdca} + T_{RLRLR}^{ecdba})+\no\\
    &~\ccoeff{10}{12}
   (T^{abdce}_{RLLLR}+T^{abecd}_{RLLLR}+T^{acdbe}_{RLLLR}+T^{acebd}_{RLLLR}+T^{adbec}_{RLLLR}+T^{adceb}_{RLLLR}+T^{aebdc}_{RLLLR}+T^{aecdb}_{RLLLR}-\no \\
   &\qquad T^{bdcea}_{RLLLR}- T^{becda}_{RLLLR}-T^{cadbe}_{RLLRR}-T^{caebd}_{RLLRR}+T^{cdabe}_{RLLRR}-T^{cdbea}_{RLLLR}+T^{ceabd}_{RLLRR}-T^{cebda}_{RLLLR}-\no \\
   &\qquad T^{dabec}_{RLLRR}-T^{daceb}_{RLLRR}+T^{dbaec}_{RLLRR}-T^{dbaec}_{RRLLR}+T^{dbeac}_{RRLLR}-T^{dbeca}_{RLLLR}+T^{dcaeb}_{RLLRR}-T^{dceba}_{RLLLR}-\no \\
   &\qquad T^{eabdc}_{RLLRR}-T^{eacdb}_{RLLRR}+T^{ebadc}_{RLLRR}-T^{ebadc}_{RRLLR}+T^{ebdac}_{RRLLR}-T^{ebdca}_{RLLLR}+T^{ecadb}_{RLLRR}-T^{ecdba}_{RLLLR})+\no \\
  &~\ccoeff{10}{13}
    (-T^{abdec}_{RLLLR}-T^{abedc}_{RLLLR}-T^{acdeb}_{RLLLR}
   -T^{acedb}_{RLLLR}-T^{adbce}_{RLLLR}-T^{adcbe}_{RLLLR}-
   T^{aebcd}_{RLLLR}-T^{aecbd}_{RLLLR}+ \no \\
   &\qquad T^{bdeca}_{RLLLR}+T^{bedca}_{RLLLR}-T^{cbdae}_{RRLLR}-T^{cbead}_{RRLLR}-T^{cdaeb}_{RLLRR}+T^{cdeba}_{RLLLR}-T^{ceadb}_{RLLRR}+T^{cedba}_{RLLLR} +\no\\
   &\qquad T^{daebc}_{RLLRR}+T^{daecb}_{RLLRR}-T^{dbace}_{RLLRR}+T^{dbace}_{RRLLR}+T^{dbcea}_{RLLLR}-T^{dcabe}_{RLLRR}+T^{dcabe}_{RRLLR}+T^{dcbea}_{RLLLR}+\no\\
    &\qquad  T^{eadbc}_{RLLRR}+T^{eadcb}_{RLLRR}-T^{ebacd}_{RLLRR}+T^{ebacd}_{RRLLR}+T^{ebcda}_{RLLLR}-T^{ecabd}_{RLLRR}+T^{ecabd}_{RRLLR}+T^{ecbda}_{RLLLR}) \no\\
   &~\ccoeff{10}{14}
   (T^{abdec}_{LLRLR}-T^{abdec}_{RLRLR}+T^{abedc}_{LLRLR}-
   T^{abedc}_{RLRLR}+T^{acdeb}_{LLRLR}-T^{acdeb}_{RLRLR}+T^{acedb}_{LLRLR}-T^{acedb}_{RLRLR}+ \no \\
   &\qquad T^{adbce}_{LLRLR}-T^{adbce}_{RLRLR}+T^{adcbe}_{LLRLR}-T^{adcbe}_{RLRLR}+T^{aebcd}_{LLRLR}-T^{aebcd}_{RLRLR}+T^{aecbd}_{LLRLR}-T^{aecbd}_{RLRLR}- \no \\
   &\qquad T^{bdeca}_{RLRLL}+T^{bdeca}_{RLRLR}-T^{bedca}_{RLRLL}+T^{bedca}_{RLRLR}-T^{cdeba}_{RLRLL}+T^{cdeba}_{RLRLR}-T^{cedba}_{RLRLL}+T^{cedba}_{RLRLR} -\no \\
   &\qquad  T^{dbcea}_{RLRLL}+T^{dbcea}_{RLRLR}-T^{dcbea}_{RLRLL}+T^{dcbea}_{RLRLR}-T^{ebcda}_{RLRLL}+T^{ebcda}_{RLRLR}-T^{ecbda}_{RLRLL}+T^{ecbda}_{RLRLR})+\no\\
   &~\ccoeff{10}{15}
   (T^{abdce}_{RLLRR}+T^{abecd}_{RLLRR}+T^{acdbe}_{RLLRR}+T^{acebd}_{RLLRR}+T^{adbec}_{RLLRR}+T^{adceb}_{RLLRR}+T^{aebdc}_{RLLRR}+T^{aecdb}_{RLLRR}-\no\\
   & \qquad T^{badce}_{RLLLR}-T^{baecd}_{RLLLR}+T^{bdcae}_{RLLLR}+T^{becad}_{RLLLR}-T^{cadbe}_{RLLLR}-T^{caebd}_{RLLLR}+T^{cdbae}_{RLLLR}+T^{cebad}_{RLLLR}-\no\\
   & \qquad T^{dabec}_{RLLLR}-T^{daceb}_{RLLLR}+T^{dbeac}_{RLLLR}-T^{dbeac}_{RLLRR}-T^{dbeca}_{RRLLR}+T^{dceab}_{RLLLR}-T^{dceab}_{RLLRR}-T^{dceba}_{RRLLR}-\no\\
   & \qquad T^{eabdc}_{RLLLR}-T^{eacdb}_{RLLLR}+T^{ebdac}_{RLLLR}-T^{ebdac}_{RLLRR}-T^{ebdca}_{RRLLR}+T^{ecdab}_{RLLLR}-T^{ecdab}_{RLLRR}-T^{ecdba}_{RRLLR})+\no \\
   &~\ccoeff{10}{16}(-T^{abcde}_{RLLLR}-T^{abced}_{RLLLR}-T^{acbde}_{RLLLR}-T^{acbed}_{RLLLR}-T^{adebc}_{RLLLR}-T^{adecb}_{RLLLR}- T^{aedbc}_{RLLLR}-T^{aedcb}_{RLLLR}+ \no \\
   & \qquad T^{bcdea}_{RLLLR}+T^{bceda}_{RLLLR}+T^{cadeb}_{RLLRR}+T^{caedb}_{RLLRR}+T^{cbade}_{RRLLR}+T^{cbaed}_{RRLLR}+T^{cbdea}_{RLLLR}+T^{cbeda}_{RLLLR}+ \no \\
   & \qquad T^{dabce}_{RLLRR}+T^{dacbe}_{RLLRR}-T^{dbcae}_{RRLLR}-T^{dcbae}_{RRLLR}-T^{deabc}_{RLLRR}-T^{deacb}_{RLLRR}+T^{debca}_{RLLLR}+ T^{decba}_{RLLLR}+\no \\
   & \qquad T^{eabcd}_{RLLRR}+T^{eacbd}_{RLLRR}-T^{ebcad}_{RRLLR}-T^{ecbad}_{RRLLR}-T^{edabc}_{RLLRR}-T^{edacb}_{RLLRR}+T^{edbca}_{RLLLR}+T^{edcba}_{RLLLR})+\no \\
   &~\ccoeff{10}{17} (T^{abcde}_{RLLRR}+T^{abced}_{RLLRR}+T^{acbde}_{RLLRR}+T^{acbed}_{RLLRR}+T^{adebc}_{RLLRR}+T^{adecb}_{RLLRR}+T^{aedbc}_{RLLRR}+T^{aedcb}_{RLLRR}  - \no \\
   & \qquad T^{badec}_{RLLLR}-T^{baedc}_{RLLLR}+T^{bdeac}_{RLLLR}+T^{bedac}_{RLLLR}-T^{cadeb}_{RLLLR}-T^{caedb}_{RLLLR}-T^{cbdea}_{RRLLR}-T^{cbeda}_{RRLLR}+ \no \\
   & \qquad  T^{cdeab}_{RLLLR} -T^{cdeab}_{RLLRR}+T^{cedab}_{RLLLR}-T^{cedab}_{RLLRR}-T^{dabce}_{RLLLR}-T^{dacbe}_{RLLLR}+T^{dbcae}_{RLLLR}+T^{dcbae}_{RLLLR}-\no \\
   & \qquad T^{eabcd}_{RLLLR}- T^{eacbd}_{RLLLR}+T^{ebcad}_{RLLLR}-T^{ebcad}_{RLLRR}+T^{ecbad}_{RLLLR}-T^{ecbad}_{RLLRR}-T^{edbca}_{RRLLR}-T^{edcba}_{RRLLR})+\no\\
     &~\ccoeff{10}{18}(-T^{abdec}_{RLLRR}-T^{abedc}_{RLLRR}-T^{acdeb}_{RLLRR}-T^{acedb}_{RLLRR}-T^{adbce}_{RLLRR}-T^{adcbe}_{RLLRR}- T^{aebcd}_{RLLRR}-T^{aecbd}_{RLLRR}+\no \\
   & \qquad T^{bacde}_{RLLLR}+T^{baced}_{RLLLR}-T^{bcdae}_{RLLLR}-T^{bcead}_{RLLLR}+ T^{cabde}_{RLLLR}+T^{cabed}_{RLLLR}-T^{cbdae}_{RLLLR}-T^{cbead}_{RLLLR}+\no \\
   & \qquad T^{daebc}_{RLLLR}+T^{daecb}_{RLLLR}+T^{dbcea}_{RRLLR}+T^{dcbea}_{RRLLR}-T^{debac}_{RLLLR}+T^{debac}_{RLLRR}-T^{decab}_{RLLLR}+T^{decab}_{RLLRR}+\no \\
   & \qquad T^{eadbc}_{RLLLR}+T^{eadcb}_{RLLLR}+T^{ebcda}_{RRLLR}+T^{ecbda}_{RRLLR}-T^{edbac}_{RLLLR}+T^{edbac}_{RLLRR}-T^{edcab}_{RLLLR}+T^{edcab}_{RLLRR}) \,, \no \\[6 pt]
    \ci{11}{abcde} =&~ \ccoeff{11}{1}( -T^{cadbe}_{LLLLL}  + T^{cadbe}_{RRRR} + 
 T^{caebd}_{LLLLL}  - T^{caebd}_{RRRR} + 
 T^{cbdae}_{LLLLL}  - T^{cbdae}_{RRRR} - 
 T^{cbead}_{LLLLL}  + T^{cbead}_{RRRR} -  \\
   & \qquad
 T^{dabce}_{LLLLL}  + T^{dabce}_{RRRR} + 
 T^{dabec}_{LLLLL}  - T^{dabec}_{RRRR} + 
 T^{dacbe}_{LLLLL}  - T^{dacbe}_{RRRR} - 
 T^{daebc}_{LLLLL}  + T^{daebc}_{RRRR} + \no \\
   & \qquad
 T^{dbace}_{LLLLL}  - T^{dbace}_{RRRR} - 
 T^{dbaec}_{LLLLL}  + T^{dbaec}_{RRRR} - 
 T^{dbcae}_{LLLLL}  + T^{dbcae}_{RRRR} + 
 T^{dbeac}_{LLLLL}  - T^{dbeac}_{RRRR} - \no \\
   & \qquad
 T^{dcabe}_{LLLLL}  + T^{dcabe}_{RRRR} + 
 T^{dcbae}_{LLLLL}  - T^{dcbae}_{RRRR} + 
 T^{eabcd}_{LLLLL}  - T^{eabcd}_{RRRR} - 
 T^{eabdc}_{LLLLL}  + T^{eabdc}_{RRRR} - \no \\
   & \qquad
 T^{eacbd}_{LLLLL}  + T^{eacbd}_{RRRR} + 
 T^{eadbc}_{LLLLL}  - T^{eadbc}_{RRRR} - 
 T^{ebacd}_{LLLLL}  + T^{ebacd}_{RRRR} + 
 T^{ebadc}_{LLLLL}  - T^{ebadc}_{RRRR} + \no \\
   & \qquad
 T^{ebcad}_{LLLLL}  - T^{ebcad}_{RRRR} - 
 T^{ebdac}_{LLLLL}  + T^{ebdac}_{RRRR} + 
 T^{ecabd}_{LLLLL}  - T^{ecabd}_{RRRR} - 
 T^{ecbad}_{LLLLL}  + T^{ecbad}_{RRRR}) \no \\
 &~ \ccoeff{11}{2} (T^{bacde}_{RLLLL} - T^{baced}_{RLLLL} - 
 T^{badce}_{RLLLL} + T^{badec}_{RLLLL} + 
 T^{baecd}_{RLLLL} - T^{baedc}_{RLLLL} + 
 T^{bcade}_{RLLLL}- T^{bcaed}_{RLLLL}-  \no \\
   & \qquad
 T^{bcdae}_{RLLLL}+ T^{bcdae}_{RLRRR} + 
 T^{bcead}_{RLLLL}- T^{bcead}_{RLRRR} - 
 T^{bdace}_{RLLLL}+ T^{bdcae}_{RLLLL}- 
 T^{bdcae}_{RLRRR} + T^{bdeac}_{RLRRR} +  \no \\
   & \qquad
 T^{becad}_{RLRRR} - T^{bedac}_{RLRRR} - 
 T^{cabde}_{RLLLL} + T^{cabed}_{RLLLL} - 
 T^{cadbe}_{LLLLR} + T^{cadbe}_{RLLLL} - 
 T^{cadeb}_{RLLLL} + T^{caebd}_{LLLLR} -  \no \\
   & \qquad
 T^{caebd}_{RLLLL} + T^{caedb}_{RLLLL} - 
 T^{cbade}_{RLLLL}+ T^{cbaed}_{RLLLL}+ 
 T^{cbdae}_{RLLLL}- T^{cbdae}_{RLRRR} - 
 T^{cbead}_{RLLLL}+ T^{cbead}_{RLRRR} +  \no \\
   & \qquad
 T^{cdabe}_{RLLLL}- T^{cdbae}_{RLLLL}+ 
 T^{cdbae}_{RLRRR} - T^{cdeab}_{RLRRR} - 
 T^{cebad}_{RLRRR} + T^{cedab}_{RLRRR} - 
 T^{dabce}_{LLLLR} + T^{dabce}_{RLLLL} +  \no \\
   & \qquad
 T^{dabec}_{LLLRL} - T^{dabec}_{RLLLL} + 
 T^{dacbe}_{LLLLR} - T^{dacbe}_{RLLLL} + 
 T^{daceb}_{RLLLL} - T^{daebc}_{LLLLR} + 
 T^{daebc}_{RLLLL} + T^{daecb}_{LLLLR} -  \no \\
   & \qquad
 T^{daecb}_{RLLLL} + T^{dbace}_{LLLLR} + 
 T^{dbace}_{RLLLL}- T^{dbaec}_{RLLLL}- 
 T^{dbcae}_{RLLLL}+ T^{dbcae}_{RLRRR} + 
 T^{dbeac}_{RLLLL}- T^{dbeac}_{RLRRR} -  \no \\
   & \qquad
 T^{dcabe}_{LLLLR} - T^{dcabe}_{RLLLL}+ 
 T^{dcbae}_{RLLLL}- T^{dcbae}_{RLRRR} + 
 T^{dceab}_{RLRRR} + T^{debac}_{RLRRR} - 
 T^{decab}_{RLRRR} + T^{eabcd}_{LLLLR} +  \no \\
   & \qquad
 T^{eabcd}_{LLLRL} - T^{eabcd}_{RLLLL} - 
 T^{eabdc}_{LLLLR} - T^{eabdc}_{LLLRL} + 
 T^{eabdc}_{RLLLL} - T^{eacbd}_{LLLLR} - 
 T^{eacbd}_{LLLRL} + T^{eacbd}_{RLLLL} +  \no \\
   & \qquad
 T^{eacdb}_{LLLLR} - T^{eacdb}_{RLLLL} + 
 T^{eadbc}_{LLLLR} - T^{eadbc}_{RLLLL} - 
 T^{eadcb}_{LLLLR} + T^{eadcb}_{RLLLL} - 
 T^{ebacd}_{LLLLR} - T^{ebacd}_{RLLLL}+  \no \\
   & \qquad
 T^{ebadc}_{LLLLR} + T^{ebadc}_{RLLLL}+ 
 T^{ebcad}_{RLLLL}- T^{ebcad}_{RLRRR} - 
 T^{ebdac}_{RLLLL}+ T^{ebdac}_{RLRRR} + 
 T^{ecabd}_{LLLLR} + T^{ecabd}_{RLLLL}-  \no \\
   & \qquad
 T^{ecadb}_{LLLLR} - T^{ecbad}_{RLLLL}+ 
 T^{ecbad}_{RLRRR} - T^{ecda b}_{RLRRR} - 
 T^{edabc}_{LLLLR} + T^{edacb}_{LLLLR} - 
 T^{edbac}_{RLRRR} + T^{edcab}_{RLRRR}) \no \\
 &~\ccoeff{11}{3}(-T^{abcde}_{RLLLL} + T^{abced}_{RLLLL} + 
 T^{abdce}_{RLLLL} - T^{abecd}_{RLLLL} + 
 T^{acbde}_{RLLLL} - T^{acbed}_{RLLLL} - 
 T^{acdbe}_{RLLLL} + T^{acebd}_{RLLLL} - \no \\
   & \qquad
 T^{adbce}_{RLLLL} + T^{adcbe}_{RLLLL} - 
 T^{cadbe}_{RLRRR} + T^{caebd}_{RLRRR} + 
 T^{cbdae}_{RRRLR} - T^{cbead}_{RRRLR} - 
 T^{dabce}_{RLRRR} + T^{dabec}_{RLRRR} +  \no \\
   & \qquad
 T^{dacbe}_{RLRRR} - T^{daebc}_{RLRRR} - 
 T^{dbace}_{LLRLL} + T^{dbace}_{RRLRR} - 
 T^{dbaec}_{RRLRR} - T^{dbcae}_{RRRLR} + 
 T^{dbeac}_{RRRLR} + T^{dbeca}_{LLLLR} +  \no \\
   & \qquad
 T^{dcabe}_{LLRLL} - T^{dcabe}_{RRLRR} + 
 T^{dcbae}_{RRRLR} - T^{dceba}_{LLLLR} + 
 T^{eabcd}_{RLRRR} - T^{eabdc}_{RLRRR} - 
 T^{eacbd}_{RLRRR} + T^{eadbc}_{RLRRR} +  \no \\
   & \qquad
 T^{ebacd}_{LLRLL} - T^{ebacd}_{RRLRR} + 
 T^{ebadc}_{RRLRR} - T^{ebcad}_{LLLRL} + 
 T^{ebcad}_{RRRLR} + T^{ebcda}_{LLLLR} - 
 T^{ebdac}_{RRRLR} - T^{ebdca}_{LLLLR} -  \no \\
   & \qquad
 T^{ecabd}_{LLRLL} + T^{ecabd}_{RRLRR} + 
 T^{ecbad}_{LLLRL} - T^{ecbad}_{RRRLR} - 
 T^{ecbda}_{LLLLR} + T^{ecdba}_{LLLLR} + 
 T^{edbca}_{LLLLR} - T^{edcba}_{LLLLR}) \no  \\
 & ~ \ccoeff{11}{4} ( -T^{bacde}_{RLLLL} + T^{baced}_{RLLLL} + 
 T^{badce}_{RLLLL} - T^{baecd}_{RLLLL} + 
 T^{cabde}_{RLLLL} - T^{cabed}_{RLLLL} - 
 T^{cadbe}_{LLRLL} - T^{cadbe}_{RLLLL} +  \no \\
   & \qquad
 T^{cadbe}_{RRLRR} - T^{cadeb}_{RRLRR} + 
 T^{caebd}_{LLRLL} + T^{caebd}_{RLLLL} - 
 T^{caebd}_{RRLRR} + T^{caedb}_{RRLRR} + 
 T^{cbade}_{RRRLR} - T^{cbaed}_{RRRLR} +  \no \\
   & \qquad
 T^{cbdae}_{LLLLR} + T^{cbdae}_{LLRLL} - 
 T^{cbdae}_{RRLRR} - T^{cbead}_{LLLLR} - 
 T^{cbead}_{LLRLL} + T^{cbead}_{RRLRR} - 
 T^{cdabe}_{RLRRR} + T^{cdaeb}_{RLRRR} +  \no \\
   & \qquad
 T^{ceabd}_{RLRRR} - T^{ceadb}_{RLRRR} - 
 T^{dabce}_{LLRLL} - T^{dabce}_{RLLLL} + 
 T^{dabce}_{RRLRR} + T^{dabec}_{RLLLL} - 
 T^{dabec}_{RRLRR} + T^{dacbe}_{LLRLL} +  \no \\
   & \qquad
 T^{dacbe}_{RLLLL} - T^{dacbe}_{RRLRR} + 
 T^{daceb}_{RRLRR} - T^{daebc}_{LLRLL} - 
 T^{daebc}_{RLLLL} + T^{daebc}_{RRLRR} - 
 T^{daecb}_{RRLRR} - T^{dbace}_{RLRRR} -  \no \\
   & \qquad
 T^{dbace}_{RRRLR} - T^{dbaec}_{LLLRL} + 
 T^{dbaec}_{RLRRR} + T^{dbaec}_{RRRLR} - 
 T^{dbcae}_{LLLLR} - T^{dbcae}_{LLRLL} + 
 T^{dbcae}_{RRLRR} + T^{dbeac}_{LLLLR} +  \no \\
   & \qquad
 T^{dbeac}_{LLRLL} - T^{dbeac}_{RRLRR} + 
 T^{dcabe}_{RLRRR} + T^{dcabe}_{RRRLR} - 
 T^{dcaeb}_{RLRRR} + T^{dcbae}_{LLLLR} + 
 T^{dcbae}_{LLRLL} - T^{dcbae}_{RRLRR} -  \no \\
   & \qquad
 T^{dceab}_{LLLLR} - T^{deabc}_{RLRRR} + 
 T^{deacb}_{RLRRR} + T^{eabcd}_{LLRLL} + 
 T^{eabcd}_{RLLLL} - T^{eabcd}_{RRLRR} - 
 T^{eabdc}_{RLLLL} + T^{eabdc}_{RRLRR} -  \no \\
   & \qquad
 T^{eacbd}_{LLRLL} - T^{eacbd}_{RLLLL} + 
 T^{eacbd}_{RRLRR} - T^{eacdb}_{RRLRR} + 
 T^{eadbc}_{LLRLL} + T^{eadbc}_{RLLLL} - 
 T^{eadbc}_{RRLRR} + T^{eadcb}_{RRLRR} -  \no \\
   & \qquad
 T^{ebacd}_{LLLRL} + T^{ebacd}_{RLRRR} + 
 T^{ebacd}_{RRRLR} + T^{ebadc}_{LLLRL} - 
 T^{ebadc}_{RLRRR} - T^{ebadc}_{RRRLR} + 
 T^{ebcad}_{LLLLR} + T^{ebcad}_{LLRLL} -  \no \\
   & \qquad
 T^{ebcad}_{RRLRR} - T^{ebdac}_{LLLLR} - 
 T^{ebdac}_{LLRLL} + T^{ebdac}_{RRLRR} + 
 T^{ecabd}_{LLLRL} - T^{ecabd}_{RLRRR} - 
 T^{ecabd}_{RRRLR} + T^{ecadb}_{RLRRR} -  \no \\
   & \qquad
 T^{ecbad}_{LLLLR} - T^{ecbad}_{LLRLL} + 
 T^{ecbad}_{RRLRR} + T^{ecda b}_{LLLLR} + 
 T^{edabc}_{RLRRR} - T^{edacb}_{RLRRR} + 
 T^{edbac}_{LLLLR} - T^{edcab}_{LLLLR}) + \no \\
 & \ccoeff{11}{5}( T^{abcde}_{RLLLR} - T^{abced}_{RLLLR} - 
 T^{abdce}_{RLLLR} + T^{abdec}_{RLLLR} + 
 T^{abecd}_{RLLLR} - T^{abedc}_{RLLLR} - 
 T^{acbde}_{RLLLR} + T^{acbed}_{RLLLR} +  \no \\
   & \qquad
 T^{acdbe}_{RLLLR} - T^{acdeb}_{RLLLR} - 
 T^{acebd}_{RLLLR} + T^{acedb}_{RLLLR} + 
 T^{adbce}_{RLLLR} - T^{adbec}_{RLLLR} - 
 T^{adcbe}_{RLLLR} + T^{adceb}_{RLLLR} +  \no \\
   & \qquad
 T^{adebc}_{RLLLR} - T^{adecb}_{RLLLR} - 
 T^{aebcd}_{RLLLR} + T^{aebdc}_{RLLLR} + 
 T^{aecbd}_{RLLLR} - T^{aecdb}_{RLLLR} - 
 T^{aedbc}_{RLLLR} + T^{aedcb}_{RLLLR} +   \no \\
   & \qquad
 T^{bcdea}_{RLLLR} - T^{bceda}_{RLLLR} - 
 T^{bdcea}_{RLLLR} + T^{bdeca}_{RLLLR} + 
 T^{becda}_{RLLLR} - T^{bedca}_{RLLLR} + 
 T^{cadbe}_{RLLRR} - T^{cadeb}_{RLLRR} -  \no \\
   & \qquad
 T^{caebd}_{RLLRR} + T^{caedb}_{RLLRR} + 
 T^{cbade}_{RRLLR} - T^{cbaed}_{RRLLR} - 
 T^{cbdae}_{RRLLR} - T^{cbdea}_{RLLLR} + 
 T^{cbead}_{RRLLR} + T^{cbeda}_{RLLLR} -\no \\
   & \qquad
 T^{cdabe}_{RLLRR} + T^{cdaeb}_{RLLRR} + 
 T^{cdbea}_{RLLLR} - T^{cdeba}_{RLLLR} + 
 T^{ceabd}_{RLLRR} - T^{ceadb}_{RLLRR} - 
 T^{cebda}_{RLLLR} + T^{cedba}_{RLLLR} + \no \\
   & \qquad
 T^{dabce}_{RLLRR} - T^{dabec}_{RLLRR} - 
 T^{dacbe}_{RLLRR} + T^{daceb}_{RLLRR} + 
 T^{daebc}_{RLLRR} - T^{daecb}_{RLLRR} - 
 T^{dbace}_{RLLRR} - T^{dbace}_{RRLLR} + \no \\
   & \qquad
 T^{dbaec}_{RLLRR} + T^{dbaec}_{RRLLR} + 
 T^{dbcae}_{RRLLR} + T^{dbcea}_{RLLLR} - 
 T^{dbeac}_{RRLLR} - T^{dbeca}_{RLLLR} + 
 T^{dcabe}_{RLLRR} + T^{dcabe}_{RRLLR} - \no \\
   & \qquad
 T^{dcaeb}_{RLLRR} - T^{dcbae}_{RRLLR} - 
 T^{dcbea}_{RLLLR} + T^{dceba}_{RLLLR} - 
 T^{deabc}_{RLLRR} + T^{deacb}_{RLLRR} + 
 T^{debca}_{RLLLR} - T^{decba}_{RLLLR} - \no \\
   & \qquad
 T^{eabcd}_{RLLRR} + T^{eabdc}_{RLLRR} + 
 T^{eacbd}_{RLLRR} - T^{eacdb}_{RLLRR} - 
 T^{eadbc}_{RLLRR} + T^{eadcb}_{RLLRR} + 
 T^{ebacd}_{RLLRR} + T^{ebacd}_{RRLLR} - \no \\
   & \qquad
 T^{ebadc}_{RLLRR} - T^{ebadc}_{RRLLR} - 
 T^{ebcad}_{RRLLR} - T^{ebcda}_{RLLLR} + 
 T^{ebdac}_{RRLLR} + T^{ebdca}_{RLLLR} - 
 T^{ecabd}_{RLLRR} - T^{ecabd}_{RRLLR} + \no \\
   & \qquad
 T^{ecadb}_{RLLRR} + T^{ecbad}_{RRLLR} + 
 T^{ecbda}_{RLLLR} - T^{ecdba}_{RLLLR} + 
 T^{edabc}_{RLLRR} - T^{edacb}_{RLLRR} - 
 T^{edbca}_{RLLLR} + T^{edcba}_{RLLLR} ) \no \\
 &~\ccoeff{11}{6} ( -T^{abcde}_{RLRLR} + T^{abced}_{RLRLR} + 
 T^{abdce}_{RLRLR} - T^{abdec}_{RLRLR} - 
 T^{abecd}_{RLRLR} + T^{abedc}_{RLRLR} + 
 T^{acbde}_{RLRLR} - T^{acbed}_{RLRLR} -  \no \\
   & \qquad
 T^{acdbe}_{RLRLR} + T^{acdeb}_{RLRLR} + 
 T^{acebd}_{RLRLR} - T^{acedb}_{RLRLR} - 
 T^{adbce}_{RLRLR} + T^{adbec}_{RLRLR} + 
 T^{adcbe}_{RLRLR} - T^{adceb}_{RLRLR} - \no \\
   & \qquad
 T^{adebc}_{RLRLR} + T^{adecb}_{RLRLR} + 
 T^{aebcd}_{RLRLR} - T^{aebdc}_{RLRLR} - 
 T^{aecbd}_{RLRLR} + T^{aecdb}_{RLRLR} + 
 T^{aedbc}_{RLRLR} - T^{aedcb}_{RLRLR} - \no \\
   & \qquad
 T^{bacde}_{RLLRL} + T^{baced}_{RLLRL} + 
 T^{badce}_{RLLRL} + T^{bcade}_{RLLRL} - 
 T^{bcaed}_{RLLRL} - T^{bcdae}_{RLRLL} - 
 T^{bcdea}_{RLRLR} + T^{bcead}_{RLRLL} + \no \\
   & \qquad
 T^{bceda}_{RLRLR} - T^{bdace}_{RLLRL} + 
 T^{bdcae}_{RLRLL} + T^{bdcea}_{RLRLR} - 
 T^{bdeca}_{RLRLR} - T^{becda}_{RLRLR} + 
 T^{bedca}_{RLRLR} + T^{cabde}_{RLLRL} - \no \\
   & \qquad
 T^{cabed}_{RLLRL} - T^{cadbe}_{LLRLR} - 
 T^{cadbe}_{RLLRL} + T^{caebd}_{LLRLR} - 
 T^{cbade}_{RLLRL} + T^{cbaed}_{RLLRL} + 
 T^{cbdae}_{RLRLL} + T^{cbdea}_{RLRLR} - \no \\
   & \qquad
 T^{cbead}_{RLRLL} - T^{cbeda}_{RLRLR} + 
 T^{cdabe}_{RLLRL} - T^{cdbae}_{RLRLL} - 
 T^{cdbea}_{RLRLR} + T^{cdeba}_{RLRLR} + 
 T^{cebda}_{RLRLR} - T^{cedba}_{RLRLR} - \no \\
   & \qquad
 T^{dabce}_{LLRLR} - T^{dabce}_{RLLRL} + 
 T^{dabec}_{RLLRL} + T^{dacbe}_{LLRLR} + 
 T^{dacbe}_{RLLRL} - T^{daebc}_{LLRLR} + 
 T^{daecb}_{LLRLR} + T^{dbace}_{RLLRL} - \no \\
   & \qquad
 T^{dbaec}_{RLLRL} - T^{dbcae}_{RLRLL} - 
 T^{dbcea}_{RLRLR} + T^{dbeac}_{RLRLL} + 
 T^{dbeca}_{RLRLR} - T^{dcabe}_{RLLRL} + 
 T^{dcbae}_{RLRLL} + T^{dcbea}_{RLRLR} - \no \\
   & \qquad
 T^{dceba}_{RLRLR} - T^{debca}_{RLRLR} + 
 T^{decba}_{RLRLR} + T^{eabcd}_{LLRLR} + 
 T^{eabcd}_{RLLRL} - T^{eabdc}_{LLRLR} - 
 T^{eabdc}_{RLLRL} - T^{eacbd}_{LLRLR} - \no \\
   & \qquad
 T^{eacbd}_{RLLRL} + T^{eacdb}_{LLRLR} + 
 T^{eadbc}_{LLRLR} - T^{eadcb}_{LLRLR} - 
 T^{ebacd}_{RLLRL} + T^{ebadc}_{RLLRL} + 
 T^{ebcad}_{RLRLL} + T^{ebcda}_{RLRLR} - \no \\
   & \qquad
 T^{ebdac}_{RLRLL} - T^{ebdca}_{RLRLR} + 
 T^{ecabd}_{RLLRL} - T^{ecbad}_{RLRLL} - 
 T^{ecbda}_{RLRLR} + T^{ecdba}_{RLRLR} + 
 T^{edbca}_{RLRLR} - T^{edcba}_{RLRLR}) \no \\
 &~\ccoeff{11}{7} (T^{bacde}_{RLRLL} - T^{badce}_{RLRLL} + 
 T^{baecd}_{RLRLL} + T^{bcade}_{RLRLR} - 
 T^{bcaed}_{RLRLR} - T^{bdace}_{RLRLR} + 
 T^{bdaec}_{RLRLR} + T^{beacd}_{RLRLR} - \no \\
   & \qquad
 T^{beadc}_{RLRLR} - T^{cabde}_{RLRLL} + 
 T^{cadbe}_{RLRLL} - T^{caebd}_{RLRLL} - 
 T^{cbade}_{RLRLR} + T^{cbaed}_{RLRLR} - 
 T^{cbdae}_{LLRLR} + T^{cbead}_{LLRLR} + \no \\
   & \qquad
 T^{cdabe}_{RLRLR} - T^{cdaeb}_{RLRLR} - 
 T^{ceabd}_{RLRLR} + T^{ceadb}_{RLRLR} + 
 T^{dabce}_{RLRLL} - T^{dacbe}_{RLRLL} + 
 T^{daebc}_{RLRLL} + T^{dbace}_{RLRLR} - \no \\
   & \qquad
 T^{dbaec}_{RLRLR} + T^{dbcae}_{LLRLR} - 
 T^{dbeac}_{LLRLR} - T^{dcabe}_{RLRLR} + 
 T^{dcaeb}_{RLRLR} - T^{dcbae}_{LLRLR} + 
 T^{dceab}_{LLRLR} + T^{deabc}_{RLRLR} - \no \\
   & \qquad
 T^{deacb}_{RLRLR} - T^{eabcd}_{RLRLL} + 
 T^{eacbd}_{RLRLL} - T^{eadbc}_{RLRLL} - 
 T^{ebacd}_{RLRLR} + T^{ebadc}_{RLRLR} - 
 T^{ebcad}_{LLRLR} + T^{ebdac}_{LLRLR} + \no \\
   & \qquad
 T^{ecabd}_{RLRLR} - T^{ecadb}_{RLRLR} + 
 T^{ecbad}_{LLRLR} - T^{ecda b}_{LLRLR} - 
 T^{edabc}_{RLRLR} + T^{edacb}_{RLRLR} - 
 T^{edbac}_{LLRLR} + T^{edcab}_{LLRLR} ) \no \\
  &~\ccoeff{11}{8} (-T^{abcde}_{RLLRL} - T^{abcde}_{RLRLL} + 
 T^{abced}_{RLLRL} + T^{abdce}_{RLLRL} + 
 T^{abdce}_{RLRLL} - T^{abecd}_{RLRLL} + 
 T^{acbde}_{RLLRL} + T^{acbde}_{RLRLL} -  \no \\
   & \qquad
 T^{acbed}_{RLLRL} - T^{acdbe}_{RLLRL} - 
 T^{acdbe}_{RLRLL} + T^{acebd}_{RLRLL} - 
 T^{adbce}_{RLLRL} - T^{adbce}_{RLRLL} + 
 T^{adcbe}_{RLLRL} + T^{adcbe}_{RLRLL} -  \no \\
   & \qquad
 T^{bacde}_{RLRLR} + T^{baced}_{RLRLR} + 
 T^{badce}_{RLRLR} - T^{badec}_{RLRLR} - 
 T^{baecd}_{RLRLR} + T^{baedc}_{RLRLR} - 
 T^{bcade}_{RLRLL} + T^{bcdae}_{RLLRL} -  \no \\
   & \qquad
 T^{bcdae}_{RLRLR} + T^{bcead}_{RLRLR} + 
 T^{bdace}_{RLRLL} - T^{bdcae}_{RLLRL} + 
 T^{bdcae}_{RLRLR} - T^{bdeac}_{RLRLR} - 
 T^{becad}_{RLRLR} + T^{bedac}_{RLRLR} +  \no \\
   & \qquad
 T^{cabde}_{RLRLR} - T^{cabed}_{RLRLR} - 
 T^{cadbe}_{RLRLR} + T^{cadeb}_{RLRLR} + 
 T^{caebd}_{RLRLR} - T^{caedb}_{RLRLR} + 
 T^{cbade}_{RLRLL} - T^{cbdae}_{RLLRL} +  \no \\
   & \qquad
 T^{cbdae}_{RLRLR} - T^{cbead}_{RLRLR} - 
 T^{cdabe}_{RLRLL} + T^{cdbae}_{RLLRL} - 
 T^{cdbae}_{RLRLR} + T^{cdeab}_{RLRLR} + 
 T^{cebad}_{RLRLR} - T^{cedab}_{RLRLR} -  \no \\
   & \qquad
 T^{dabce}_{RLRLR} + T^{dabec}_{RLRLR} + 
 T^{dacbe}_{RLRLR} - T^{daceb}_{RLRLR} - 
 T^{daebc}_{RLRLR} + T^{daecb}_{RLRLR} - 
 T^{dbace}_{LLRLR} - T^{dbace}_{RLRLL} +  \no \\
   & \qquad
 T^{dbcae}_{RLLRL} - T^{dbcae}_{RLRLR} + 
 T^{dbeac}_{RLRLR} + T^{dbeca}_{LLRLR} + 
 T^{dcabe}_{LLRLR} + T^{dcabe}_{RLRLL} - 
 T^{dcbae}_{RLLRL} + T^{dcbae}_{RLRLR} -  \no \\
   & \qquad
 T^{dceab}_{RLRLR} - T^{dceba}_{LLRLR} - 
 T^{debac}_{RLRLR} + T^{decab}_{RLRLR} + 
 T^{eabcd}_{RLRLR} - T^{eabdc}_{RLRLR} - 
 T^{eacbd}_{RLRLR} + T^{eacdb}_{RLRLR} +  \no \\
   & \qquad
 T^{eadbc}_{RLRLR} - T^{eadcb}_{RLRLR} + 
 T^{ebacd}_{LLRLR} + T^{ebacd}_{RLRLL} - 
 T^{ebadc}_{LLRLR} - T^{ebcad}_{RLLRL} + 
 T^{ebcad}_{RLRLR} + T^{ebcda}_{LLRLR} -  \no \\
   & \qquad
 T^{ebdac}_{RLRLR} - T^{ebdca}_{LLRLR} - 
 T^{ecabd}_{LLRLR} - T^{ecabd}_{RLRLL} + 
 T^{ecadb}_{LLRLR} + T^{ecbad}_{RLLRL} - 
 T^{ecbad}_{RLRLR} - T^{ecbda}_{LLRLR} +  \no \\
   & \qquad
 T^{ecda b}_{RLRLR} + T^{ecdba}_{LLRLR} + 
 T^{edabc}_{LLRLR} - T^{edacb}_{LLRLR} + 
 T^{edbac}_{RLRLR} + T^{edbca}_{LLRLR} - 
 T^{edcab}_{RLRLR} - T^{edcba}_{LLRLR}) \no \\
& ~\ccoeff{11}{9} (-T^{bacde}_{RLLLR} + T^{bacde}_{RRLLR} + 
 T^{baced}_{RLLLR} - T^{baced}_{RRLLR} + 
 T^{badce}_{RLLLR} - T^{badce}_{RRLLR} - 
 T^{badec}_{RLLLR} + T^{badec}_{RRLLR} -  \no \\
   & \qquad
 T^{baecd}_{RLLLR} + T^{baecd}_{RRLLR} + 
 T^{baedc}_{RLLLR} - T^{baedc}_{RRLLR} - 
 T^{bcdae}_{RLLLR} + T^{bcdae}_{RLLRR} + 
 T^{bcead}_{RLLLR} - T^{bcead}_{RLLRR} +  \no \\
   & \qquad
 T^{bdcae}_{RLLLR} - T^{bdcae}_{RLLRR} - 
 T^{bdeac}_{RLLLR} + T^{bdeac}_{RLLRR} - 
 T^{becad}_{RLLLR} + T^{becad}_{RLLRR} + 
 T^{bedac}_{RLLLR} - T^{bedac}_{RLLRR} +  \no \\
   & \qquad
 T^{cabde}_{RLLLR} - T^{cabde}_{RRLLR} - 
 T^{cabed}_{RLLLR} + T^{cabed}_{RRLLR} - 
 T^{cadbe}_{RLLLR} + T^{cadbe}_{RRLLR} + 
 T^{cadeb}_{RLLLR} - T^{cadeb}_{RRLLR} +  \no \\
   & \qquad
 T^{caebd}_{RLLLR} - T^{caebd}_{RRLLR} - 
 T^{caedb}_{RLLLR} + T^{caedb}_{RRLLR} + 
 T^{cbdae}_{RLLLR} - T^{cbdae}_{RLLRR} - 
 T^{cbead}_{RLLLR} + T^{cbead}_{RLLRR} -  \no \\
   & \qquad
 T^{cdbae}_{RLLLR} + T^{cdbae}_{RLLRR} + 
 T^{cdeab}_{RLLLR} - T^{cdeab}_{RLLRR} + 
 T^{cebad}_{RLLLR} - T^{cebad}_{RLLRR} - 
 T^{cedab}_{RLLLR} + T^{cedab}_{RLLRR} -  \no \\
   & \qquad
 T^{dabce}_{RLLLR} + T^{dabce}_{RRLLR} + 
 T^{dabec}_{RLLLR} - T^{dabec}_{RRLLR} + 
 T^{dacbe}_{RLLLR} - T^{dacbe}_{RRLLR} - 
 T^{daceb}_{RLLLR} + T^{daceb}_{RRLLR} -  \no \\
   & \qquad
 T^{daebc}_{RLLLR} + T^{daebc}_{RRLLR} + 
 T^{daecb}_{RLLLR} - T^{daecb}_{RRLLR} - 
 T^{dbcae}_{RLLLR} + T^{dbcae}_{RLLRR} + 
 T^{dbeac}_{RLLLR} - T^{dbeac}_{RLLRR} +  \no \\
   & \qquad
 T^{dcbae}_{RLLLR} - T^{dcbae}_{RLLRR} - 
 T^{dceab}_{RLLLR} + T^{dceab}_{RLLRR} - 
 T^{debac}_{RLLLR} + T^{debac}_{RLLRR} + 
 T^{decab}_{RLLLR} - T^{decab}_{RLLRR} +  \no \\
   & \qquad
 T^{eabcd}_{RLLLR} - T^{eabcd}_{RRLLR} - 
 T^{eabdc}_{RLLLR} + T^{eabdc}_{RRLLR} - 
 T^{eacbd}_{RLLLR} + T^{eacbd}_{RRLLR} + 
 T^{eacdb}_{RLLLR} - T^{eacdb}_{RRLLR} +  \no \\
   & \qquad
 T^{eadbc}_{RLLLR} - T^{eadbc}_{RRLLR} - 
 T^{eadcb}_{RLLLR} + T^{eadcb}_{RRLLR} + 
 T^{ebcad}_{RLLLR} - T^{ebcad}_{RLLRR} - 
 T^{ebdac}_{RLLLR} + T^{ebdac}_{RLLRR} -  \no \\
   & \qquad
 T^{ecbad}_{RLLLR} + T^{ecbad}_{RLLRR} + 
 T^{ecda b}_{RLLLR} - T^{ecda b}_{RLLRR} + 
 T^{edbac}_{RLLLR} - T^{edbac}_{RLLRR} - 
 T^{edcab}_{RLLLR} + T^{edcab}_{RLLRR})\no \\
& ~\ccoeff{11}{10}( T^{bcade}_{RLLLR} - T^{bcaed}_{RLLLR} - 
 T^{bcdea}_{RLLRR} + T^{bceda}_{RLLRR} - 
 T^{bdace}_{RLLLR} + T^{bdaec}_{RLLLR} + 
 T^{bdcea}_{RLLRR} - T^{bdeca}_{RLLRR} +  \no \\
   & \qquad
 T^{beacd}_{RLLLR} - T^{beadc}_{RLLLR} - 
 T^{becda}_{RLLRR} + T^{bedca}_{RLLRR} - 
 T^{cbade}_{RLLLR} + T^{cbaed}_{RLLLR} + 
 T^{cbdea}_{RLLRR} - T^{cbeda}_{RLLRR} +  \no \\
   & \qquad
 T^{cdabe}_{RLLLR} - T^{cdaeb}_{RLLLR} - 
 T^{cdbea}_{RLLRR} + T^{cdeba}_{RLLRR} - 
 T^{ceabd}_{RLLLR} + T^{ceadb}_{RLLLR} + 
 T^{cebda}_{RLLRR} - T^{cedba}_{RLLRR} +  \no \\
   & \qquad
 T^{dbace}_{RLLLR} - T^{dbaec}_{RLLLR} - 
 T^{dbcea}_{RLLRR} + T^{dbeca}_{RLLRR} - 
 T^{dcabe}_{RLLLR} + T^{dcaeb}_{RLLLR} + 
 T^{dcbea}_{RLLRR} - T^{dceba}_{RLLRR} +  \no \\
   & \qquad
 T^{deabc}_{RLLLR} - T^{deacb}_{RLLLR} - 
 T^{debca}_{RLLRR} + T^{decba}_{RLLRR} - 
 T^{ebacd}_{RLLLR} + T^{ebadc}_{RLLLR} + 
 T^{ebcda}_{RLLRR} - T^{ebdca}_{RLLRR} +  \no \\
   & \qquad
 T^{ecabd}_{RLLLR} - T^{ecadb}_{RLLLR} - 
 T^{ecbda}_{RLLRR} + T^{ecdba}_{RLLRR} - 
 T^{edabc}_{RLLLR} + T^{edacb}_{RLLLR} + 
 T^{edbca}_{RLLRR} - T^{edcba}_{RLLRR}) \,.  \no
  \end{align}
}
\subsection{Solution of Wess-Zumino conditions in the bosonic sector}
\label{app:solWZ}
\subsubsection{P-even}
\label{app:solWZPeven}

{\footnotesize
\begin{align}
& \ccoeff{3}{RLL}-  i \ccoeff{0}{} - \ccoeff{3}{LLR}=0   \\
&  \ccoeff{3}{LLL}= i \ccoeff{0}{} \no \\ 
& \ccoeff{2}{LLR}-4 \ccoeff{6}{} + \ccoeff{2}{LLL}+ 2 \ccoeff{2}{RLL} =0 \no \\
& \ccoeff{3}{LLR}= -i \ccoeff{0}{} + 2 \ccoeff{6}{} \no \\ 
& \ccoeff{5}{}+ \ccoeff{4}{} - 2 \ccoeff{6}{} =0 \no  \\
& \ccoeff{7}{LRRR}= -i \ccoeff{6}{} +  \frac{i}{2} \ccoeff{2}{RLL} + \ccoeff{7}{LRLL} \no \\
& \ccoeff{7}{LLRL}= -\frac{i}{2}  \ccoeff{2}{RLL} + \ccoeff{7}{LLLR} \no \\
&\ccoeff{7}{LRRR}= 2 i \ccoeff{6}{} - i \ccoeff{2}{RLL} + \ccoeff{7 \prime}{LLRL}\no \\
& \ccoeff{8 \prime}{LRLL} = 4 i \ccoeff{6}{} -  i\ccoeff{2}{RLL} - \ccoeff{8}{LLLR}+ \ccoeff{8}{LLRL}  + \ccoeff{8 \prime}{LLRL},\no \\
&\ccoeff{8\prime \prime}{LLLL}= \ccoeff{8 \prime}{LLLL}\no \\
& \ccoeff{8\prime \prime}{LLLR} -2 i \ccoeff{4}{} + 2 i \ccoeff{6}{} + i \ccoeff{2}{RLL} - 2 \ccoeff{7}{LRLL} +  \ccoeff{8}{LLRL} - \ccoeff{8}{LRLL}\no \\
 &\ccoeff{8 \prime}{LLLL} -2 i \ccoeff{6}{} + 4 \ccoeff{7}{LLLR} + 2 \ccoeff{7}{LLRR}+    4 \ccoeff{7}{LRLL} + 2 \ccoeff{7}{LRLR} + 2 \ccoeff{7}{LRRL}\no \\
& \ccoeff{8}{LRRL}= -4 i \ccoeff{6}{} + 2 i \ccoeff{2}{RLL} - 
    4 \ccoeff{7 \prime}{LLRL} - \ccoeff{8}{LLLL} - \ccoeff{8}{LLLR} - 
    \ccoeff{8}{LLRR} - \ccoeff{8}{LRLL} - \ccoeff{8}{LRLR} \no \\
 & \ccoeff{8}{LLLL}+ 4 i \ccoeff{6}{} - 2 i \ccoeff{2}{RLL} + 4 \ccoeff{7\prime}{LLLR} + 4 \ccoeff{7 \prime}{LLRL} + 4 \ccoeff{7\prime}{LLRR}+ 2 \ccoeff{7\prime}{LRLR} = 0 \no \\
      & \ccoeff{8 \prime \prime}{LLRL} -2 i \ccoeff{4}{} + 2 I \ccoeff{2}{RLL} - 4 \ccoeff{7}{LLLR} + \ccoeff{8 \prime}{LLLR} \no \\
&\ccoeff{8 \prime \prime}{LLRR}+2 i \ccoeff{4}{} - 4 \ccoeff{7}{LRRL}+ \ccoeff{8 \prime}{LRRL} =0 \no \\
& \ccoeff{8 \prime \prime}{LRLL}-4 \ccoeff{7}{LLLR} + \ccoeff{8\prime}{LLRL}=0 \no \\
 & \ccoeff{8}{LRLR}= 2 \ccoeff{7\prime}{LRLR} \no \\
 &\ccoeff{8 \prime \prime}{LRRL} -4 \ccoeff{7}{LLRR} + \ccoeff{8\prime}{LLRR}  = 0 \no \\
 &\ccoeff{8}{LRLL}= -2 i \ccoeff{4}{} + 4 \ccoeff{7 \prime}{LLLR} - 
    \ccoeff{8}{LLLR} \no \\
& \ccoeff{8 \prime \prime}{LRLR}-4 \ccoeff{7}{LRLR} +\ccoeff{8\prime}{LRLR} =0 \no \\
& 2 \ccoeff{8}{LLLR}= -2 i \ccoeff{2}{RLL} + 4 \ccoeff{7 \prime}{LLLR} \no \\
& \ccoeff{8 \prime}{LRRL}= -8 i \ccoeff{6}{} + 4 \ccoeff{7}{LLLR} + 
    2 \ccoeff{7}{LLRR} + 2 \ccoeff{7}{LRLL} + 2 \ccoeff{7}{LRLR} + 
    2 \ccoeff{7}{LRRL}+ 4 \ccoeff{7 \prime}{LLLR} \no \\
    & \qquad \qquad  - 2 \ccoeff{8}{LLRL} - 
    \ccoeff{8 \prime}{LLLR} - 2 \ccoeff{8 \prime}{LLRL} - \ccoeff{8 \prime}{LLRR}- \ccoeff{8 \prime}{LRLR} \no \\
    & 4 \ccoeff{7 \prime}{LLRL}= -8 i \ccoeff{6}{} + 2 i \ccoeff{2}{RLL} + 4 \ccoeff{7 \prime}{LLLR} \no \\
& 2 \ccoeff{7}{LRLL}= 2 i \ccoeff{6}{} - i \ccoeff{2}{RLL} + 2 \ccoeff{7}{LLLR} \no \\
& \ccoeff{8 \prime}{LLLR}= 2 i \ccoeff{4}{} - i \ccoeff{2}{RLL} + 2 \ccoeff{7}{LLLR} \no \\
& \ccoeff{8 \prime}{LLRL}+ 4 i \ccoeff{6}{} - 2 \ccoeff{7}{LLLR} - 
    2 \ccoeff{7 \prime}{LLLR} + \ccoeff{8}{LLRL} \no \\
& \ccoeff{8 \prime}{LRLR}= 2 \ccoeff{7}{LRLR}\no \\
& \ccoeff{8 \prime}{LLRR}= -2 i \ccoeff{4}{} + 2 I \ccoeff{6}{} + 2 \ccoeff{7}{LLRR} - 2 \ccoeff{7 \prime}{LLRR} + \ccoeff{8}{LLRR}\no \\
 & \ccoeff{8}{LLRL}= -4 i  \ccoeff{6}{} + i \ccoeff{2}{RLL} + 2 \ccoeff{7\prime}{LLLR} \no \\
& \ccoeff{8}{LLRR}= 2 i  \ccoeff{4}{} - i \ccoeff{2}{RLL} + 2 \ccoeff{7\prime}{LLRR} \no \\
& 4 \ccoeff{10}{11}-   i \ccoeff{7 \prime}{LRLR}=0 \no \\
& 4 \ccoeff{10}{8}+ i \ccoeff{7}{LLLR} = 0 \no \\
&  \ccoeff{10}{7}=\ccoeff{10}{13} = - \ccoeff{10}{8} \no  \\
& \ccoeff{10}{1}= \ccoeff{10}{2} = \ccoeff{10}{3} =\ccoeff{10}{4}=\ccoeff{10}{5}=\ccoeff{10}{6} =0 \no \\
& 8 \ccoeff{10}{9} =    \ccoeff{2}{RLL} + 2 i \ccoeff{7 \prime}{LLLR} =0 \no \\
& 4 \ccoeff{10}{10} = 4 \ccoeff{10}{14}  =-  i \ccoeff{7}{LRLR} =0 \no \\
& 4 \ccoeff{10}{16} -  i  \ccoeff{7}{LLLR}=0\no \\
 &4 \ccoeff{10}{17}+ i \ccoeff{7}{LLRR}=0 \no \\
& 4 \ccoeff{10}{12} + \frac12 \ccoeff{2}{RLL} +  i \ccoeff{7 \prime}{LLLR} =0\no  \\
  &  4 \ccoeff{10}{18}= \ccoeff{4}{} - \frac12 \ccoeff{2}{RLL} + i \ccoeff{7}{LRRL} \no \\
 & 4 \ccoeff{10}{15}=  \ccoeff{4}{} - \frac12 \ccoeff{2}{RLL} - i \ccoeff{7 \prime}{LLRR}\no
\end{align}
}
\subsubsection{P-odd}
\label{app:solWZPodd}

{\footnotesize
\begin{align}
&\ccoeff{9\prime}{LLLL}= i \ccoeff{1}{LLL} -\ccoeff{9}{LLLL} \\
& \ccoeff{9\prime}{LRRR}= \ccoeff{9}{LLLR} \no\\
 &\ccoeff{9\prime}{LLLR}= i \ccoeff{1}{RLL} + \ccoeff{9}{LRRR}\no\\
 & \ccoeff{9}{LRRR}= -i \ccoeff{1}{RLL} - \ccoeff{9}{LLRL}\no \\
 &\ccoeff{9\prime}{LLRL}= i \ccoeff{1}{LLR} - \ccoeff{9}{LRLL} \no \\
& \ccoeff{9}{LRRL} =i \ccoeff{1}{LLR}  \no \\
& \ccoeff{9\prime}{LLRR}= -i \ccoeff{1}{LLR} \no \\
& \ccoeff{9\prime}{LRLR}= \ccoeff{9}{LRLR} \no \\
& 2 \ccoeff{9}{LLLL}- i \ccoeff{1}{LLL} = 0 \no \\
& \ccoeff{9}{LLRL}+ i \ccoeff{1}{RLL} + \ccoeff{9}{LLLR} =0\no \\
& \ccoeff{9}{LRLL}+  i \ccoeff{1}{LLR} - \ccoeff{9}{LLLR}=0\no \\
& \ccoeff{1}{RLL}+ 2 \ccoeff{1}{LLR} =0 \no \\
& \ccoeff{9}{LLRR}= 0 \no \\
& \ccoeff{11}{1}=\ccoeff{11}{3}= \ccoeff{11}{7}=\ccoeff{11}{10}=\ccoeff{11}{8}=\ccoeff{11}{2}=\ccoeff{11}{9}= 0 \no \\
& 12 \ccoeff{1}{4} + i \ccoeff{9}{LLLR} = 0\no \\
& 12 \ccoeff{1}{6}+ i \ccoeff{9}{LRLR} = 0\no  \\
 & 12 \ccoeff{1}{5} - i \ccoeff{9}{LLLR}  = 0 \no
\end{align}
}

\newpage 

\section{Heat kernel}
\label{app:heatkernel}

The heat kernel method was pioneered by Schwinger and then developed by De Witt and Seeley. For a lucid review, we refer the reader to Ref.~\cite{DeWitt}.

This method allows us to write the matrix element of ${\slashed{D}}^{-2}$ in position space as
\ba\label{HK}
\langle x|\frac{1}{{\slashed{D}}^2}|y\rangle=i\int dt~\langle x|e^{-i{\slashed{D}}^2t}|y\rangle=i\int_0^\infty dt~H(x,y;t).
\ea
The solution $H(x,y;t)$, which is referred to as the heat kernel, can be calculated perturbatively in the limit $t\to0$. We write the ansatz $H(x,y;t)=H_0(x,y;t)U(x,y;t)$, with the ``free" solution $H_0$ being the solution of \eqref{HK} with ${\slashed{D}}^2$ replaced by $\partial^2$, namely 
\ba
H_0(x,y;t)=\frac{i}{(4\pi it)^{d/2}}e^{i\frac{(x-y)^2}{4t}-\varepsilon t},
\ea
and 
\ba\label{Uxy}
U(x,y;t)=\sum_{n=0,1,2,\cdots}a_n(x,y)(it)^n.
\ea
The heat kernel coefficients $a_n(x,y)$ are smooth in the limit $y\to x$, and satisfy the boundary condition $a_0(x,y)=1$, i.e. $H(x,y;0)=\delta^{(d)}(x-y)$. The parameter $\varepsilon>0$ follows from the $i\varepsilon$ prescription in the Feynman propagator and should not be confused with $\epsilon=(4-d)/2$. 

Employing the above expansion we get 
\ba\label{HKinitial}
{\rm Tr}\left[\alpha_a{T}^a_A\left\{{\slashed{D}},\gamma_5\right\}\frac{1}{{\slashed{D}}}\right]
&=&{\rm Tr}\left[\alpha_a{T}^a_A\left\{{\slashed{D}},\gamma_5\right\}{\slashed{D}}\frac{1}{{\slashed{D}}^2}\right]\\\no
&=&i\int d^dx~\alpha_a(x)\lim_{y\to x}\int_0^\infty dt~{\rm tr}\left[T_A^a\left(\left\{{\slashed{D}},\gamma_5\right\}{\slashed{D}}\right)_yH_0(y,x)U(x,y)\right]\\\no
&=&i\int d^dx~\alpha_a(x)\lim_{y\to x}\int_0^\infty dt~{\rm tr}\left[T_A^a\left(2\gamma^{\hat\mu}\gamma_5\gamma^\nu\right)\partial_{\hat\mu,y}[(\partial_{\nu,y}H_0) U+H_0D_{\nu,y}U]\right]\\\no
&=&i\int d^dx~\alpha_a(x)\lim_{y\to x}\int_0^\infty dt~{\rm tr}\left[T_A^a\left(2\gamma^{\hat\mu}\gamma_5\gamma^\nu\right)(\partial_{\hat\mu,y}\partial_{\nu,y}H_0)U\right]+{\rm Eva}\\\no
&=&-(d-4)\int d^dx~\alpha_a(x)\lim_{y\to x}\int_0^\infty \frac{dt}{t}H_0(x,y)~{\rm tr}\left[T_A^a\gamma_5U(x,y)\right]+{\rm Eva}.
\ea
In the second equality of \eqref{HKinitial} we merely used the definition of heat kernel, and in the third applied the derivatives. In the fourth step we took advantage of the fact that all terms with a single derivative of $H_0$ vanish in the limit ${y\to x}$. The non-vanishing contributions come from the second derivative $\lim_{y\to x}\partial_\mu\partial_\nu H_0=+i g_{\mu\nu}H_0/(2t)$ as well as evanescent terms proportional to $D_{\hat\nu}U$. These latter can be neglected, as explained around Eq.~\eqref{resultDelta1}. The last equality follows from the identity $\gamma^{\hat\mu}\gamma_{\hat\mu}=(d-4)$. Finally, the integral in $dt$ can be performed explicitly for any order $n$ of the perturbative expansion of $U$, and is proportional to $\Gamma(d/2-n)$.\footnote{IR-divergences at large $t$ are cutoff by the factor $e^{-\varepsilon t}$ in $H_0(x,y)$.} There is a unique contribution that survives the $d\to4$ limit. This emerges from an UV divergence $t\to0$ that results in a factor $\Gamma(d/2-n)\sim1/(d/2-n)$. The latter can exactly compensate the $(d-4)$ in front of the integral, and thus lead to a non-trivial result, only at the $n=2$ order of the expansion \eqref{Uxy}. The existence of a compensation between the evanescent $(d-4)$ factor and UV divergent Feynman integrals is typical of quantum anomalies. The result is 
\ba\label{4dfermionAnomaly1}
\lim_{d\to4}{\rm Tr}\left[\alpha_a{T}^a_A\left\{{\slashed{D}},\gamma_5\right\}\frac{1}{{\slashed{D}}}\right]
&=&-\frac{i}{8\pi^2}\int d^dx~\alpha_a(x)~{\rm tr}\left[T_A^a\gamma_5a_2(x,x)\right],
\ea
where the $4$-dimensional limit is formally defined such that $\lim_{d\to4}{\rm Eva}=0$ for all evanescent operators.

The $4-$dimensional limit of the heat-kernel coefficients $a_n(x,x)$ can be obtained recursively. We first observe that ${\slashed{D}}^2={\cal D}_\mu{\cal D}^\mu+X$, with ${\cal D}_\mu=\partial_\mu+iP_\mu$, and where $P_\mu, X$ explicitly read
\ba
P^\mu&=&{\cal V}^{\mu}+\frac{1}{2}[\gamma^{\mu},\gamma^{\nu}]\gamma_5{\cal A}_{\nu}\\\no
X
&=&
\frac{i}{4}[\gamma^{\mu},\gamma^{\nu}]\left({\cal V}_{\mu\nu}+\gamma_5{\cal A}_{\mu\nu}\right)+
i\gamma_5\gamma^{\mu}\gamma^{\nu}\left(\partial_\nu {\cal A}_{\mu}+i[{\cal V}_{\nu},{\cal A}_{\mu}]\right)
-2{\cal A}_\alpha{\cal A}^\alpha.
\ea
The field strengths of the vector and axial components were previously introduced in \eqref{AVfs}. Now, the heat kernel, defined in \eqref{HK}, satisfies $i\frac{d}{dt}H(x,y;t)={\slashed{D}}^2_xH(x,y;t)$. Inserting the ansatz $H=H_0U$ this becomes $idU/dt=-i(x-y)_\mu{\cal D}^\mu U/t+[{\cal D}^2+X]U$. Equating order by order in $t^n$ gives the recursive relations 
\ba\label{recursion12}
&&(x-y)_\mu{\cal{D}}^\mu_xa_{0}(x,y)=0,~~~~~~\\\no
&&[n+1+(x-y)_\mu{\cal{D}}^\mu_x]a_{n+1}(x,y)=-[{\cal D}_x^2+X]a_n(x,y)~~~~~~~(n>0).
\ea
The first definition, along with $\lim_{y\to x}a_0(x,y)=1$ defines $a_0(x,y)$. We are interested in $a_2(x,x)=-\frac{1}{2}\lim_{y\to x}[{\cal{D}}^2_x+X]a_1(x,y)$, but to find its explicit expression we need $a_1(x,y)$ and its second derivative:
\ba
&&\lim_{y\to x}a_1(x,y)=-\lim_{y\to x}[{\cal D}^2_x+X]a_0(x,y)\\\no
&&\lim_{y\to x}{\cal D}^2_xa_1(x,y)=-\frac{1}{3}\lim_{y\to x}{\cal D}^2_x[{\cal D}^2_x+X]a_0(x,y)
\ea
The first relation follows directly from the second equation in \eqref{recursion12}. Differentiating twice the same relation with $n=0$ with respect to ${\cal D}$ we obtain the other one. Similarly, differentiating the first relation in \eqref{recursion12} we derive $\lim_{y\to x}{\cal D}_\alpha a_0(x,y)=\lim_{y\to x}{\cal D}^2 a_0(x,y)=0$. This leads us to
\ba\label{a2General}
a_2(x,x)&=&\lim_{y\to x}\frac{1}{6}{\cal D}_x^2{\cal D}_x^2a_0(x,y)+\frac{1}{6}{\cal D}^2X+\frac{1}{2}X^2\\\no
&=&\frac{1}{12}[{\cal D}_\mu,{\cal D}_\nu][{\cal D}^\mu,{\cal D}^\nu]+\frac{1}{6}{\cal D}^2X+\frac{1}{2}X^2\\\no
&=&-\frac{1}{12}P_{\mu\nu}P^{\mu\nu}+\frac{1}{6}{\cal D}^2X+\frac{1}{2}X^2,
\ea
where $P_{\mu\nu}=\partial_{\mu}P_{\nu}-\partial_{\nu}P_{\mu}+i[P_{\mu},P_{\nu}]$. In evaluating $a_2$ we used the linearity of the derivative, namely ${\cal D}_\mu[Xa_0]=[{\cal D}_\mu X]a_0+X[{\cal D}_\mu a_0]$. The relation ${\cal D}_x^2{\cal D}_x^2a_0(x,y)=\frac{1}{2}[{\cal D}_\mu,{\cal D}_\nu][{\cal D}^\mu,{\cal D}^\nu]a_0(x,y)$ is proven differentiating four times the first equation in \eqref{recursion12}, and contracting with the metric tensor. Because $[{\cal D}_\mu,{\cal D}_\nu]$ is not a differential operator, the limit ${y\to x}$ can be performed trivially and \eqref{a2General} follows.

\end{document}